\documentclass{lmcs} 
\pdfoutput=1

\usepackage[utf8]{inputenc}

\usepackage{lastpage}
\lmcsdoi{17}{4}{1}
\lmcsheading{}{\pageref{LastPage}}{}{}%
{Dec.~21,~2020}{Oct.~01,~2021}{}

\usepackage{xspace}
\usepackage{amsmath,amsthm}

\usepackage{mathpartir}
\usepackage{amssymb}
\usepackage{stmaryrd}
\usepackage{mathtools}
\usepackage{xcolor}
\usepackage{float}
\usepackage{graphicx}
\newenvironment{mdframed}{}{}

\usepackage{caption}
\usepackage{hyperref}

\hypersetup{%
  colorlinks=true, final, linkcolor = red,
            urlcolor  = blue,
            citecolor = blue
  }


\theoremstyle{plain} 

\begin{document}

\title[Causal Consistency for Reversible Multiparty Protocols]{Causal Consistency for Reversible Multiparty Protocols}

\author[C.A.~Mezzina]{Claudio Antares Mezzina}	
\address{Dipartimento di Scienze Pure e Applicate (DiSPeA), Universit\`a di Urbino, Italy
}	
\email{claudio.mezzina@uniurb.it}  

\author[J.A.~P\'erez]{Jorge A. P\'erez}	
\address{University of Groningen \& CWI, Amsterdam, The Netherlands}	
\email{j.a.perez@rug.nl}  


\newtheorem{innercustomlem}[thm]{Lemma}
\newenvironment{lm}[1]
  {\renewcommand\theinnercustomlem{#1}\innercustomlem}
  {\endinnercustomlem}
\newtheorem{innercustomtm}[thm]{Theorem}
\newenvironment{tm}[1]
  {\renewcommand\theinnercustomlem{#1}\innercustomtm}
  {\endinnercustomlem}

\theoremstyle{definition}
\newtheorem{remark}[thm]{Remark}
\newtheorem{example}[thm]{Example}




\begin{abstract}
In programming models with a \emph{reversible semantics}, computational steps can be undone. 
This paper addresses the integration of reversible semantics into 
process languages for communication-centric systems equipped with \emph{behavioral types}. 
In prior work, we introduced a \emph{monitors-as-memories} approach to seamlessly integrate
 reversible semantics into a process model in which concurrency is governed by \emph{session types} (a class of behavioral types), 
  covering binary (two-party) protocols with syn\-chro\-nous communication.
  The applicability and expressiveness of the binary setting, however, is limited. 
Here we extend our approach, and use it to define reversible semantics for 
an  expressive process model that accounts for 
\emph{multiparty} ($n$-party)  protocols,
asynchronous communication,
 decoupled rollbacks,
 and abstraction passing.
As main result, we prove that our  reversible semantics for multiparty protocols is \emph{causally-consistent}.
A key technical ingredient in our developments is an alternative reversible semantics with \emph{atomic rollbacks}, which is conceptually simple and is shown to characterize decoupled rollbacks. 
\end{abstract}

\maketitle





%
\definecolor[named]{ABlue}{cmyk}{1,0.1,0,0.1}
\definecolor[named]{ARed}{cmyk}{0,0.90,0.86,0}
\definecolor[named]{APurple}{cmyk}{0.55,1,0,0.15}
\definecolor[named]{AOrange}{cmyk}{0,0.42,1,0.01}

\newcommand{\semicolon}{:}
\newcommand{\lrangle}[1]{\langle #1 \rangle}
\newcommand{\blrangle}[1]{\big\langle #1 \big\rangle}

\newcommand{\parenthtext}[1]{(\textrm{\small #1})}
\newcommand{\brtext}[1]{[\textrm{\small #1}]}
\newcommand{\textinmath}[1]{\textrm{#1}}
\newcommand{\srule}[1]{\parenthtext{#1}}
\newcommand{\strule}[1]{\textrm{#1}}
\newcommand{\stypes}[1]{{\footnotesize \parenthtext{#1}}}
\newcommand{\ltsrule}[1]{{\footnotesize \lrangle{\textrm{#1}}}}
\newcommand{\eltsrule}[1]{{\footnotesize [\textrm{#1}]}}
\newcommand{\trule}[1]{{\footnotesize\brtext{#1}}}
\newcommand{\orule}[1]{{\scriptsize{\brtext{#1}}}}
\newcommand{\mrule}[1]{{\footnotesize{\parenthtext{#1}}}}

\newcommand{\iftag}{{\textrm{if }}}

\newcommand{\noi}{\noindent}
\newcommand{\Hline}{\rule{\linewidth}{.5pt}}
\newcommand{\Hlinefig}{\rule{\linewidth}{.5pt}\vspace{-4mm}}
\newcommand{\myparagraph}[1]{\noindent{\textbf{#1}\ }}
\newcommand{\jparagraph}[1]{\paragraph{\textbf{#1}}}



\newcommand{\bnfis}{\;\;::=\;\;}
\def\bnfbar{\; \mbox{\large{$\parallel$}}\;}
\def\sbnfbar{\;\mbox{\Large{$\mid$}}\;}

\newcommand{\Case}[1]{\noi {\bf Case: }#1\\}
\newcommand{\proofend}{\qed}
\newcommand{\Proof}{\noi {\bf Proof: }}

\newcommand{\LogAnd}{\texttt{ and }}
\newcommand{\LogOr}{\texttt{ or }}


\newcommand{\basic}{\noi {\bf Basic Step:}\\}
\newcommand{\inductive}{\noi {\bf Inductive Hypothesis:}\\}
\newcommand{\induction}{\noi {\bf Induction Step:}\\}


\newcommand{\tree}[2]{
\ensuremath{\displaystyle
		\frac
		{
			#1
		}{
			#2
		}
	}
}

\newcommand{\treeusing}[3]{
\begin{prooftree}
	#1
	\justifies
	#2
	\using
	#3
\end{prooftree}}

\newcommand{\vect}[1]{\tilde{#1}}
\newcommand{\mytilde}[1]{\widetilde{#1}}

\newcommand{\set}[1]{\{#1\}}
\newcommand{\es}{\emptyset}
\newcommand{\maxset}[1]{\max(#1)}
\newcommand{\setbar}{\ \ |\ \ }
\newcommand{\tuple}[2]{(#1, #2)}
\newcommand{\suchthat}{\cdot}
\newcommand{\powerset}[1]{\mathcal{P}(#1)}
\newcommand{\product}{\times}

\newcommand{\eval}{\downarrow}

\newcommand{\setsubtr}[2]{#1 \backslash #2}

\newcommand{\func}[2]{#1(#2)}
\newcommand{\dom}[1]{\mathtt{dom}(#1)}
\newcommand{\codom}[1]{\mathtt{codom}(#1)}

\newcommand{\funcbr}[2]{#1\lrangle{#2}}

\newcommand{\entails}{\text{implies}}


\newcommand{\freev}[1]{\lrangle{#1}}
\newcommand{\boundv}[1]{(#1)}

\newcommand{\send}[1]{\overline{#1}}
\newcommand{\ol}[1]{\overline{#1}}
\newcommand{\receive}[1]{#1.}
\newcommand{\inact}{\mathbf{0}}
\newcommand{\If}{\sessionfont{if}\ }
\newcommand{\Then}{\sessionfont{then}\ }
\newcommand{\Else}{\sessionfont{else}\ }
\newcommand{\ifthen}[2]{\If #1\ \Then #2\ }
\newcommand{\Par}{\;|\;}
\newcommand{\news}[1]{(\nu\, #1)}
\newcommand{\newsp}[2]{(\nu\, #1)(#2)}
\newcommand{\varp}[1]{#1}
\newcommand{\rvar}[1]{#1}
\newcommand{\recp}[2]{\mu \rvar{#1}. #2}

\newcommand{\Def}{\sessionfont{def}\ }

\newcommand{\defeq}{\stackrel{\Def}{=}}

\newcommand{\repl}{\ast\,}
\newcommand{\parcomp}[2]{\prod_{#1}{#2}}

\newcommand{\bn}[1]{\mathtt{bn}(#1)}
\newcommand{\fn}[1]{\mathtt{fn}(#1)}
\newcommand{\sn}[1]{\mathtt{sn}(#1)}
\newcommand{\ofn}[1]{\mathsf{ofn}(#1)}
\newcommand{\fv}[1]{\mathtt{fv}(#1)}
\newcommand{\bv}[1]{\mathtt{bv}(#1)}
\newcommand{\fs}[1]{\mathtt{fs}(#1)}
\newcommand{\fpv}[1]{\mathtt{fpv}(#1)}
\newcommand{\nam}[1]{\mathtt{n}(#1)}

\newcommand{\subj}[1]{\mathtt{subj}(#1)}
\newcommand{\obj}[1]{\mathtt{obj}(#1)}

\newcommand{\relfont}[1]{\mathcal{#1}}
\newcommand{\rel}[3]{#1\ \relfont{#2}\ #3}

\newcommand{\scong}{\equiv}
\newcommand{\acong}{\scong_{\alpha}}
\newcommand{\wb}{\approx}
\newcommand{\fwb}{\approx^\mathtt{C}}
\newcommand{\hwb}{\approx^\mathtt{H}}
\newcommand{\swb}{\approx^{s}}
\newcommand{\wbc}{\approx}
\newcommand{\WB}{\approx}

\newcommand{\red}{\longrightarrow}
\newcommand{\Red}{\rightarrow\!\!\!\!\!\rightarrow}
\newcommand{\Redleft}{\leftarrow\!\!\!\!\!\leftarrow}

\def\subst#1#2{\{\raisebox{.5ex}{\small$#1$}\! / \mbox{\small$#2$}\}}

\newcommand{\hole}{-}
\newcommand{\context}[2]{#1[#2]}
\newcommand{\Ccontext}[1]{\C[#1]}

\newcommand{\Econtext}[1]{\E[#1]}

\newcommand{\barb}[1]{\downharpoonright_{#1}}
\newcommand{\Barb}[1]{\Downarrow_{#1}}
\newcommand{\nbarb}[1]{\not\downarrow_{#1}}
\newcommand{\nBarb}[1]{\not\Downarrow_{#1}}

\newcommand{\ESP}{\text{ESP}}
\newcommand{\ESPsel}{\ESP^+}


\newcommand{\sessionfont}[1]{\mathtt{#1}}
\newcommand{\vart}[1]{\mathsf{#1}}

\newcommand{\ssep}{;}
\newcommand{\shsep}{.}
\newcommand{\outses}{!}
\newcommand{\inpses}{?}
\newcommand{\selses}{\triangleleft}
\newcommand{\brases}{\triangleright}
\newcommand{\dual}[1]{\overline{#1}}
\newcommand{\cat}{\cdot}

\newcommand{\allstypes}{\mathcal{S}}


\newcommand{\bacc}[2]{#1 \boundv{#2} \shsep}
\newcommand{\breq}[2]{\send{#1} \freev{#2} \shsep}
\newcommand{\bareq}[2]{\send{#1} \freev{#2}}

\newcommand{\breqt}[3]{\send{#1} \boundv{#2:#3} \shsep}
\newcommand{\bacct}[3]{#1 \boundv{#2:#3} \shsep}

\newcommand{\bout}[2]{#1 \outses \freev{#2} \shsep}
\newcommand{\bbout}[2]{#1 \outses \blrangle{#2} \shsep}
\newcommand{\binp}[2]{#1 \inpses \boundv{#2} \shsep}
\newcommand{\bselold}[2]{#1 \selses #2 \shsep}
\newcommand{\bsel}[2]{#1 \selses \set{#2}}

\newcommand{\bbra}[2]{#1 \brases \set{#2}}
\newcommand{\bbras}[2]{#1 \brases #2}
\newcommand{\bbraP}[1]{#1 \brases \lPi}


\newcommand{\role}[1]{[#1]}

\newcommand{\srole}[2]{#1\role{#2}}
\newcommand{\sqrole}[2]{#1^{[]}\role{#2}}

\newcommand{\fromto}[2]{\role{#1} \role{#2}}
\newcommand{\sfromto}[3]{#1\fromto{#2}{#3}}

\newcommand{\sout}[3]{\srole{#1}{#2} \outses \freev{#3} \ssep}
\newcommand{\sinp}[3]{\srole{#1}{#2} \inpses \boundv{#3} \ssep}
\newcommand{\sdel}[4]{\srole{#1}{#2} \outses \freev{\srole{#3}{#4}} \ssep}
\newcommand{\ssel}[3]{\srole{#1}{#2} \selses #3 \ssep}
\newcommand{\sbra}[3]{\srole{#1}{#2} \brases \set{#3}}
\newcommand{\sbras}[3]{\srole{#1}{#2} \brases #3}
\newcommand{\sbraP}[2]{\srole{#1}{#2} \brases \lPi}

\newcommand{\acc}[3]{#1 \role{#2} \boundv{#3} \shsep}
\newcommand{\req}[3]{\send{#1} \role{#2} \boundv{#3} \shsep}
\newcommand{\areq}[3]{\send{#1} \role{#2} \freev{#3}}

\newcommand{\out}[4]{\sfromto{#1}{#2}{#3} \outses \freev{#4} \ssep}
\newcommand{\inp}[4]{\sfromto{#1}{#2}{#3} \inpses \boundv{#4} \ssep}
\newcommand{\del}[5]{\sfromto{#1}{#2}{#3} \outses \freev{\srole{#4}{#5}} \ssep}
\newcommand{\sel}[4]{\sfromto{#1}{#2}{#3} \selses #4 \ssep}
\newcommand{\bra}[4]{\sfromto{#1}{#2}{#3} \brases \set{#4}}
\newcommand{\bras}[4]{\sfromto{#1}{#2}{#3} \brases #4}
\newcommand{\braP}[3]{\sfromto{#1}{#2}{#3} \brases \lPi}

\newcommand{\arrivetext}{\mathtt{arrive}}
\newcommand{\arrive}[1]{\arrivetext\ #1}
\newcommand{\arrivem}[2]{\arrivetext\ #1\ #2}

\newcommand{\typecasetext}{\mathtt{typecase}}
\newcommand{\oftext}{\mathtt{of}}
\newcommand{\typecase}[2]{\typecasetext\ #1\ \oftext\ \set{#2}}

\newcommand{\inputsym}{\mathtt{i}}
\newcommand{\outputsym}{\mathtt{o}}


\newcommand{\typingred}{\red}
\newcommand{\typingRed}{\Red}

\newcommand{\wellconf}[1]{\mathtt{wc}(#1)}
\newcommand{\cohses}[2]{\mathtt{co}(#1(#2))}
\newcommand{\coherent}[1]{\mathtt{co}(#1)}
\newcommand{\fcoherent}[1]{\mathtt{fco}(#1)}




\newcommand{\emp}{\epsilon}
\newcommand{\squeue}[3]{\srole{#1}{#2}:#3}
\newcommand{\srqueue}[4]{\srole{#1}{#2}[\inputsym: #3, \outputsym: #4]}
\newcommand{\srqueuei}[3]{\srole{#1}{#2}[\inputsym: #3]}
\newcommand{\srqueueo}[3]{\srole{#1}{#2}[\outputsym: #3]}
\newcommand{\srqueueio}[4]{\srole{#1}{#2}[\inputsym: #3, \outputsym: #4]}
\newcommand{\sgqueue}[2]{\srole{#1}:#2}

\newcommand{\shqueue}[2]{#1[#2]}

\newcommand{\squeueio}[3]{#1[\inputsym: #2, \outputsym: #3]}
\newcommand{\squeuei}[2]{#1[\inputsym: #2]}
\newcommand{\squeueo}[2]{#1[\outputsym: #2]}

\newcommand{\squeuetio}[4]{#1[#2, \inputsym: #3, \outputsym: #4]}
\newcommand{\squeueto}[3]{#1[#2, \outputsym: #3]}
\newcommand{\squeueti}[3]{#1[#2, \inputsym: #3]}
\newcommand{\squeuet}[2]{#1[#2]}


\newcommand{\queuev}[2]{\role{#1}(#2)}
\newcommand{\queuel}[2]{\role{#1} #2}
\newcommand{\queues}[3]{\role{#1}(\srole{#2}{#3})}


\newcommand{\impl}[2]{\{\!\{#1\}\!\}_{\mathcal{#2}}}
\newcommand{\impll}[2]{\{\!\{#1\}\!\}_{\mathcal{#2}}^\fw}

\newcommand{\gtfont}[1]{\mathtt{#1}}
\newcommand{\gsep}{.}

\newcommand{\globaltype}[1]{\lrangle{#1}}
\newcommand{\parties}[1]{\mathtt{\pa}(#1)}
\newcommand{\roles}[1]{\mathtt{roles}(#1)}

\newcommand{\fromtogt}[2]{#1 \rightarrow #2 \semicolon}

\newcommand{\valuegt}[3]{\fromtogt{#1}{#2} \lrangle{#3} \gsep}
\newcommand{\selgt}[3]{\fromtogt{#1}{#2} \set{#3}}
\newcommand{\selgtG}[2]{\fromtogt{#1}{#2} \lGi}
\newcommand{\recgt}[2]{\mu \vart{#1}. #2}
\newcommand{\vargt}[1]{\vart{#1}}
\newcommand{\inactgt}{\gtfont{end}}

\newcommand{\swap}{\ensuremath{\approx_{\mathtt{sw}}}\xspace}


\newcommand{\projset}[1]{\mathtt{proj}(#1)}
\newcommand{\aprojset}[1]{\mathtt{aproj}\ #1 }
\newcommand{\gcong}{\equiv}
\newcommand{\govcong}{\cong_g}
\newcommand{\gperm}{\simeq}


\newcommand{\projsymb}{\lceil}
\newcommand{\proj}[2]{#1 \projsymb #2}


\newcommand{\tfont}[1]{\mathtt{#1}}
\newcommand{\tsep}{;}

\newcommand{\chtype}[1]{\lrangle{#1}}
\newcommand{\chtypei}[1]{\inputsym \lrangle{#1}}
\newcommand{\chtypeo}[1]{\outputsym \lrangle{#1}}
\newcommand{\chtypeio}[1]{\inputsym \outputsym \lrangle{#1}}

\newcommand{\outtype}{\outses}
\newcommand{\inptype}{\inpses}
\newcommand{\seltype}{\selses}
\newcommand{\bratype}{\brases}

\newcommand{\trec}[2]{\mu\vart{#1}.#2}
\newcommand{\tvar}[1]{\vart{#1}}
\newcommand{\tset}[1]{\set{#1}}
\newcommand{\tinact}{\tfont{end}}

\newcommand{\sminus}[1]{#1^{\text{--}}}

\newcommand{\subt}{\leq}
\newcommand{\supt}{\geq}

\newcommand{\tout}[2]{\role{#1} \outtype \lrangle{#2} \tsep}
\newcommand{\tinp}[2]{\role{#1} \inptype (#2) \tsep}
\newcommand{\tsel}[2]{\role{#1} \seltype \set{#2}}
\newcommand{\tsels}[2]{\role{#1} \seltype #2}
\newcommand{\tselT}[1]{\role{#1} \seltype \lTi}
\newcommand{\tbra}[2]{\role{#1} \bratype \set{#2}}
\newcommand{\tbras}[2]{\role{#1} \bratype #2}
\newcommand{\tbraT}[1]{\role{#1} \bratype \lTi}

\newcommand{\btout}[1]{\outtype \lrangle{#1} \tsep}
\newcommand{\bbtout}[1]{\outtype \big\langle{#1}\big\rangle \tsep}
\newcommand{\btinp}[1]{\inptype (#1) \tsep}
\newcommand{\bbtinp}[1]{\inptype \big({#1}\big) \tsep}
\newcommand{\btsel}[1]{\oplus \set{#1}}
\newcommand{\btselS}{\oplus \lSi}
\newcommand{\btbra}[1]{\& \set{#1}}
\newcommand{\btbraS}{\& \lSi}


\newcommand{\mtout}[2]{\role{#1} \outtype \lrangle{#2}}
\newcommand{\mtinp}[2]{\role{#1} \inptype (#2)}
\newcommand{\mtsel}[2]{\role{#1} \seltype #2}
\newcommand{\mtbra}[2]{\role{#1} \bratype #2}


\newcommand{\bmtout}[1]{\outtype \lrangle{#1}}
\newcommand{\bmtinp}[1]{\inptype (#1)}
\newcommand{\bmtsel}[1]{\seltype #1}
\newcommand{\bmtbra}[1]{\bratype #1}

\newcommand{\mcat}{\;*\;}
\newcommand{\icat}{\;\circ\;}


\newcommand{\Ga}{\Gamma}
\newcommand{\De}{\Delta}
\newcommand{\proves}{\vdash}
\newcommand{\hastype}{\triangleright}

\newcommand{\Decat}[1]{\De \cat #1}
\newcommand{\Gacat}[1]{\Ga \cat #1}

\newcommand{\tcat}{\circ}

\newcommand{\typed}[1]{#1:}
\newcommand{\typedrole}[2]{\typed{\srole{#1}{#2}}}

\newcommand{\typedprocess}[3]{#1 \proves #2 \hastype #3}

\newcommand{\Eproves}[3]{#1 \proves \typed{#2} #3}
\newcommand{\Gproves}[2]{\Eproves{\Ga}{#1}{#2}}

\newcommand{\tprocess}[3]{#1 \proves #2 \hastype #3}
\newcommand{\Gtprocess}[2]{\tprocess{\Ga}{#1}{#2}}
\newcommand{\Gptprocess}[2]{\tprocess{\Ga'}{#1}{#2}}

\newcommand{\noGtprocess}[2]{#1 \hastype #2}


\newcommand{\globalenv}[1]{\set{#1}}
\newcommand{\globalenvI}{E}
\newcommand{\globalenvJ}{\set{\typed{s_j} \G_j}_{j \in J}}

\newcommand{\Gltprocess}[4]{\tprocess{#1}{#2}{#3, #4}}

\newcommand{\geI}{\globalenvI}
\newcommand{\geJ}{\globalenvJ}

\newcommand{\Stprocess}[3]{\tprocess{\geI, #1}{#2}{#3}}
\newcommand{\SGtprocess}[2]{\Gtprocess{#1}{#2, \globalenvI}}
\newcommand{\SJGtprocess}[2]{\globalenvJ, \Gtprocess{#1}{#2}}

\newcommand{\Observer}[2]{\mathsf{Observer}(#1, #2)}
\newcommand{\ObserverG}[1]{\mathsf{Observer}(\globalenvI, #1)}

\newcommand{\Obs}{\ensuremath{\mathsf{Obs}}}


\newcommand{\fromtolts}[2]{\fromto{#1}{#2}}

\newcommand{\outlts}{\outses}
\newcommand{\inplts}{\inpses}
\newcommand{\sellts}{\oplus}
\newcommand{\bralts}{\&}

\newcommand{\actreq}[3]{\send{#1} \role{#2} \boundv{#3}}

\newcommand{\actreqs}[3]{\send{#1} \role{\set{#2}} \boundv{#3}}

\newcommand{\actacc}[3]{#1 \role{#2} \boundv{#3}}
\newcommand{\actaccs}[3]{#1 \role{\set{#2}} \boundv{#3}}

\newcommand{\actout}[4]{#1 \fromtolts{#2}{#3} \outlts \freev{#4}}
\newcommand{\actqout}[4]{#1^{[]} \fromtolts{#2}{#3} \outlts \freev{#4}}
\newcommand{\actbout}[4]{#1 \fromtolts{#2}{#3} \outlts \boundv{#4}}

\newcommand{\actdel}[5]{#1 \fromtolts{#2}{#3} \outlts \freev{\srole{#4}{#5}}}
\newcommand{\actbdel}[5]{#1 \fromtolts{#2}{#3} \outlts \boundv{\srole{#4}{#5}}}
\newcommand{\actqdel}[5]{#1^{[]} \fromtolts{#2}{#3} \outlts \boundv{\srole{#4}{#5}}}

\newcommand{\actinp}[4]{#1 \fromtolts{#2}{#3} \inplts \freev{#4}}
\newcommand{\actqinp}[4]{#1^{[]} \fromtolts{#2}{#3} \inplts \freev{#4}}

\newcommand{\actsel}[4]{#1 \fromtolts{#2}{#3} \sellts #4}
\newcommand{\actqsel}[4]{#1^{[]} \fromtolts{#2}{#3} \sellts #4}

\newcommand{\actbra}[4]{#1 \fromtolts{#2}{#3} \bralts #4}
\newcommand{\actqbra}[4]{#1^{[]} \fromtolts{#2}{#3} \bralts #4}

\newcommand{\actval}[4]{#1: #2 \rightarrow #3:#4}
\newcommand{\actgsel}[4]{#1: #2 \rightarrow #3:#4}

\newcommand{\bactreq}[2]{\send{#1} \freev{#2}}
\newcommand{\bactbreq}[2]{\send{#1} \boundv{#2}}
\newcommand{\bactacc}[2]{#1 \freev{#2}}

\newcommand{\bactout}[2]{#1 \outlts \freev{#2}}
\newcommand{\bactbout}[2]{#1\outlts \boundv{#2}}
\newcommand{\bactinp}[2]{#1 \inplts \freev{#2}}
\newcommand{\bactsel}[2]{#1 \sellts #2}
\newcommand{\bactbra}[2]{#1 \bralts #2}

\newcommand{\By}[1]{\stackrel{#1}{\Longrightarrow}}

\newcommand{\hby}[1]{\stackrel{#1}{\longmapsto}}
\newcommand{\Hby}[1]{\stackrel{#1}{\Longmapsto}}

\newcommand{\barbreq}[1]{\barb{#1}}
\newcommand{\barbout}[3]{\barb{\sfromto{#1}{#2}{#3}}}

\newcommand{\Barbreq}[2]{\Barb{\send{#1}\role{#2}}}
\newcommand{\Barbout}[3]{\Barb{\sfromto{#1}{#2}{#3}}}

\newcommand{\bbarbreq}[1]{\barb{#1}}
\newcommand{\bbarbout}[1]{\barb{#1}}

\newcommand{\bBarbreq}[1]{\Barb{\send{#1}}}
\newcommand{\bBarbout}[1]{\Barb{#1\outses}}

\newcommand{\comp}{\asymp}
\newcommand{\coh}{\asymp}
\newcommand{\bistyp}{\rightleftharpoons}

\newcommand{\typingbeh}{\leftrightarrow}

\newcommand{\bufrel}{\succ}

\newcommand{\ordercup}{\bowtie}


\newcommand{\envtrans}[1]{\by{#1}}
\newcommand{\envTrans}[1]{\by{#1}}
\newcommand{\typedtrans}[1]{\by{#1}}
\newcommand{\typedTrans}[1]{\By{#1}}
\newcommand{\typedred}{\red}
\newcommand{\typedRed}{\Red}

\newcommand{\benv}[2]{(#1, #2)}
\newcommand{\bGenv}[1]{\benv{\Ga}{#1}}
\newcommand{\bGDenv}{\envtyp{\Ga}{\De}}

\newcommand{\envby}[5]{\benv{#1}{#2} \envtrans{#3} \benv{#4}{#5}}
\newcommand{\Genvby}[3]{\envby{\Ga}{#1}{#2}{\Ga}{#3}}


\newcommand{\env}[3]{(#1, #2, #3)}
\newcommand{\Genv}[2]{\env{\globalenvI}{#1}{#2}}
\newcommand{\GGenv}[1]{\env{\globalenvI}{\Ga}{#1}}
\newcommand{\GGDenv}{\env{\globalenvI}{\Ga}{\De}}


\newcommand{\ftby}[7]{\tprocess{#1}{#2}{#3} \typedtrans{#4} \tprocess{#5}{#6}{#7}}
\newcommand{\ftBy}[7]{\tprocess{#1}{#2}{#3} \typedTrans{#4} \tprocess{#5}{#6}{#7}}

\newcommand{\tpby}[6]{\tprocess{#1}{#2}{#3} \typedtrans{#4} \noGtprocess{#5}{#6}}
\newcommand{\tpBy}[6]{\tprocess{#1}{#2}{#3} \typedTrans{#4} \noGtprocess{#5}{#6}}
\newcommand{\Gtpby}[5]{\Gtprocess{#1}{#2} \typedtrans{#3} \noGtprocess{#4}{#5}}
\newcommand{\GtpBy}[5]{\Gtprocess{#1}{#2} \typedTrans{#3} \noGtprocess{#4}{#5}}

\newcommand{\GGtpby}[5]{\tprocess{\globalenvI, \Ga}{#1}{#2} \typedtrans{#3} \noGtprocess{#4}{#5}}
\newcommand{\GGtpBy}[5]{\tprocess{\globalenvI, \Ga}{#1}{#2} \typedTrans{#3} \noGtprocess{#4}{#5}}


\newcommand{\ftpred}[6]{\tprocess{#1}{#2}{#3} \typedred \tprocess{#4}{#5}{#6}}
\newcommand{\ftpRed}[6]{\tprocess{#1}{#2}{#3} \typedRed \tprocess{#4}{#5}{#6}}

\newcommand{\tpred}[5]{\tprocess{#1}{#2}{#3} \typedred \noGtprocess{#4}{#5}}
\newcommand{\tpRed}[5]{\tprocess{#1}{#2}{#3} \typedRed \noGtprocess{#4}{#5}}
\newcommand{\Gtpred}[4]{\Gtprocess{#1}{#2} \typedred \noGtprocess{#3}{#4}}
\newcommand{\GtpRed}[4]{\Gtprocess{#1}{#2} \typedRed \noGtprocess{#3}{#4}}


\newcommand{\obsred}{\red_{obs}}
\newcommand{\obsRed}{\Red_{obs}}


\newcommand{\fulltrel}[7]{\rel{\typedprocess{#1}{#2}{#3}}{#4}{\typedprocess{#5}{#6}{#7}}}
\newcommand{\treld}[6]{\rel{\typedprocess{#1}{#2}{#3}}{#4}{\noGtypedprocess{#5}{#6}}}
\newcommand{\trel}[5]{\rel{#1 \proves #2}{#3}{\noGtypedprocess{#4}{#5}}}

\newcommand{\tcong}{\cong}
\newcommand{\twb}{\approx}
\newcommand{\govwb}{\approx_g}
\newcommand{\tequiv}{\approx}


\newcommand{\confpair}[2]{(#1, #2)}
\newcommand{\uptoconfpair}[2]{[#1, #2]}


\newcommand{\sesstrans}[1]{\stackrel{#1}{\longrightarrow_{s}}}
\newcommand{\sessTrans}[1]{\stackrel{#1}{\Longrightarrow_{s}}}

\newcommand{\fulltypedsesstrans}[7]{\typedprocess{#1}{#2}{#3} \sesstrans{#4} \typedprocess{#5}{#6}{#7}}
\newcommand{\fulltypedsessTrans}[7]{\typedprocess{#1}{#2}{#3} \sessTrans{#4} \typedprocess{#5}{#6}{#7}}

\newcommand{\typedsesstrans}[6]{\typedprocess{#1}{#2}{#3} \sesstrans{#4} \noGtypedprocess{#5}{#6}}
\newcommand{\typedsessTrans}[6]{\typedprocess{#1}{#2}{#3} \sessTrans{#4} \noGtypedprocess{#5}{#6}}
\newcommand{\Gtypedsesstrans}[5]{\Gtypedprocess{#1}{#2} \sesstrans{#3} \noGtypedprocess{#4}{#5}}
\newcommand{\GtypedsessTrans}[5]{\Gtypedprocess{#1}{#2} \sessTrans{#3} \noGtypedprocess{#4}{#5}}

\newcommand{\confact}[2]{#1 \lfloor #2}


\newcommand{\map}[1]{[\!\![#1]\!\!]}
\newcommand{\umap}[1]{[\!\![#1]\!\!]^u}
\newcommand{\pmap}[2]{\ensuremath{[\!\![#1]\!\!]^#2}}
\newcommand{\pmapp}[3]{\ensuremath{[\!\![#1]\!\!]^#2_#3}}
\newcommand{\auxmap}[2]{\ensuremath{\{\!\{#1\}\!\}^#2}}
\newcommand{\tauxmap}[2]{\ensuremath{\{\!|#1|\!\}^#2}}
\newcommand{\auxmapp}[3]{\ensuremath{\big\lfloor\!\!\big\lfloor#1\big\rfloor\!\!\big\rfloor^#2_#3}}
\newcommand{\tmap}[2]{\ensuremath{(\!\!\langle#1\rangle\!\!)^{#2}}}
\newcommand{\vtmap}[2]{{\ensuremath{\big\lfloor #1\big\rfloor^{#2}}}}
\newcommand{\mapt}[1]{\ensuremath{(\!\!\langle#1\rangle\!\!)}}
\newcommand{\mapa}[1]{\ensuremath{\{\!\!\{#1\}\!\!\}}}
\newcommand{\namemap}[2]{#1\map{#2}}

\newcommand{\enc}[2]{\big\langle\map{#1}, \mapt{#2}\big\rangle}
\newcommand{\enco}[1]{\big\langle #1\big\rangle}
\newcommand{\encod}[3]{\lrangle{\map{#1}^{#3}, \mapt{#2}^{#3}}}
\newcommand{\fencod}[4]{\lrangle{\map{#1}^{#3}_{#4} \, , \, \mapt{#2}^{#3}}}

\newcommand{\calc}[5]{\lrangle{#1, #2, #3, #4, #5}}
\newcommand{\tyl}[1]{\ensuremath{\mathcal{#1}}}


\newcommand{\constrtype}[1]{\mathtt{#1}}

\newcommand{\Let}{\constrtype{let}\ }
\newcommand{\In}{\constrtype{in}\ }
\newcommand{\To}{\constrtype{to}\ }
\newcommand{\new}{\constrtype{new}\ }
\newcommand{\from}{\constrtype{from}\ }
\newcommand{\select}{\constrtype{select}\ }
\newcommand{\register}{\constrtype{register}\ }
\newcommand{\Update}{\constrtype{update}\ }

\newcommand{\selectfrom}[2]{\select #1\ \from #2\ \In}
\newcommand{\registerto}[2]{\register #1\ \To #2\ \In}

\newcommand{\newselector}[1]{\new \constrtype{sel}\ #1\ \In}
\newcommand{\newselectorT}[2]{\new \constrtype{sel}\lrangle{#2}\ #1\ \In}
\newcommand{\selecttype}[1]{\dual{\constrtype{sel}}\lrangle{#1}}
\newcommand{\sselecttype}[1]{\constrtype{sel}\lrangle{#1}}

\newcommand{\update}[3]{\Update(#1, #2, #3)\ \In}

\newcommand{\newenv}[1]{\new \mathtt{env}\ #1\ \In\ }
\newcommand{\Letin}[2]{\Let #1 = #2\ \In}

\newcommand{\selqueue}[2]{#1\lrangle{#2}}

\newcommand{\dualof}{\ \mathsf{dual}\ }


\newcommand{\labs}[2]{\lambda #1. #2}

\newcommand{\HOp}{\ensuremath{\mathsf{HO}\pi}\xspace}
\newcommand{\sessp}{\ensuremath{\pi}\xspace}
\newcommand{\haskp}{\ensuremath{\pi^{\lambda}}\xspace}
\newcommand{\pHOp}{\ensuremath{\mathsf{HO}\tilde{\pi}}\xspace}
\newcommand{\HO}{\ensuremath{\mathsf{HO}}\xspace}
\newcommand{\HOpp}{\ensuremath{\mathsf{HO\pi^{+}}}\xspace}
\newcommand{\PHOp}{\ensuremath{\mathsf{HO}\,{\widetilde{\pi}}}\xspace}
\newcommand{\PHOpp}{\ensuremath{\mathsf{HO}\,{\widetilde{\pi}}^{\,+}}\xspace}
\newcommand{\PHO}{\ensuremath{\vec{\mathsf{HO}}}\xspace}
\newcommand{\Psessp}{\ensuremath{\vec{\pi}}\xspace}

\newcommand{\CAL}{\ensuremath{\mathsf{C}}\xspace}

\newcommand{\ST}{\mathsf{ST}}

\newcommand{\Proc}{\ensuremath{\diamond}}

\newcommand{\appl}[2]{#1\, {#2}}
\newcommand{\abs}[2]{\lambda #1.\,#2}

\newcommand{\lollipop}{\multimap}
\newcommand{\sharedop}{\rightarrow}
\newcommand{\logicop}{\multimapdot}

\newcommand{\lhot}[1]{#1\! \lollipop\! \diamond}
\newcommand{\shot}[1]{#1\! \sharedop\! \diamond}
\newcommand{\thunkt}{\ensuremath\{\!\{\diamond\}\!\}}
\newcommand{\thunkp}[1]{\ensuremath\{\!\{#1\}\!\}}
\newcommand{\dummyn}{\ensuremath{\ast}}
\newcommand{\hot}[1]{#1 \logicop \diamond}

\newcommand{\vmap}[1]{|\!|#1|\!|}
\newcommand{\smap}[1]{(\!|\!|#1|\!|\!)^s}
\newcommand{\svmap}[1]{(\!|\!|#1|\!|\!)^{s\rightarrow v}}
\newcommand{\amap}[1]{\mathcal{A}\map{#1}}
\newcommand{\absmap}[2]{\mathcal{A}\map{#1}^{#2}}


\newcommand{\hotrigger}[2]{\binp{#1}{x} \newsp{s}{\appl{x}{s} \Par \bout{\dual{s}}{#2} \inact}}
\newcommand{\fotrigger}[5]{\binp{#1}{#2} \newsp{#3}{\map{#4}^{#3} \Par \bout{\dual{#3}}{#5} \inact}}


\newcommand{\horel}[6]{#1; #2 \proves #3 #4 #5 \proves #6}

\newcommand{\mhorel}[7]{
	\begin{array}{rcll}
		#1; \es; #2 &#4& #5 \proves& #3\\
			&#4& #6 & #7
	\end{array}
}


\newcommand{\Loop}{\mathsf{Loop}}
\newcommand{\CodeBlocks}{\mathsf{CodeBlocks}}

\newcommand{\lnmap}[1]{\namemap{LN}{#1}}
\newcommand{\lnrmap}[1]{\namemap{LNR}{#1}}
\newcommand{\lnblockmap}[1]{\namemap{\mathcal{B}}{#1}}
\newcommand{\lnnonblockmap}[2]{\map{#1, #2}}

\newcommand{\mapenv}[2]{\map{#1}_{#2}}


\newcommand{\true}{\sessionfont{tt}}
\newcommand{\false}{\sessionfont{ff}}

\newcommand{\bool}{\sessionfont{bool}}
\newcommand{\nat}{\sessionfont{nat}}


\newcommand{\PP}{\ensuremath{P}}
\newcommand{\Q}{\ensuremath{Q}}
\newcommand{\R}{\ensuremath{R}}
\newcommand{\OP}{\ensuremath{\mathsf{O}}}

\newcommand{\En}{\ensuremath{En}}

\newcommand{\s}{\ensuremath{s}}
\newcommand{\ds}{\ensuremath{\dual{s}}}

\newcommand{\sd}{\mathtt{sd}}
\newcommand{\shd}{\mathtt{shd}}

\newcommand{\Ia}{\ensuremath{a}}
\newcommand{\Iu}{\ensuremath{u}}

\newcommand{\x}{\ensuremath{x}}
\newcommand{\y}{\ensuremath{y}}
\newcommand{\ks}{\ensuremath{k}}
\newcommand{\va}{\ensuremath{v}}
\newcommand{\n}{\ensuremath{n}}


\newcommand{\X}{\varp{X}}
\newcommand{\Y}{\varp{Y}}

\newcommand{\p}{\ensuremath{\mathtt{p}}\xspace}
\newcommand{\pa}{\ensuremath{\mathtt{pa}}}
\newcommand{\q}{\ensuremath{\mathtt{q}}\xspace}
\newcommand{\er}{\ensuremath{\mathtt{r}}}
\newcommand{\A}{\ensuremath{A}}

\newcommand{\gG}{\globaltype{\G}}
\newcommand{\So}{\ensuremath{S}}
\newcommand{\T}{\ensuremath{T}}

\newcommand{\M}{\ensuremath{M}}
\newcommand{\I}{\ensuremath{M_\inputsym}}
\newcommand{\Om}{\ensuremath{M_\outputsym}}
\newcommand{\Typ}{\ensuremath{\mathsf{T}}}

\newcommand{\h}{\ensuremath{h}}

\newcommand{\m}{\ensuremath{\mu}}

\newcommand{\E}{\ensuremath{E}}

\newcommand{\TT}{\ensuremath{T}}
\newcommand{\suc}{\textrm{succ}}
\newcommand{\fail}{\textrm{fail}}

\newcommand{\lPi}{\set{l_i:\PP_i}_{i \in I}}
\newcommand{\lGi}{\set{l_i:\G_i}_{i \in I}}
\newcommand{\lTi}{\set{l_i:\T_i}_{i \in I}}
\newcommand{\lSi}{\set{l_i:\So_i}_{i \in I}}


\newcommand{\SEL}{P_\mathit{Sel}}
\newcommand{\DSEL}{P_\mathit{DSel}}
\newcommand{\Sel}{\mathsf{Sel}}
\newcommand{\IfSel}{\mathsf{IfSel}}
\newcommand{\DSel}{\mathsf{DSel}}
\newcommand{\PSel}{\mathsf{PermSel}}
\newcommand{\PIfSel}{\mathsf{PermIfSel}}
\newcommand{\PDSel}{\mathsf{PermDSel}}

\newtheorem{notation}[thm]{Notation}

\newcommand{\nonhosyntax}[1]{\colorbox{lightgray}{\ensuremath{#1}}}
\newcommand{\nonpisyntax}[1]{\fcolorbox{black}{white}{\ensuremath{#1}}}
\newcommand{\rtsyn}[1]{\fcolorbox{black}{white}{\ensuremath{#1}}}

\newenvironment{mytheorem}{
	\begin{theorem}
}{
	\end{theorem}
}

\newenvironment{myproposition}{
	\begin{proposition}
}{
	\end{proposition}
}

\newenvironment{mycorollary}{
	\begin{corollary}
}{
	\end{corollary}
}

\newenvironment{mylemma}{
	\begin{lemma}
}{
	\end{lemma}
}

\newenvironment{mydefinition}{
	\begin{definition}
}{
	\end{definition}
}



\newcommand{\Appendix}[1]{Appendix \ref{#1}}

\newcommand{\dimcom}[1]{{\bf Comment: #1 \\}}

\newcommand{\hintcom}[1]{{\bf Hint: #1 \\}}

\newif\ifny\nyfalse
\newcommand{\NY}[1]
{\ifny{\color{purple}{#1}}\else{#1}\fi}

\newcommand{\KH}[1]
{\ifny{\color{brown}{#1}}\else{#1}\fi}

\newif\ifdm\dmtrue
\newcommand{\dk}[1]
{\ifdm{\color{blue}{#1}}\else{#1}\fi}

\newif\ifrhu\rhutrue
\newcommand{\rh}[1]
{\ifdm{\color{red}{#1}}\else{#1}\fi}

\newif\ifjp\jptrue

\newif\ifjp\jptrue
\jpfalse
\newcommand{\jpc}[1]
{\ifjp{\color{red}{#1}}\else{#1}\fi}

\newcommand{\newj}[1]{{#1}}
\newcommand{\newjb}[1]{{#1}}

\newcommand{\ENCan}[1]{\langle #1 \rangle}
\newcommand{\NI}{\noindent}

\newcommand{\syntaxvspace}{\\[1mm]}

\newcommand{\TO}[2]{#1\to #2}
\newcommand{\GS}[3]{\TO{#1}{#2}\colon \!\ENCan{#3}}

\newcommand{\ASET}[1]{\{#1\}}
\newcommand{\participant}[1]{\mathtt{#1}}
\newcommand{\CODE}[1]{{\tt #1}}

\newcommand{\AT}[2]{#1 \! : \! #2}

\newcommand{\myrm}{}

\newcommand{\secref}[1]{\S\,\ref{#1}}
\newcommand{\defref}[1]{Def.~\ref{#1}}
\newcommand{\notref}[1]{Not.~\ref{#1}}
\newcommand{\defsref}[1]{Defs.~\ref{#1}}
\newcommand{\figref}[1]{Fig.~\ref{#1}}
\newcommand{\thmref}[1]{Thm.~\ref{#1}}
\newcommand{\thmsref}[1]{Thms.~\ref{#1}}
\newcommand{\exref}[1]{Ex.~\ref{#1}}
\newcommand{\propref}[1]{Prop.~\ref{#1}}
\newcommand{\propsref}[1]{Props.~\ref{#1}}
\newcommand{\appref}[1]{App.~\ref{#1}}
\newcommand{\lemref}[1]{Lem.~\ref{#1}}

\newcommand{\stytra}[6]{\ensuremath{#1; #3 \proves #4 \hby{#2} #5 \proves #6 }}
\newcommand{\stytraarg}[7]{\ensuremath{#1; #3 \proves_{#7} #4 \hby{#2} #5 \proves_{#7} #6 }}
\newcommand{\stytraargi}[8]{\ensuremath{#1; #3 \proves_{#7} #4 \hby{#2}_{#8} #5 \proves_{#7} #6 }}
\newcommand{\wtytra}[6]{\ensuremath{#1; #3 \proves #4 \Hby{#2}  #5 \proves #6}}
\newcommand{\wtytraarg}[7]{\ensuremath{#1; #3 \proves_{#7} #4 \Hby{#2}  #5 \proves_{#7} #6 }}
\newcommand{\wtytraargi}[8]{\ensuremath{#1; #3 \proves_{#7} #4 \Hby{#2}_{#8}  #5 \proves_{#7} #6 }}
\newcommand{\wbb}[6]{\ensuremath{#1; #3 \proves #4 \wb #5 \proves #6 }}
\newcommand{\wbbarg}[7]{\ensuremath{#1; #3 \proves_{#7} #4 \wb_{#7} #5 \proves_{#7} #6 }}

\newcommand{\minussh}{\ensuremath{\mathsf{-sh}}\xspace}

\definecolor{lightgray}{gray}{0.75}

\newcommand\greybox[1]{%
  \vskip\baselineskip%
  \par\noindent\colorbox{lightgray}{%
    \begin{minipage}{\textwidth}#1\end{minipage}%
  }%
  \vskip\baselineskip%
}

\newcommand{\mapchar}[2]{\ensuremath{[\!\!(#1)\!\!]^{#2}}}
\newcommand{\omapchar}[1]{\ensuremath{[\!\!(#1)\!\!]_{\mathsf{c}}}}

\newcommand{\trigger}[3]{#1 \leftarrow\!\!\!\!\!\!\!\leftarrow #2:#3 }
\newcommand{\ftrigger}[3]{#1 \Leftarrow \AT{#2}{#3}}
\newcommand{\htrigger}[2]{#1 \hookleftarrow #2}

\newcommand{\btau}{\tau_{\beta}}
\newcommand{\stau}{\tau_{s}}
\newcommand{\dtau}{\tau_{d}}


\newcommand{\procs}{\mathcal{P}}
\newcommand{\confs}{\mathcal{M}}
\newcommand{\agents}{\mathcal{A}}
\newcommand{\newg}[1]{\keyword{new }\,(#1)}
\newcommand{\cm}[1]{\textcolor{blue}{[CM: #1]}}
\newcommand{\jp}[1]{\textcolor{red}{[JP: #1]}}
\newcommand{\added}[1]{{#1}}
\newcommand{\modif}[1]{#1}

\newcommand{\fwcolor}[1]{\textcolor{ABlue}{#1}}
\newcommand{\bkcolor}[1]{\textcolor{ARed}{#1}}
\newcommand{\sepcolor}[1]{\textcolor{AOrange}{#1}}

\newcommand{\rmark}{\bkcolor{\blacklozenge}}
\newcommand{\normark}{\lozenge}
\newcommand{\hnormark}{\bkcolor{\circ}}

\newcommand{\key}[2]{{#1}_{[#2]}}
\newcommand{\np}[2]{#1:#2}
\newcommand{\ep}[2]{#1_{[#2]}}
\newcommand{\mysepp}{\cdot}
\newcommand{\mysep}{\,,\,}
\newcommand{\store}{\sigma}
\newcommand{\queue}[1]{\mbox{{\Large $\lfloor$}} #1 \mbox{{\Large $\rfloor$}}}
\newcommand{\mytagg}{\spadesuit}

\newcommand{\monig}[4]{\ensuremath{{#1\queue{\textcolor{black}{#2}\mysepp #3\mysepp #4}^{\mytagg}}}}
\newcommand{\moni}[4]{\ensuremath{{#1\queue{\textcolor{black}{#2}\mysepp #3\mysepp #4}}}}
\newcommand{\hmoni}[4]{\ensuremath{{#1\queue{\textcolor{black}{#2}\mysepp #3\mysepp #4}^{\normark}}}}
\newcommand{\monir}[4]{\ensuremath{{#1\queue{\textcolor{black}{#2}\mysepp #3\mysepp #4}^{\bkcolor{\rmark}}}}}

\newcommand{\mem}[3]{\ensuremath{#1\queue{#2\mysep #3}}}
\newcommand{\past}{\,\sepcolor{\text{\textbf{\textasciicircum\!\!\textasciicircum}}}}
\newcommand{\typeOut}[2]{#1\outtype \langle #2\rangle}
\newcommand{\typeIn}[2]{#1\inptype ( #2)}

\newcommand{\coda}[2]{#1:#2}
\newcommand{\history}{\sepcolor{\mathbf{\star}}}
\newcommand{\codah}[4]{\coda{#1}{(#2\,\history\,#3)#4}}
\newcommand{\len}{\mathtt{len}}
\newcommand{\anyv}{m}
\newcommand{\equivq}{\equiv_\mathsf{q}}

\newcommand{\singleQ}[1]{[#1]\!\! \downharpoonleft \!\downharpoonright}

\newcommand{\cons}{\circ}
\newcommand{\consh}{\circledwedge}
\newcommand{\valueq}[3]{(#1 \,,\, #2 \,,\,#3)}
\newcommand{\myeval}[2]{#2(#1)}
\newcommand{\ctx}[1]{\mathbb{#1}}
\newcommand{\myctx}[2]{\ctx{#1}[#2]}
\newcommand{\myctxb}[2]{{#1\left[\textcolor{black}{#2}\right]}}
\newcommand{\myctxr}[2]{{#1\left[\textcolor{black}{#2}\right]}}
\newcommand{\freeze}[1]{\left\langle #1 \right\rangle}
\newcommand{\upd}[2]{[#1\mapsto #2]}
\newcommand{\rup}[2]{#1\setminus #2}
\newcommand{\restrict}[1]{\downarrow_{#1}}
\newcommand{\loc}{\ell}	
\newcommand{\myloc}[2]{#1\left\{#2\right\}}	
\newcommand{\lbl}{l}	

\newcommand{\epS}{\ep{s}{\mathtt{V}}}
\newcommand{\pS}{\ensuremath{\mathtt{V}}\xspace}

\newcommand{\epA}{\ep{s}{\mathtt{A}}}
\newcommand{\pA}{\ensuremath{\mathtt{A}}\xspace}
\newcommand{\epB}{\ep{s}{\mathtt{B}}}
\newcommand{\pB}{\ensuremath{\mathtt{B}}\xspace}
\newcommand{\epC}{\ep{s}{\mathtt{C}}}
\newcommand{\pC}{\ensuremath{\mathtt{C}}\xspace}
\newcommand{\exBook}{\text{`Logicomix'}}

\newcommand{\gend}{\mathtt{end}}
\newcommand{\lend}{\mathsf{end}}
\newcommand{\gpart}[1]{\mathtt{#1}}
\newcommand{\gtact}[5]{\gpart{#1}\to\gpart{#2}:\{#3\langle #4\rangle.#5\}_{i \in I}}
\newcommand{\gtactp}[5]{\gpart{#1}\to\!\past \gpart{#2}:#3\langle #4\rangle.#5}
\newcommand{\gtactpp}[3]{\gpart{#1}\to \gpart{#2}: \past#3}
\newcommand{\gtcho}[4]{\gpart{#1}\to\gpart{#2}:\{#3:#4\}_{i \in I}}
\newcommand{\gtchoi}[5]{\gpart{#1}\to\gpart{#2}:\{#3{:}#4\}_{#5}}
\newcommand{\gtcom}[4]{\gpart{#1}\to\gpart{#2}:\langle #3\rangle.#4}

\newcommand{\lsend}[4]{\gpart{#1}!\{#2\langle #3\rangle.#4\}_{i \in I}}
\newcommand{\lrecv}[4]{\gpart{#1}?\{#2\langle #3\rangle.#4\}_{i \in I}}

\newcommand{\ltout}[3]{\gpart{#1}!\langle#2\rangle.#3}
\newcommand{\ltinp}[3]{\gpart{#1}?\langle#2\rangle.#3}
\newcommand{\ltsel}[5]{\gpart{#1}\!\oplus\!\{#2: #3\}_{#4 \in #5}}
\newcommand{\ltbra}[5]{\gpart{#1}\&\{#2: #3\}_{#4 \in #5}}

\newcommand{\ltoutp}[3]{\gpart{#1}!\langle#2\rangle.\past #3}
\newcommand{\ltinpp}[3]{\gpart{#1}?\langle#2\rangle.\past #3}
\newcommand{\ltselp}[4]{\gpart{#1}\!\oplus\!\{#2\}_{#3 \in #4}}
\newcommand{\ltbrap}[4]{\gpart{#1}\&\{#2\}_{#3 \in #4}}

\newcommand{\tproj}[2]{#1\!\downarrow_{#2}}
\newcommand{\pproj}[3]{[\![#1]\!]^{#3}\downarrow_{#2}}

\newcommand{\confa}[2]{\lbag #1 \; \sepcolor{\textbf{;}}\; #2 \rbag} 
\newcommand{\conf}[2]{\lbag #1 \; \sepcolor{\textbf{;}}\; #2 \rbag} 
\newcommand{\stack}[1]{\mathtt{#1}}


\newcommand{\invertlooparrow}{%
\mathrel{\reflectbox{\rotatebox[origin=c]{180}{$\looparrowright$}}}}

\newcommand{\invertcurvearrow}{%
\mathrel{\reflectbox{\rotatebox[origin=c]{180}{$\curvearrowright$}}}}

\newcommand{\fwg}{\ensuremath{\fwcolor{\,\invertlooparrow\,}}}
\newcommand{\fwgs}{\ensuremath{\fwcolor{\,\invertlooparrow^*\,}}}
\newcommand{\bkg}{\ensuremath{\bkcolor{\,\looparrowright\,}}}
\newcommand{\bkgs}{\ensuremath{\bkcolor{\,\looparrowright^*\,}}}

\newcommand{\fw}{\ensuremath{\fwcolor{\,\curvearrowright\,}}}
\newcommand{\fwn}[1]{\ensuremath{\fwcolor{\,\curvearrowright_{#1}\,}}}
\newcommand{\fws}{\ensuremath{\fwcolor{\,\curvearrowright^*\,}}}

\newcommand{\bk}{\ensuremath{\bkcolor{\,\invertcurvearrow\,}}}
\newcommand{\bkn}[1]{\ensuremath{\bkcolor{\,\invertcurvearrow_{#1}\,}}}
\newcommand{\bks}{\ensuremath{\bkcolor{\,\invertcurvearrow^*\,}}}

\newcommand{\bkk}{\ensuremath{\bkcolor{\,\invertcurvearrow^{\,j}\,}}}
\newcommand{\trans}[1]{#1^{*}}
\newcommand{\refl}[1]{#1^{+}}

\newcommand{\inverthookrightarrow}{%
\mathrel{\reflectbox{\rotatebox[origin=c]{180}{$\hookrightarrow$}}}}

\newcommand{\fwa}{\ensuremath{\fwcolor{\hookrightarrow}}}
\newcommand{\fwas}{\ensuremath{\fwcolor{\hookrightarrow^*}}}

\newcommand{\bka}{\ensuremath{\bkcolor{\inverthookrightarrow}}}

\newcommand{\reda}{\rightarrowtail}

\newcommand{\lbka}{\xrightleftharpoons}

\newcommand{\lfwa}[1]{\xRightarrow{#1}}
\newcommand{\lreda}[1]{\xRightarrow{#1}}

\newcommand{\fwalt}{\ensuremath{\fwcolor{\rightharpoonup}\,}}

\newcommand{\bkalt}{\ensuremath{\bkcolor{\rightharpoondown}\,}}

\newcommand{\bfb}{\sim}
\newcommand{\bfbw}{\approx}
\newcommand{\congruence}[1]{\stackrel{\cdot}{#1}}

\newcommand{\ltrans}[1]{\xleftrightarrow{#1}}
\newcommand{\memstamp}{\eta}

\newcommand{\initC}[3]{\ensuremath{\mathcal{I}_{\,#3}^{#2}(#1)}}
\newcommand{\trad}[1]{\llbracket#1\rrbracket}

\newcommand{\stable}{\mathtt{sb}}
\newcommand{\stables}[1]{#1\text{-}\stable}

\newcommand{\trace}{\rho}
\newcommand{\etrace}{\varepsilon}
\newcommand{\causeq}{\asymp}
\newcommand{\nottrace}[1]{\overline{#1}}
\newcommand{\tsource}[1]{\mathtt{src}(#1)}
\newcommand{\ttarget}[1]{\mathtt{trg}(#1)}

\newcommand{\st}{\,|\,}
\newcommand{\adeq}[2]{#1 \bowtie #2}
\newcommand{\initadeq}[2]{#1 \stackrel{\mathsf{o}}{\bowtie} #2}
\newcommand{\ladeq}[3]{#1 \stackrel{#3}{\ltimes} #2}
\newcommand{\gth}[1]{\mathsf{#1}}
\newcommand{\ifk}{\mathbf{if}}
\newcommand{\thenk}{\mathbf{then}}
\newcommand{\elsek}{\mathbf{else}}
\newcommand{\ifproc}[3]{\ifk\,#1\, \thenk\, #2\,\elsek\,#3 }
\newcommand{\party}[1]{\ensuremath{\mathtt{#1}}\xspace}
\newcommand{\bigvect}[1]{\widehat{#1}}
\newcommand{\names}[1]{\mathtt{\mathsf{roles}}(#1)}


\section{Introduction}
This paper is about \emph{reversible computation} in the context of  models of concurrency for \emph{communi\-ca\-tion-centric} software systems, i.e., collections of distributed software components in which  
concurrent interactions are governed by reciprocal dialogues or \emph{protocols}.

Building upon process calculi techniques, these 
models  provide a rigorous footing for 
message-passing concurrency; on top of them, many  analysis techniques based on \emph{behavioral types}  and \emph{contracts} have been put forward  
to enforce key safety and liveness properties (see, e.g., the survey~\cite{DBLP:journals/csur/HuttelLVCCDMPRT16}). Reversibility is an appealing notion in concurrency at large~\cite{DBLP:journals/eatcs/Lanese14}, but especially so in communication-centric scenarios: it may elegantly abstract fault-tolerant communicating systems that react to  unforeseen circumstances (say, local failures) by ``undoing'' computational steps so as to reach a previous consistent  state.

In communication-centric software systems, protocols
specify the intended communication structures among interacting components. 
We focus on process calculi equipped with behavioral types, which use those protocols as \emph{types} to enforce communication correctness. The interest is in  \emph{protocol conformance}, the property that ensures that each component respects its ascribed protocol. The integration of reversibility in models of communication-centric systems has been addressed from various angles (cf. \cite{DBLP:journals/jlp/TiezziY15,DBLP:conf/rc/TiezziY16,DBLP:journals/fac/BarbaneraDd16,DBLP:journals/corr/MezzinaP16}). Focusing on \emph{session types}~\cite{honda.vasconcelos.kubo:language-primitives,HYC08}---a well established class of behavioral types---, Tiezzi and Yoshida~\cite{DBLP:journals/jlp/TiezziY15} were the first to integrate reversibility into a session-typed $\pi$-calculus, following the seminal approach of Danos and Krivine~\cite{DanosK04}; in their approach, however, session types are not used in the definition of reversible communicating systems, nor play a r\^{o}le in establishing their properties. 

Triggered by this observation,  our prior work~\cite{DBLP:journals/corr/MezzinaP16,MezzinaP17} develops a \emph{monitors-as-memories} approach. The idea is to use \emph{monitors} (i.e., run-time entities that enact protocol actions) as the \emph{memories} needed to record  and  eventually undo communication steps. 
There is a monitor for each protocol participant, which includes a session type that describes the intended protocol. 
We use a so-called \emph{cursor}  to ``mark'' the current protocol state in the type; the cursor can move to the future (enacting protocol actions) but also to the past (reversing protocol actions). 

The monitors-as-memories approach induces a streamlined process framework in which the key properties of a reversible semantics can be established with simple proofs,
because session types narrow down the spectrum of possible process behaviors, allowing only those forward and backward actions that adhere to the declared protocols.
The most significant of such properties is 
\emph{causal consistency}~\cite{DanosK04},   considered as the ``right''  criterion for reversing concurrent processes in distributed systems~\cite{wg2}.
Intuitively, causal consistency ensures that reversible steps lead to system states that could have been reached by performing forward steps only. That is, causally consistent reversibility does not lead to extraneous states, not reachable through ordinary computations.

The reversible  framework in~\cite{DBLP:journals/corr/MezzinaP16,MezzinaP17}, however,  accounts only for  $\pi$-calculus processes implementing \emph{binary sessions}, which represent protocols between exactly two partners. Also, it considers \emph{synchronous communication} instead of the more general \emph{asynchronous (queue-based) communication}. 
Hence, our prior work 
rules out an important class of real-life protocols, namely those that describe  interaction scenarios among multiple parties without a single point of control. 
In \emph{multiparty session types}~\cite{HYC08}, these protocols are represented by a \emph{global type} that can be projected as \emph{local types} to obtain each participant's contribution to the entire interaction. 
Moving from binary to multiparty sessions is a significant jump in expressiveness; global types offer a convenient declarative description of the entire communication scenario.
However, the multiparty case also entails added challenges, as two levels of abstraction, global and local, should be considered for (reversible) protocols and their implementations. Hence, it is far from obvious that our monitors-as-memories approach to (causal consistent) reversibility extends to the multiparty case.

\subsubsection*{Contributions} 
Given this context, in this paper we make the following contributions:
\begin{enumerate}
\item We introduce a process model for reversible, multiparty sessions with asynchrony (as in~\cite{DBLP:journals/mscs/KouzapasYHH16}), abstraction passing (i.e., functions from names to processes) ~\cite{SangiorgiD:expmpa,KPY2016,DBLP:journals/iandc/KouzapasPY19},  and decoupled rollbacks~(\S\,\ref{sec:calculus}). 
We define forward and backward semantics for multiparty processes by extending the monitors-as-memories approach to both
global types and their implementations. 

\item  We prove that reversibility in our model is causally consistent (Theorem~\ref{t:causal}). The proof is challenging as we must appeal to an alternative reversible semantics with 
\emph{atomic rollbacks}, which we show to coincide with the semantics with decoupled rollbacks (Theorem~\ref{t:bf}). 

\item We formally connect reversibility at two distinct levels: the (declarative) level of global types and the (operational) level of processes monitored by local types with cursors (Theorem~\ref{t:corrgc}).
\end{enumerate}

\noindent 
We stress that asynchrony, abstraction passing, and decoupled rollbacks are not considered in prior works~\cite{DBLP:conf/rc/TiezziY16,DBLP:journals/corr/Dezani-Ciancaglini16a,DBLP:journals/corr/MezzinaP16,MezzinaP17}.
Asynchrony  and decoupled rollbacks are delicate issues in a reversible multiparty setting---we do not know of other asynchronous calculi with reversible semantics, nor featuring the same combination of constructs. 
The formal connection between global and local levels of abstraction (Theorem~\ref{t:corrgc}) is also unique to our multiparty setting.

\subsubsection*{Organization}
In Section\,\ref{sec:calculus}, we introduce our process model of reversible multiparty protocols, and  illustrate it with examples.
In Section\,\ref{s:res} we establish causal consistency by relating decoupled and atomic semantics, and connect reversibility at global and local levels. 
Section\,\ref{s:rw} discusses an alternative decoupled semantics and related works.
Section\,\ref{s:conc} collects some concluding remarks. 

This paper is a revised and extended version of the conference paper~\cite{ppdp}. 
In this presentation we consider a language with \emph{labeled choices} (not treated in~\cite{ppdp}), provide additional examples, streamline the presentation of the decoupled and atomic semantics, extend comparisons with related works, and include technical details (definitions and proofs).

\section{Reversible Multiparty Protocols}
\label{sec:calculus}

\begin{figure}
\begin{mdframed}
\begin{center}
    \includegraphics[width=10.0cm]{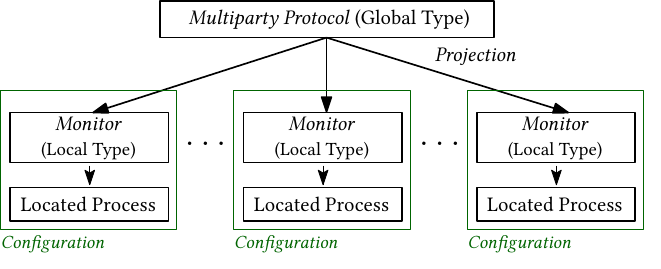}
\end{center}
\vspace{-2mm}
\end{mdframed}
\caption{Our process model of multiparty communications.}\label{f:model}
\end{figure}
Fig.~\ref{f:model} depicts the ingredients of our
two-level model of \emph{protocols} and \emph{configurations/processes}.
Multiparty protocols are defined in terms of \emph{global types}, which declaratively describe a protocol among two or more participants. 
A global type can be \emph{projected} onto each participant so as to obtain 
its corresponding  \emph{local type}, i.e., a session type that abstracts a participant's contribution to the global protocol. 

The semantics of global types is given in terms of forward and backward  transition systems (Fig.~\ref{f:gts}). 
There is a \emph{configuration} for each protocol participant: it includes a \emph{located process}
that 
 specifies asynchronous communication behavior, subject to a \emph{monitor}
that enables forward/backward steps at run-time
based on the local type.
The semantics of configurations is given in terms of forward and backward reduction relations (Figs.~\ref{fig:fw_1},~\ref{fig:fw_2},~\ref{fig:bk_1}, \ref{fig:bk_2}, and~\ref{fig:bk_3}).
Figure~\ref{f:reds} in \secref{s:res} summarizes our notations for semantics of global types and configurations.

We illustrate our model of reversible  protocols with two examples.
As a running example, we develop a reversible variant of the \emph{Three-Buyer protocol}   (see, e.g.,~\cite{CDYP2015})
with abstraction passing (code mobility), one of the 
distinctive traits of our framework. This example comes in three parts---cf. \S\,\ref{sss:buysel1},  ~\S\,\ref{sss:buysel2}, and \S\,\ref{sss:examp-tb}.
As second example, in \S\,\ref{app:choices} we present a protocol with labeled choices.

\begin{remark}[Colors]
Throughout the paper, we  use colors to improve readability. 
In particular, 
elements in \fwcolor{blue} belong to a forward semantics;
elements in \bkcolor{red} belong to a backward semantics.
Also, we use \sepcolor{orange} to highlight the cursor and other syntactic entities. 
\end{remark}

\subsection{Global and Local Types}

\subsubsection{Syntax}
Let us write $\p, \q, \gpart{r}, \pA, \pB \ldots$ to denote protocol \emph{participants}.
The syntax of global types ($G, G', \ldots$) and local types  ($T, T', \ldots$) is standard~\cite{HYC08} and defined as follows:
\begin{align*}
			G, G'  \bnfis & \gtcom{p}{q}{U}{G} \sbnfbar \gtcho{p}{q}{\lbl_i}{G_i} 
			\sbnfbar  \mu X. G \sbnfbar X \sbnfbar \gend 
			\\
			U, U'  \bnfis & \bool \sbnfbar \nat \sbnfbar \cdots 
			\sbnfbar \shot{T} 
			\\
	    	T, T'  \bnfis & \ltout{p}{U}{T} \sbnfbar \ltinp{p}{U}{T} 
		  \sbnfbar  \ltsel{p}{\lbl_i}{T_i}{i}{I} \sbnfbar \ltbra{p}{\lbl_i}{T_i}{i}{I}   
		\sbnfbar \mu X. T \sbnfbar X \sbnfbar \lend 
\end{align*}
The global type $\gtcom{p}{q}{U}{G}$ says that \p may send a value of type $U$ to \q, and then continue as $G$.
Given a finite index set $I$ and pairwise different \emph{labels} $\lbl_i$, the 
global type $\gtcho{p}{q}{\lbl_i}{G_i}$
 specifies a \emph{labeled choice}: 
 \p may choose 
label $\lbl_i$, communicate this selection to \q, and then continue as $G_i$. 
In these two types 
we assume that $\p \neq \q$.
Global 
recursive and terminated protocols are denoted $\mu X. G$ and $\gend$, respectively.
We write $\parties{G}$ to denote the set of participants in $G$.

Value types $U$ include basic first-order values (constants),   but also \emph{higher-order} values: abstractions from names to processes. 
(We write $\Proc$ to denote the type of processes.)
Local types 
$\ltout{p}{U}{T}$ and $\ltinp{p}{U}{T}$ denote, respectively, an output and input of value of type $U$ by \p.
Type $\ltbra{p}{\lbl_i}{T_i}{i}{I}$ says that \p 
offers different behaviors, available as labeled alternatives;
conversely, type $\ltsel{\p}{\lbl_i}{T_i}{i}{I}$ 
says that \p may select one of such alternatives. Terminated and recursive local types are denoted $\lend$ and $\mu X. T$, respectively. 
We use  $\alpha$ to denote type prefixes  $\typeIn{\p}{U}$, $\typeOut{\p}{U}$.

As usual, we consider only recursive types $\mu X. G$ (and $\mu X. T$) in which $X$ occurs guarded in $G$ (and $T$).
 We shall take an equi-recursive view of (global and local) types, and so we consider two  types with the same regular tree as equal.
 
Global and local types are connected by  \emph{projection}:
following~\cite{HYC08},
the projection of $G$ onto participant $\p$, written $\tproj{G}{\p}$, is defined in Fig.~\ref{f:proj}. 
Projection for $\gtcho{p}{q}{l_i}{G_i}$ is noteworthy: 
the projections of the participants not involved in the choice (different from $\p, \q$) should correspond to the same identical local type.
\begin{figure}[t!]
\begin{mdframed}
{
\begin{align*}
\tproj{(\gtcom{p}{q}{U}{G})}{\gpart{r}} & = 
\begin{cases}
\ltout{\q}{U}{(\tproj{G}{\gpart{r}})} & \text{if $\gpart{r} = \gpart{p}$} \\
\ltinp{\p}{U}{(\tproj{G}{\gpart{r}})} & \text{if $\gpart{r} = \gpart{q}$} \\
(\tproj{G}{\gpart{r}}) &  \text{if $\gpart{r} \neq \gpart{q}, \gpart{r} \neq \gpart{p}$}
\end{cases}
\\
\tproj{(\gtcho{p}{q}{l_i}{G_i})}{\gpart{r}}  & = 
\begin{cases}
\ltsel{\q}{\lbl_i}{(\tproj{G_i}{\gpart{r}})}{i}{I}  & \text{ if $\gpart{r} = \gpart{p}$} \\
\ltbra{\p}{\lbl_i}{\tproj{G_i}{\gpart{r}}}{i}{I}  & \text{ if $\gpart{r} = \gpart{q}$} \\
(\tproj{{G_1}}{\gpart{r}}) &  \text{ if $\gpart{r} \neq \gpart{q}, \gpart{r} \neq \gpart{p}$ and}  
\text{~~$\forall i, j \in I.\, \tproj{{G_i}}{\gpart{r}} = \tproj{{G_j}}{\gpart{r}}$}
\end{cases}
\\
\tproj{(\mu X. G)}{\gpart{r}} &= 
\begin{cases}
\mu X. \tproj{G}{\gpart{r}} & \text{if $\gpart{r}$ occurs in $G$}
\\
\lend & \text{otherwise}
\end{cases}
\\
\tproj{X}{\gpart{r}} & = X
\qquad
\tproj{\gend}{\gpart{r}}\, = \, \lend
\end{align*}
}
\vspace{-2mm}
\end{mdframed}
\caption{Projection of a global type $G$ onto a participant $\gpart{r}$.\label{f:proj}}
\end{figure}

\subsubsection{Example: The Three-Buyer Seller Protocol (I)}
\label{sss:buysel1}
The Three-Buyer Seller Protocol 
involves three buyers---Alice ($\pA$), Betty ($\pB$), and Carol ($\pC$)---who interact with a Seller ($\pS$) as follows:

\begin{enumerate}[1.]
\item 
Alice sends a book title to Seller, which replies back to Alice and Betty with a quote. Alice tells Betty how much she can contribute.
\item Betty notifies Seller and Alice that she agrees with the {price}, and asks Carol to assist her in completing the protocol. 
To delegate her remaining interactions with Alice and Seller to Carol, Betty sends her 
the code she must execute.
\item Carol continues the rest of the protocol with Seller and Alice as if she were Betty. 
She sends Betty's address (contained in the mobile code she received) to Seller.
\item Seller answers to Alice and Carol (who represents Betty) with the delivery date.
\end{enumerate}


We formalize this protocol as a global type denoted $G$ (see below).
We first define some convenient notation. 
\begin{itemize}
	\item 
We write
$\gtcom{\p}{\{\q_1,\q_2\}}{U}{G}$
as a shorthand notation for 
$\gtcom{\p}{\q_1}{U}{\gtcom{\p}{\q_2}{U}{G}}$
(and similarly for local types).
\item 
We write $\thunkt$ to denote the type $\shot{\lend}$.
As we will see, this is the type of a \emph{thunk process} $\abs{x}{P}$ with $x \not \in \fn{P}$, written
$\thunkp{P}$. A thunk is an inactive process; it can be activated by applying to it a dummy name of type $\lend$ (which we will denote $\dummyn$).
\end{itemize}

The global type $G$ 
between $\pA$, $\pB$, and $\pC$
is as follows:
\begin{align*}
G = ~&  \gtcom{A}{V}{\mathsf{title}}{\gtcom{V}{\{A,B\}}{\mathsf{{price}}}{\gtcom{A}{B}{\mathsf{share}}{
\\
& \quad \gtcom{B}{\{A,V\}}{\mathsf{OK}}{
\\
& \quad \gtcom{B}{C}{\mathsf{share}}{\gtcom{B}{C}{\thunkt}{
\\
& \quad \gtcom{B}{V}{\mathsf{address}}{\gtcom{V}{B}{\mathsf{date}}{\gend}}}}}}}}
\end{align*}
where  $\mathsf{{price}}$ and $\mathsf{share}$ are base types treated as integers; also, $\mathsf{title}$, $\mathsf{OK}$, $\mathsf{address}$, and $\mathsf{date}$ are base types treated as strings.

Then, following the function defined in Fig.~\ref{f:proj}, we have the projections of $G$ onto local types:
\begin{align*}
\tproj{G}{\pS} & = \ltinp{A}{\mathsf{title}}{\ltout{\{A,B\}}{\mathsf{{price}}}{\ltinp{B}{\mathsf{OK}}{\ltinp{B}{\mathsf{address}}{\ltout{B}{\mathsf{date}}{\lend}}}}}
\\
\tproj{G}{\pA} & = \ltout{V}{\mathsf{title}}{\ltinp{V}{\mathsf{{price}}}{\ltout{B}{\mathsf{share}}{\ltinp{B}{\mathsf{OK}}{\lend}}}}
\\
\tproj{G}{\pB} & = \ltinp{V}{\mathsf{{price}}}{\ltinp{A}{\mathsf{share}}{\ltout{\{A,V\}}{\mathsf{OK}}{
\ltout{C}{\mathsf{share}}{\ltout{C}{\thunkt}{
\ltout{V}{\mathsf{address}}{\ltinp{V}{\mathsf{date}}{\lend}}}}}}}
\\
\tproj{G}{\pC} & = \ltinp{B}{\mathsf{share}}{\ltinp{B}{\thunkt}{\lend}}
\end{align*}	

\subsubsection{Semantics of Protocols}\label{ss:chore}

\begin{figure}[t!]
\begin{mdframed}
{
\begin{mathpar}
\inferrule*[left=\fwcolor{(FVal1)}]{}
{\ctx{G}[ \past \gtcom{p}{q}{U}{G}] 
\fwg 
\ctx{G}[\gtcom{p}{\!\!\past q}{U}{G} ]}
\\
\inferrule*[left=\fwcolor{(FVal2)}]{}
{\ctx{G}[  \gtcom{p}{\!\!\past q}{U}{G}] 
\fwg 
\ctx{G}[\gtcom{p}{q}{U}{\past G} ]}
\\
\inferrule*[left=\fwcolor{(FCho1)}]{}
{
\ctx{G}[\past \gtcho{p}{q}{l_i}{G_i}] 
\fwg  
\ctx{G}[\gpart{p}\to\!\!\past\gpart{q}:\{\lbl_i:G_i \,;\, \lbl_j:G_j\}_{i \in I\setminus j} ] 
}
\\
\inferrule*[left=\fwcolor{(FCho2)}]{}
{
\ctx{G}[\gpart{p}\to\!\!\past\gpart{q}:\{\lbl_i:G_i \,;\, \lbl_j:G_j\}_{i \in I\setminus j} ] 
\fwg
\ctx{G}[\gpart{p}\to\gpart{q}:\{\lbl_i:G_i \,;\, \lbl_j:\past G_j\}_{i \in I\setminus j} ] 
}
\\
\inferrule*[left=\bkcolor{(BVal1)}]{}
{\ctx{G}[\gtcom{p}{\!\!\past q}{U}{G} ]
\bkg
\ctx{G}[ \past \gtcom{p}{q}{U}{G}] 
}
\\
\inferrule*[left=\bkcolor{(BVal2)}]{}
{
\ctx{G}[\gtcom{p}{q}{U}{\past G} ]
\bkg
\ctx{G}[  \gtcom{p}{\!\!\past q}{U}{G}] 
}
\\
\inferrule*[left=\bkcolor{(BCho1)}]{}
{
\ctx{G}[\gpart{p}\to\!\!\past\gpart{q}:\{\lbl_i:G_i \,;\, \lbl_j:G_j\}_{i \in I\setminus j} ] 
\bkg  
\ctx{G}[\past \gpart{p}\to\gpart{q}:\{\lbl_i:G_i\}_{i \in I} ]
}
\\
\inferrule*[left=\bkcolor{(BCho2)}]{}
{
\ctx{G}[\gpart{p}\to\gpart{q}:\{\lbl_i:G_i \,;\, \lbl_j:\past G_j\}_{i \in I\setminus j} ]
\bkg 
\ctx{G}[\gpart{p}\to\!\!\past\gpart{q}:\{\lbl_i:G_i \}_{i \in I} ] 
}
\end{mathpar}
}
\vspace{-4mm}
\end{mdframed}
\caption{Semantics of Global Types (Forward \& Backwards).\label{f:gts}}
\label{fig:prot_sem}
\end{figure}

The semantics of global types 
comprises forward and backward transition relations, denoted $\fwg$ and $\bkg$, respectively (Fig.~\ref{fig:prot_sem}). 

To formalize backward steps, we require some auxiliary notions.
We use \emph{global contexts}, ranged over by $\ctx{G}, \ctx{G}', \ldots$ with holes `$\bullet$',  to record
previous actions, including the choices discarded and committed:
$$
\ctx{G}
\bnfis   
\bullet 
\sbnfbar  
\ctx{G}[\gtcom{p}{q}{U}{\ctx G}]
\sbnfbar
\ctx{G}[\gpart{p}\to\gpart{q}:\{\lbl_i:G_i \,;\, \lbl_j:\ctx{G}\}_{i \in I\setminus j} ]
$$
We also use \emph{global types with history}, ranged over by $\mathsf{H}, \mathsf{H}', \ldots$, 
  to record the current protocol state. 
  This state is denoted by the \emph{cursor}~$\past$, which we introduced in~\cite{DBLP:journals/corr/MezzinaP16}.
  
  \begin{defi}[Global Types with History]
  \label{d:gth}
  The syntax of \emph{global types with history} is defined as follows:
      \begin{align*}
			\gth{H}, \gth{H}'  \bnfis & 
			\past G \sbnfbar	G \past 
			\sbnfbar 
			\gtcom{p}{\!\!\past q}{U}{G} 
			\sbnfbar 
			\gtcom{p}{q}{U}{\past G} \\
			\sbnfbar & 
			\gpart{p}\to\!\!\past\gpart{q}:\{\lbl_i:G_i \,;\, \lbl_j:G_j\}_{i \in I\setminus j} 
			\sbnfbar 
 			\gpart{p}\to\gpart{q}:\{\lbl_i:G_i \,;\, \lbl_j:\past G_j\}_{i \in I\setminus j}  
\end{align*}
We write 
$\parties{\gth{H}}$ to denote the set of participants in a global type with history $\gth{H}$. 
  \end{defi}

The syntax of global types with history follows some basic intuitions.
A directed exchange such as $\gtcom{p}{q}{U}{G}$ has three \emph{intermediate states},
characterized by the decoupled involvement of \p and \q in the intended asynchronous model:
\begin{enumerate}
	\item 
The \emph{first state}, denoted $\past \gtcom{p}{q}{U}{G}$, describes the situation prior to the exchange.
\item 
The \emph{second state} represents the point in which \p has sent a value of type $U$ but this message has not yet
reached \q; this is denoted as $\gtcom{p}{\!\!\past q}{U}{G}$.
\item 
The \emph{third state} represents the point in which \q has received the message from \p and
the continuation $G$ is ready to execute; this is denoted as $\gtcom{p}{q}{U}{\past G}$.
\end{enumerate}

These intuitions extend similarly to $\gtcho{p}{q}{l_i}{G_i}$, with the following caveat:
the second state should distinguish the choice made by $\p$ from the discarded alternatives; 
  we write $\gpart{p}\to\!\!\past\gpart{q}:\{\lbl_i:G_i \,;\, \lbl_j:G_j\}_{i \in I\setminus j}$ to denote that $\p$ has selected $\lbl_j$ and that this choice is still to be received by \q. Once this occurs, a state
  $\gpart{p}\to\gpart{q}:\{\lbl_i:G_i \,;\, \lbl_j:\past G_j\}_{i \in I\setminus j}$ is reached.

We may now describe the forward and backward transition rules for global types, given 
in Fig.~\ref{fig:prot_sem}. 
For a forward directed exchange of a value, 
Rule~\fwcolor{\textsc{(FVal1)}} formalizes the transition from the first   to the second state;
Rule~\fwcolor{\textsc{(FVal2)}} denotes the transition from the second to the third state.
Rules~\fwcolor{\textsc{(FCho1)}} and~\fwcolor{\textsc{(FCho2)}} are their analogues for the forward directed communication of a label.
Rules~\bkcolor{\textsc{(BVal1)}} and~\bkcolor{\textsc{(BVal2)}} undo the step performed by 
Rules~\fwcolor{\textsc{(FVal1)}} and~\fwcolor{\textsc{(FVal2)}}, respectively. 
Also, 
Rules~\bkcolor{\textsc{(BCho1)}} and~\bkcolor{\textsc{(BCho2)}} undo the step performed by 
Rules~\fwcolor{\textsc{(FCho1)}} and~\fwcolor{\textsc{(FCho2)}}, respectively.

\subsection{Processes and Configurations}
\label{subsec:syntax}

	\begin{figure}[t!]
	\begin{mdframed}
		\begin{align*}
			u,w  \bnfis& n \sbnfbar x,y,z
			\qquad \quad
			n,n'\bnfis a,b \sbnfbar \ep{s}{\p}
			\qquad \quad
			 {v},  {v}'  \bnfis   \true \sbnfbar \false \sbnfbar \cdots
			\\
			V,W \bnfis & {a,b} \sbnfbar  x,y,z \sbnfbar  v, v' \sbnfbar {\abs{x}{P}}
			\\
			P,Q
			 \bnfis &
			\bout{u}{V}{P}  \sbnfbar  \binp{u}{x}{P} \sbnfbar
			 \bsel{u}{\lbl_i. P_i}_{i\in I} \sbnfbar \bbra{u}{\lbl_i:P_i}_{i \in I}  
			 \\
			 & \sbnfbar  P \Par Q \sbnfbar  {\rvar{X} \sbnfbar \recp{X}{P}} 
						\sbnfbar  
			 {\appl{V}{u}}  
			\sbnfbar \news{n} P \sbnfbar \inact
\\[2mm]
	\mytagg \bnfis & \rmark \sbnfbar \normark
	 \qquad 
	 m \bnfis V \sbnfbar \lbl
\qquad
	 h, k \bnfis  \emp   \sbnfbar h \cons \valueq{\p}{\q}{m} 
\\
%
M,N		 \bnfis &\inact 
\sbnfbar 
\myloc{\loc}{\bout{a}{x}{P}}
\sbnfbar 
\myloc{\loc}{\binp{a}{x}{P}}
\sbnfbar 
M \Par N 
\sbnfbar 
\news{n} M
\\
& \sbnfbar 
\rtsyn{\np{\loc}{\conf{\stack C}{P}}} 
\sbnfbar 
\rtsyn{\codah{s}{h}{k}{}
} 
\sbnfbar 
\rtsyn{\mem{\kappa}{(\appl{V}{u})}{\loc}} 
\sbnfbar 
\rtsyn{\monig{s}{H}{\mytilde{x}}{\store}}  
	\\
	\stack{C}, \stack{C}' \bnfis  & \inact \sbnfbar \bsel{u}{\lbl_i. P_i}_{i\in I} \sbnfbar 
\bbra{u}{ \lbl_i:P_i}_{i \in I} \sbnfbar \stack{C}_1, \stack{C}_2
\\[2mm]
    \alpha \bnfis &   \typeIn{\q}{U} \sbnfbar  \typeOut{\q}{U} 
    \\
	T,S		 \bnfis & \lend \sbnfbar	\alpha.S \sbnfbar \ltsel{\q}{\lbl_i}{S_i}{i}{I} \sbnfbar \ltbra{\q}{\lbl_i}{S_i}{i}{I}
	\\
	H,K		 \bnfis & \past S \sbnfbar S\past \sbnfbar \alpha_1.\cdots .\alpha_n.\past S  \sbnfbar  
	\ltselp{\q}{\lbl_i:S_i \;;\; \lbl_j:H_j}{i}{I} \sbnfbar \ltbrap{\q}{\lbl_i:S_i \, , \, \lbl_j:H_j}{i}{I}
		\end{align*}
	\end{mdframed}
	\caption{Syntax of processes ($P, Q$), configurations ($M, N$),
	  stacks ($\stack{C}, \stack{C}'$)
	local types ($T, S$), and local types with history ($H, K$).
    Constructs in \rtsyn{\text{boxes}} appear only at run-time.}
	\label{fig:syntax}
\end{figure}

\subsubsection{Syntax}
The syntax of processes and configurations is given in \figref{fig:syntax}.
For processes $P, Q, \ldots$ we
follow  the syntax of \HOp, the core higher-order session $\pi$-calculus studied in~\cite{KPY2016,DBLP:journals/iandc/KouzapasPY19}.
(Actually, our syntax of processes is related to \HO, the sub-language of \HOp without name-passing.)
The syntax of configurations builds upon that of processes.

\emph{Names} $a,b,c$ (resp.~$s, s'$) 
range over shared (resp. session) names. 
We use session names indexed by  {participants}, denoted $\ep{s}{\p}, \ep{s}{\q}$. 
Names $n, n'$ are session or shared names.
First-order values  $v, v'$ include base values and constants.
Variables are denoted by $x, y$ and recursion variables are denoted by $\varp{X}, \varp{Y}$.
We write $\mytilde{x}$ to denote a sequence of variables, sometimes treated as a set.
To define configurations,   we use fresh name identifiers (keys), denoted $\kappa, \kappa', \ldots$, 
and also identifiers $\loc, \loc', \ldots$, which denote    a  process \emph{location} 
or  \emph{site} (as in, e.g., the distributed $\pi$-calculus~\cite{Hennessy07}).

The syntax of values $V$ includes shared names, first-order values, but also 
abstractions $\abs{x}{P}$, where $P$ is a process.
Abstractions are {higher-order} values, as they denote functions from names to processes.
As shown in~\cite{KPY2016,DBLP:journals/iandc/KouzapasPY19}, abstraction passing suffices to express name passing (\emph{delegation}).



Process terms include prefixes for sending and receiving values $V$, written 
$\bout{u}{V} P$ and $\binp{u}{x} P$, respectively.
Given a finite index set $I$,
processes $\bsel{u}{\lbl_i. P_i}_{i\in I}$ and $\bbra{u}{l_i: P_i}_{i \in I}$ implement 
selection and branching (internal and external labeled choices, respectively). 
The selection  $\bsel{u}{\lbl_i. P_i}_{i\in I}$ is actually 
a non-deterministic choice over $I$.
In an improvement with respect to~\cite{DBLP:journals/corr/MezzinaP16,MezzinaP17}, here we consider parallel composition of processes $P \Par Q$ 
and recursion $\recp{X}{P}$ (which binds  $\varp{X}$ in process $P$).
Process $\appl{V}{u}$ is the application which leads to substitute name $u$ on the abstraction~$V$. 
Constructs for  restriction $\news{n} P$ and 
inaction $\inact$ are standard.

Session restriction $\news{s} P$ simultaneously binds all the participant endpoints in $P$.
We write $\fv{P}$ and $\fn{P}$ 
to denote the sets of free 
variables 
and names in $P$.
We assume $V$ in $\bout{u}{V}{P}$ does not include free recursion 
variables $\rvar{X}$.
If $\fv{P} = \emptyset$, we call $P$ {\em closed}.

The syntax of configurations $M, N, \ldots$, includes  inaction $\inact$, the parallel composition  $M \Par N$,
and name restriction $\news{n} M$. 
Also, it includes constructs for \emph{session initiation}: 
configuration $\myloc{\loc}{\bout{a}{x}{P}}$ denotes the \emph{request} of a service identified with $a$ implemented in 
$P$ as $x$; conversely, configuration 
$\myloc{\loc}{\binp{a}{x}{P}}$
denotes service \emph{acceptance}. 

Configurations also include the following \emph{run-time elements}: 
\begin{enumerate}[$\bullet$]
\item \emph{Running processes} are of the form $\np{\loc}{\conf{\stack C}{P}}$, where $\loc$ is a location that 
hosts a process $P$ and a \emph{(process) stack} $\stack{C}$.
A stack is simply a list of processes, useful 
  to record/reinstate the discarded alternatives   in a labeled choice.

\item \emph{Monitors} are of the form $\monig{s}{H}{\mytilde{x}}{\store}$, 
where 
$s$ is the session being monitored, 
$H$ is a local type with history (i.e. in which the cursor $\past$ acts as a ``memory''), 
$\mytilde{x}$ is a set of free variables, 
 the \emph{store} $\store$  records the value of such variables (see Def.~\ref{d:store}), 
 and $\mytagg$ is the monitor's \emph{tag} (see next).

These five elements allow us to track the current protocol
and  state of the   monitored process.
The {tag} $\mytagg$ can be either \emph{empty} (denoted `$\normark$') or \emph{full} (denoted `$\rmark$'). 
When first created, all monitors have an empty tag; a full tag indicates that the running process associated to the monitor
is currently involved in a  decoupled reversible step.
We often omit the empty tag 
(so we write 
$\moni{s}{H}{\mytilde{x}}{\store}$ instead of 
$\hmoni{s}{H}{\mytilde{x}}{\store}$)
and write $\monir{s}{H}{\mytilde{x}}{\store}$ to emphasize the reversible (red) nature of a 
monitor with full tag.

\item Following~\cite{DBLP:journals/mscs/KouzapasYHH16}, we have \emph{message queues}
$\codah{s}{h}{k}{}$,
where $s$ is a session, $h$ is the input part of the queue, $k$ is the output part of the queue, 
and `$\history$' acts as a delimiter between the two.

Each queue contains messages of the form $\valueq{\p}{\q}{\anyv}$, which is read: ``message $\anyv$ is sent from $\p$ to $\q$''.
As we will see, an output prefix in a process places the message in its corresponding output queue;
conversely, an input prefix retrieves the first message from its input queue. 
Messages in the queue are \textit{never consumed}: a process reads a message $\valueq{\p}{\q}{\anyv}$ by moving it from the (tail of) queue $k$ to the (top of) queue $h$. 
This way, the    delimiter $\history$ distinguishes the \textit{past} of the queue from its \textit{future}.

\item We use \emph{running functions}  $\mem{\kappa}{(\appl{V}{u})}{\loc}$ to reverse an 
 application $\appl{V}{u}$. While $\kappa$ is a fresh identifier (key) for this term, $\loc$ is the location of 
the running process that contains the application.

\end{enumerate}


As customary, we write $\prod_{i \in \{1..n\}} P_i$ to stand for the process $P_1 \Par P_2 \Par \cdots \Par P_n$ (and similarly for configurations).
We shall write $\procs$ and $\confs$ to indicate the set of processes and configurations, respectively. 
We call \emph{agent} an element of the set $\agents = \confs \cup \procs$. 
We let $P, Q$  to range over $\procs$; also, we use $L,M,N$ to range over $\confs$ and $A,B,C$ to range over $\agents$.

\subsubsection{Example: The Three-Buyer Seller Protocol (II)}
\label{sss:buysel2}
Continuing with the example in \S\ref{sss:buysel1}, 
we now give processes for each participant:
\begin{align*}
\text{Seller} & =  \bout{d}{x:\tproj{G}{\pS}}\binp{x}{t}\bout{x}{\mathit{price}(t)}\bout{x}{\mathit{price}(t)}
\binp{x}{ok}\binp{x}{a} \bout{x}{date}\inact  
\\
\text{Alice} & =  \binp{d}{y:\tproj{G}{\pA}}\bout{y}{\exBook}\binp{y}{p}\binp{y}{s}\binp{y}{ok}\inact  
\\
\text{Betty} & =  \binp{d}{z:\tproj{G}{\pB}}\binp{z}{p}\binp{z}{s}\bout{z}{ok}\bout{z}{ok}\bout{z}{s}
  \bbout{z}{\thunkp{\bout{z}{\text{`Urbino, 61029'}}\binp{z}{d}\inact}}\inact
  \\
\text{Carol} & =  \binp{d}{w:\tproj{G}{\pC}}\binp{w}{s}\binp{w}{code}(\appl{code}{\dummyn})
\end{align*}
where we assume $\mathit{price}(\cdot)$ returns a value of type $\mathsf{{price}}$ given a $\mathsf{title}$.
Observe how Betty's implementation sends part of its protocol to Carol in the form of a thunk containing 
her session name $z$ and address. This is how abstraction passing implements session delegation.

The whole system, given by configuration $M$ below, is obtained by placing these process implementations in appropriate locations:
\begin{equation}
\label{eq:buyer}
	M = \myloc{\loc_1}{\text{Seller}} 
\Par
\myloc{\loc_2}{\text{Alice}} 
\Par
\myloc{\loc_3}{\text{Betty}} 
\Par 
\myloc{\loc_4}{\text{Carol}} 
\end{equation}

\subsection{A Decoupled Semantics for Configurations}\label{ss:semconf}
We define a reduction relation on configurations, coupled with a structural congruence
 on processes and configurations.
Our reduction semantics defines a \emph{decoupled} treatment for reversing communication actions within a protocol.
Reduction is thus defined as $\red \subset \confs \times \confs$, whereas structural congruence is defined as $\scong\, \subset \procs^{2} \cup \confs^{2}$.  

\subsubsection{Preliminaries}
We require auxiliary definitions for \emph{contexts}, \emph{stores}, and \emph{type contexts}.

\emph{Evaluation contexts} are configurations with one hole 
 `$\bullet$', as defined by the following grammar:
$$\ctx{E}\bnfis \bullet \sbnfbar M\Par \ctx{E} \sbnfbar \news{n}\, \ctx{E}$$
\emph{General contexts} $\ctx{C}$ are processes or configurations with one hole~$\bullet$: they are obtained  by replacing one occurrence of $\inact$ (either as a process or as a configuration) with $\bullet$.
A congruence on processes and configurations is an equivalence relation $\Re$ that is closed under general contexts: $P \,\Re\,  Q \Longrightarrow \myctx{C}{P}\,\Re\,  \myctx{C}{Q}$
and $M\Re N \Longrightarrow \myctx{C}{M}\,\Re\,  \myctx{C}{N}$.

We define $\equiv$  as the smallest congruence on processes and configurations
that satisfies the rules in Fig.~\ref{fig:str} \added{and is closed under the equivalence on queues defined below. 
\begin{defi}[Equivalence on message queues]\label{eq:queue}
We define the structural equivalence on queues, denoted $\equivq$, as follows:
\begin{align*}
h\cons \valueq{\p_1}{\q_1}{\anyv_1} \cons \valueq{\p_2}{\q_2}{\anyv_2}  \cons h'
\equivq 
h\cons \valueq{\p_2}{\q_2}{\anyv_2} \cons \valueq{\p_1}{\q_1}{\anyv_1} \cons h'
\end{align*}
whenever $\p_1 \neq \p_2 \land \q_1 \neq \q_2$.
The equivalence $\equivq$ extends  to configurations as expected.
\end{defi}
}
A relation $\Re$ on configurations is \emph{evaluation-closed} if it satisfies the   following rules:
\begin{mathpar}
 \inferrule*[left=(Ctx)]{M\,\Re\, N}{\myctx{E}{M}\,\Re\, \myctx{E}{N}} \and 
 \inferrule*[left=(Eqv)]{M \scong M' \and M' \,\Re\, N' \and N'\scong N}{M \,\Re\, N} \and 
\end{mathpar}

\begin{figure}[t!]
\begin{mdframed}
{
\[
	\begin{array}{c}
		A \Par \inact \scong A
		\quad
		A \Par B \scong B \Par A
		\quad
		A \Par (B \Par C) \scong (A \Par B) \Par C
		\\[1mm]
		A \Par \news{n} B \scong \news{n}(A \Par B)
		\ (n \notin \fn{A})
				\quad
		\news{n} \inact \scong \inact
		\\[1mm]
		\recp{X}{P} \scong P\subst{\recp{X}{P}}{\rvar{X}}
		\quad
		A \scong B \textrm{ if } A \scong_\alpha B
	\end{array}
\]
}
\end{mdframed}
\caption{Structural Congruence}
\label{fig:str}
\end{figure}
%

The state of monitored processes is formalized as follows:

\begin{defi}[Store] \label{d:store}
The \emph{store} $\store$ is a mapping from variables to values. Given a store $\store$, a variable $x$, and a value $V$, 
the \emph{update} $\store\upd{x}{V}$ and the 
 \emph{reverse update} $\rup{\store}{x}$ are defined as follows:
\begin{align*}
\store\upd{x}{V} & = \begin{cases}
\store\cup \{(x,V)\} &\text{ if } x\not\in \mathtt{dom}(\store) \\ 
\text{ undefined }  & \text{otherwise}
\end{cases} 
\\
\rup{\store}{x} & =  \begin{cases}
\store_1 &\text{ if } \store=\store_1\cup \{(x,V)\} \\ 
\store  & \text{otherwise}
\end{cases} 
\end{align*}
The \emph{evaluation} of value $V$ under store $\store$, written 
$\myeval{V}{\sigma}$, is defined as follows:
$$
\myeval{V}{\sigma} = \begin{cases}
	W & \text{if $V = x$ and $\store = \store' \cup \{(x,W)\}$}
	\\
	V & \text{otherwise}
\end{cases}
$$
\end{defi}

\noindent Together with local types with history, the following notion of type context allows us to record the current protocol state:
\begin{defi} 
 We define 
\textit{type contexts} as (local) types with one hole, denoted `$\bullet$':
\begin{align*}
 \ctx{T},\ctx{S}   \bnfis  &  \bullet  \sbnfbar
 \q\btsel{\lbl_w:\ctx{T} \;;\; \lbl_i:S_i}_{i \in I\setminus w} \sbnfbar \q\btbra{\lbl_w:\ctx{T} \, , \, \lbl_i:S_i}_{i \in I\setminus w}
  \sbnfbar   \alpha.\ctx{T} \sbnfbar    \kappa.\ctx{T} \sbnfbar (\loc,\loc_1,\loc_2).\ctx{T} 
\end{align*}
\end{defi}
\noindent Type contexts $\kappa.\ctx{T}$ 
and $(\loc,\loc_1,\loc_2).\ctx{T}$ 
will be instrumental in formalizing reversibility of
name applications and thread spawning, respectively, which are not described by local types.


As already mentioned, abstraction passing can represent name passing in a fully abstract way (cf.~\cite{KPY2016,DBLP:journals/iandc/KouzapasPY19}).
Such a representation suffices to implement a form of \emph{session delegation}, by including free session names (indexed by participant identities) in the body of an abstraction
(cf. Betty's implementation, discussed above).
The following definition identifies those names:

\begin{defi}
Let 
$h$ and $\p$ be a queue 
and a participant, respectively. 
Also, let  
$\{\valueq{\q_1}{\p}{\abs{x_1}{P_1}},$ $\ldots, \valueq{\q_k}{\p}{\abs{x_k}{P_k}} \}$ 
 denote the (possibly empty) set of messages in $h$ containing abstractions sent to $\p$.
We write 
$\names{\p, h}$ to denote the set of participant identities occurring in 
$P_1, \ldots, P_k$.
\end{defi}

\begin{figure}[!t]
\begin{mdframed}
\begin{align*}
& \inferrule[\fwcolor{(Init)}]
{\parties{G}= \{\p_1,\cdots,\p_n\} \and 
T_1 = \tproj{G}{\gpart{p}_1} ~~ \cdots~~   T_n = \tproj{G}{\gpart{p}_n}
}{
\prod_{i \in \{1 .. n\}} L_i ~
\fw 
\news{s}\Big(\codah{s}{\emp}{\emp}{} \Par  \prod_{i \in \{1 .. n\}} M_i \Par N_i 
\Big)}
\\
& \text{where:}
\\
& \begin{array}{ll}
L_1  = \myloc{\loc_1}{\bout{a}{x_1:T_1}{P_{1}}} 
&
M_i  = \np{\key{\loc_i}{\p_i}}{ \conf{\inact}{P_i\subst{\ep{s}{\p_i}}{x_i}}} \text{~for $i = 1..n$} 
\\
L_j  = \myloc{\loc_j}{\binp{a}{x_j:T_j}{P_{j}}} \text{~for $j = 2..n$} 
&
N_i  = \moni{s_{\p_i}}{\past T_{i}}{x_i}{\upd{x_i}{a }} \text{~for $i = 1..n$} 
\end{array}
\end{align*}
\begin{align*}
& \inferrule[\fwcolor{(Out)}]
{\p = \er \,\vee\, \p \in \names{\er, h}
}{
M \Par N  \Par \codah{s}{h}{k}{}  
~ \fw 
M' \Par N' \Par \codah{s}{h}{k'}{}
}
\\
& \text{where:}
\\
& 
\begin{array}{lll}
M  = \np{\key{\loc}{\er}}{ \conf{\stack{C}}{\bout{\ep{s}{\p}}{V}{P}}}
&
N  = \moni{s_\p}{\myctxr{\ctx{T}}{\past  \ltout{\q}{U}{S}}}{\mytilde x}{\store}
& 
k'  = k\cons \valueq{\p}{\q}{\myeval{V}{\sigma}}
\\
M'  = \np{\key{\loc}{\er}}{\conf{\stack{C}}{P}}
&
N'  =\moni{s_\p}{\myctxr{\ctx{T}}{ \ltoutp{\q}{U}{S}}}{\mytilde x}{\store}
&
\end{array}
\end{align*}
\begin{align*}
& \inferrule[\fwcolor{(In)}]
{\p = \er \,\vee\, \p \in \names{\er, h}
}{
M \Par N  \Par  \codah{s}{h}{k}{}
~ \fw 
M' \Par N' \Par \codah{s}{h'}{k'}{} 
}
\\
& \text{where:}
\\
& \begin{array}{ll}
M  = \np{\key{\loc}{\er}}{\conf{\stack{C}}{\binp{\ep{s}{\p}}{y}{P}}} 
&
N  = \moni{s_\p}{\myctxr{\ctx{T}}{\past  \ltinp{\q}{U}{S}}}{\mytilde x}{\store} 
\\
M'  = \np{\key{\loc}{\er}}{\conf{\stack{C}}{P}} 
&
N'  =\moni{s_\p}{ \myctxr{\ctx{T}}{\ltinpp{\q}{U}{S}}}{\mytilde x, y}{\store\upd{y}{V}} 
\\
k  = \valueq{\q}{\p}{V}\cons k' 
&
h'  = h \cons \valueq{\q}{\p}{V}
\end{array}
\end{align*}
	
\end{mdframed}

\caption{Decoupled semantics for configurations: Forward reduction ($\fw$) - Part~1/2.}
\label{fig:fw_1}
\end{figure}

\begin{figure}[!t]
 \begin{mdframed}
\begin{align*}
& \inferrule[\fwcolor{(Sel)}]
{\p = \er \,\vee\, \p \in \names{\er, h}
\and
w\in J \and J\subseteq I 
}{
M \Par N  \Par  \codah{s}{h}{k}{}
~ \fw 
M' \Par N' \Par \codah{s}{h}{k'}{} 
}
\\
& \text{where:}  
\\
& \begin{array}{ll}
M  = \np{\key{\loc}{\er}}{ \conf{\stack C}{\bsel{\ep{s}{\p}}{\lbl_i. P_i}_{i\in I}}} 
&
N  =\moni{s_\p}{\myctxr{\ctx{T}}{\past \ltsel{\q}{\lbl_j}{S_j}{j}{J}} }{\mytilde x}{\store} 
\\
M'  = \np{\key{\loc}{\er}}{ \conf{\stack C , \bsel{\ep{s}{\p}}{\lbl_i.P_i}_{i\in I\setminus w} }{P_w} }
&
N'  = \moni{s_\p}{ \myctxr{\ctx{T}}{ \ltselp{\q}{\lbl_j:S_j \, , \, \lbl_w:\past S_w}{j}{J\setminus w}} }{\mytilde x}{\store} 
\\
&  
k'  = k \cons  \valueq{\p}{\q}{\lbl_w} 
\end{array}
\end{align*}
\begin{align*}
& \inferrule[\fwcolor{(Bra)}]
{\p = \er \,\vee\, \p \in \names{\er, h}
\and
w\in J \and J\subseteq I 
}{
M \Par N  \Par  \codah{s}{h}{k}{}
~ \fw 
M' \Par N' \Par \codah{s}{h'}{k'}{} 
}
\\
& \text{where:}  
\\
& \begin{array}{ll}
M  = \np{\key{\loc}{\er}}{ \conf { \stack C}{\bbra{\ep{s}{\p}}{\lbl_i:P_i}_{i \in I} } } 
&
N  =\moni{s_\p}{\myctxr{\ctx{T}}
{\past \ltbra{\q}{\lbl_j}{S_j}{j}{J}} }{\mytilde x}{\store} 
\\
M'  = \np{\key{\loc}{\er}}{ \conf { \stack C , \bbra{\ep{s}{\p}}{\lbl_i:P_i}_{i \in I\setminus w}}{ P_w} } 
&
N'  = \moni{s_\p}{\myctxr{\ctx{T}}{\ltbrap{\q}{\lbl_j:S_j\,,\, \lbl_w:\past S_w}{j}{J\setminus w}} }{\mytilde x}{\store}
\\
k  = \valueq{\q}{\p}{\lbl_w}\cons k'
& 
h'  = h\cons  \valueq{\q}{\p}{\lbl_w} 
\end{array}
\end{align*}
\begin{align*}
& \inferrule[\fwcolor{(Beta)}]
{\myeval{V}{\sigma} =\abs{x}{P} 
}{
M \Par N  
~ \fw 
\news{\kappa}\left(  
M' \Par N' \Par \mem{\kappa}{(\appl{V}{w})}{\loc} \right)  
}
\\
& \text{where:}  
\\
& \begin{array}{ll}
M  = \np{\key{\loc}{\p}}{\conf{\stack{C}}{(\appl{V}{w})}}
&
N  =\moni{s_\p}{\myctxr{\ctx{T}}{\past S}}{\mytilde x}{\store} 
\\
M'  = {\np{\key{\loc}{\p}}{\conf{\ctx{\stack {C}}}{P\subst{\myeval{w}{\sigma}}{x}}}} 
&
N'  = \moni{s_\p}{\myctxr{\ctx T}{ \kappa. \past S}}{\mytilde x}{\store}
\end{array}
\end{align*}
\begin{align*}
& \inferrule[\fwcolor{(Spawn)}]
{\myeval{V}{\sigma} =\abs{x}{P} 
}{
M \Par N  
~ \fw 
\news{\loc_1,\loc_2} \left(  
M' \Par 
\np{\key{\loc_1}{\p}}{\conf{\inact}{P}}
\Par 
\np{\key{\loc_2}{\p}}{\conf{\inact}{Q}} 
\Par N'  \right)  
}
\\
& \text{where:}  
\\
& \begin{array}{ll}
M  = \np{\key{\loc}{\p}}{\conf{\stack{C}}{P \Par Q}} 
&
N  =\moni{s_\p}{\myctxr{\ctx{T}}{\past S}}{\mytilde x}{\store} 
\\
M'  = \np{\key{\loc}{\p}}{\conf{\stack{C}}{\inact}} 
&
N'  = \moni{s_\p}{\myctxr{\ctx T}{ (\loc,\loc_1,\loc_2). \past S}}{\mytilde x}{\store} 
\end{array}
\end{align*}
 \end{mdframed}
\caption{Decoupled semantics for configurations: Forward reduction ($\fw$) - Part~2/2.}
\vspace{1cm} 
\label{fig:fw_2}
\end{figure}



\subsubsection{Reduction}
We define $\red$  as the union of two relations: 
the forward and backward
reduction relations, denoted $\fw$  and $\bk$, respectively. That is, $\red = \fw \cup \bk$. Relations $\fw$ and 
$\bk$ are the smallest evaluation-closed relations satisfying the rules in Figs.~\ref{fig:fw_1} -- \ref{fig:bk_2}. We indicate with $ \trans{\red} $, $ \fws $, and $\bks $  the reflexive and transitive closure of $\red$, $\fw$, and $\bk$, respectively.

We now discuss the forward reduction rules (Fig.~\ref{fig:fw_1} and Fig.~\ref{fig:fw_2}), omitting empty tags $\normark$:

\begin{itemize}[$\fwcolor{\blacktriangleright}$]
\item Rule \fwcolor{\textsc{(Init)}} initiates a given protocol  $G$ with $n$ participants.
Given the composition of one service request and $n-1$ service accepts (all along $a$, available in different locations~$\loc_i$),
this rule establishes the session by setting up the run-time elements:
running processes and  monitors---one for each participant, with empty tag (omitted)---and the empty session queue. 
A unique session identifier  ($s$ in the rule) is also created.
The processes are inserted in 
their respective running structures, and instantiated with an appropriate session name.
The local types for each participant are inserted in their respective monitor, with the cursor `$\past$' at the beginning.

\item Rule \fwcolor{\textsc{(Out)}} starts the output of value $V$ from $\p$ to $\q$. 
Given an output-prefixed process as running process, and a monitor with a local type supporting an output action, reduction adds the message $\valueq{\p}{\q}{\myeval{V}{\sigma}}$ to the output part of the session queue (where $\sigma$ is the current store). 
Also, the cursor within the local type is moved accordingly.
In this rule (but also in several other rules), the premise $\p = \er \,\vee\, \p \in \names{\er, h}$ allows performing actions on names previously received via abstraction passing.

\item Rule \fwcolor{\textsc{(In)}} allows a participant $\p$ to receive a value $V$ from $\q$: it  takes the first element of the output part of the queue and places it in the input part.
The cursor of the local type and the state in the  monitor for $\p$ are adjusted accordingly.

\item Rule \fwcolor{\textsc{(Sel)}} is the forward rule for labeled selection, which in our case entails a non-deterministic choice between 
pairwise different labels indexed by $I$. 
\added{We require that $I$ (the set that indexes the choice in the process) is contained in $J$ (the set that indexes the choice in the protocol)}. 
After reduction, the selected label ($\lbl_w$ in the rule) is added to the output part of the queue, and the continuation $P_w$ is kept in the running process;
to support reversibility,
alternatives different from $\lbl_w$ are stored in the stack $\stack{C}$ with their continuations.
The cursor is also adjusted in the  monitor accordingly.

\item Rule~\fwcolor{\textsc{(Bra)}} is similar to Rule \fwcolor{\textsc{(Sel)}}: it takes a message containing a label $\lbl_w$ as the first element in the output part of the queue, and places it into the input part. 
This entails a selection between the options indexed by $I$; the continuation $P_w$ is kept in the running process, and all those options different from $\lbl_w$ are kept in the stack. Also, the local type in the monitor is adjusted accordingly.

\item Rule~\fwcolor{\textsc{(Beta)}} handles applications, which in our setting are always name applications.
 Reduction creates a fresh identifier ($\kappa$ in the rule) for the running function, which keeps (i) the structure of the process prior to application, and (ii) the identifier of the running process that ``invokes'' the application. Notice that $\kappa$ is recorded also in the monitor: this is needed to undo applications in the right order. We use the store  $\store$ to determine the actual abstraction and the name applied.

\item Rule~\fwcolor{\textsc{(Spawn)}} handles parallel composition. 
Location $\loc$ is ``split'' into running processes with fresh identifiers ($\loc_1, \loc_2$ in the rule). This split is recorded in the monitor.
\end{itemize}


\begin{figure}[!h]
\begin{mdframed}
\begin{align*}
& \inferrule[\bkcolor{(RInit)}]
{\parties{G}= \{\p_1,\cdots,\p_n\} \and 
T_1 = \tproj{G}{\gpart{p}_1} ~~ \cdots~~   T_n = \tproj{G}{\gpart{p}_n}
}{
\news{s}\Big(\codah{s}{\emp}{\emp}{} \Par \prod_{i \in \{1 .. n\}} M_i \Par N_i 
\Big)
\bk
\prod_{i \in \{1 .. n\}} L_i 
}
\\
& \text{where:}
\\
& \begin{array}{ll}
L_1  = \myloc{\loc_1}{\bout{a}{x_1:T_1}{P_{1}}} 
&
M_i  = \np{\key{\loc_i}{\p_i}}{ \conf{\inact}{P_i\subst{\ep{s}{\p_i}}{x_i}}} \text{~for $i = 1..n$} 
\\
L_j  = \myloc{\loc_j}{\binp{a}{x_j:T_j}{P_{j}}} \text{~for $j = 2..n$} 
&
N_i  = \hmoni{s_{\p_i}}{\past T_{i}}{x_i}{\upd{x_i}{a }} \text{~for $i = 1..n$} 
\end{array}
\end{align*}
\begin{align*}
& \inferrule[\bkcolor{(ROut)}]
{\p = \er \,\vee\, \p \in \names{\er, h}
}{
M \Par N \Par \codah{s}{h}{k}{}
~\bk 
M' \Par N'  \Par \codah{s}{h}{k'}{}  
}
\\
& \text{where:}
\\
& \begin{array}{lll}
M  = \np{\key{\loc}{\er}}{\conf{\stack{C}}{P}}
&
N  =\monir{s_\p}{\myctxr{\ctx{T}}{ \ltoutp{\q}{U}{S}}}{\mytilde x}{\store}
& 
k  =  \valueq{\p}{\q}{V} \cons k'
\\
M'  = \np{\key{\loc}{\er}}{ \conf{\stack{C}}{\bout{\ep{s}{\p}}{V}{P}}}
&
N'  = \hmoni{s_\p}{\myctxr{\ctx{T}}{\past  \ltout{\q}{U}{S}}}{\mytilde x}{\store}
& 
\end{array}
\end{align*}
\begin{align*}
& \inferrule[\bkcolor{(RIn)}]
{\p = \er \,\vee\, \p \in \names{\er, h}
}{
M \Par N  \Par  \codah{s}{h}{k}{}
~ \bk 
M' \Par N' \Par \codah{s}{h'}{k'}{} 
}
\\
& \text{where:}
\\
& \begin{array}{lll}
M  = \np{\key{\loc}{\er}}{\conf{\stack{C}}{P}} 
&
N  =\monir{s_\p}{ \myctxr{\ctx T} {\ltinpp{\q}{U}{S}} }{\mytilde x,y}{\store} 
&
h  = \valueq{\q}{\p}{V}\cons h' 
\\
M'  = \np{\key{\loc}{\er}}{\conf{\stack{C}}{\binp{\ep{s}{\p}}{y}{P}}} 
&
N'  = \hmoni{s_\p}{ \myctxr{\ctx T}{\past\ltinp{\q}{U}{S}}}{\mytilde x}{\rup{\store}{y}}
&
k'  =  \valueq{\q}{\p}{V} \cons k 
\end{array}
\end{align*}
\end{mdframed}
\caption{Decoupled semantics for configurations: Backwards reduction ($\bk$) - Part~1/3.}
\label{fig:bk_1}
\end{figure}

\begin{figure}[!h]
\begin{mdframed}
\begin{align*}
& \inferrule[\bkcolor{(RollS)}]
{ 
}{
N_1^{\normark} \Par N_2^{\normark} 
\Par \codah{s}{h}{k}{}
~ \bk 
N_1^{\rmark} \Par N_2^{\rmark} 
\Par \codah{s}{h}{k}{}  }
\\
& \text{where:}
\\
& \begin{array}{ll}
N_1  = 
\moni{s_\p}{ \myctxr{\ctx T}{\ltinpp{\q}{U}{T}}}{\mytilde x}{\store_1} 
&
N_2  = \moni{s_\q}{ \myctxr{\ctx S}{\ltoutp{\p}{U}{S}}}{\mytilde y}{\store_2}  
\end{array}
\end{align*}
\begin{align*}
& \inferrule[\bkcolor{(RollC)}]
{ 
}{
N_1^{\normark} \Par N_2^{\normark} 
\Par \codah{s}{h}{k}{}
~ \bk 
N_1^{\rmark} \Par N_2^{\rmark} 
\Par \codah{s}{h}{k}{}  }
\\
& \text{where:}
\\
& \begin{array}{l}
N_1  = 
\moni{s_\p}{ \myctxr{\ctx T}{\ltbrap{\q}{\lbl_z:\past S_z\,,\,\lbl_w:S_w}{z}{J\setminus w}}}{\mytilde x}{\store_1}
\\
N_2  = \moni{s_\q}{\myctxr{\ctx S}{\ltselp{\p}{\lbl_z:\past S_z \, , \, \lbl_w: S_w}{z}{J\setminus w}}}{\mytilde y}{\store_2}  
\end{array}
\end{align*}
\end{mdframed}
\caption{Decoupled semantics for configurations: Backwards reduction ($\bk$) - Part~2/3.}
\label{fig:bk_2}
\end{figure}

\begin{figure}[!t]
\begin{mdframed}
\begin{align*}
& \inferrule[\bkcolor{(RSel)}]
{\p = \er \,\vee\, \p \in \names{\er, h}
\and
w\in J \and J\subseteq I 
}{
M \Par N  \Par  \codah{s}{h}{k}{}
~ \bk
M' \Par N' \Par \codah{s}{h}{k'}{} 
}
\\
& \text{where:}
\\
& \begin{array}{ll}
M  = \np{\key{\loc}{\er}}{ \conf{\stack C , \bsel{\ep{s}{\p}}{\lbl_i.P_i}_{i\in I\setminus w} }{P_w} }
&
N  = \monir{s_\p}{ \myctxr{\ctx T}{ \ltselp{\q}{\lbl_j:S_j \, , \, \lbl_w:\past S_w}{j}{J\setminus w} } }{\mytilde x}{\store}
\\
M'  = \np{\key{\loc}{\er}}{ \conf{\stack C}{\bsel{\ep{s}{\p}}{\lbl_i. P_i}_{i\in I}}} 
&
N'  = \hmoni{s_\p}{ \myctxr{\ctx T}{ \past \ltsel{\q}{\lbl_j}{S_j}{j}{J} } }{\mytilde x}{\store} 
\\
k  = \valueq{\p}{\q}{\lbl_w}\cons k' &
\end{array}
\end{align*}
\begin{align*}
& \inferrule[\bkcolor{(RBra)}]
{\p = \er \,\vee\, \p \in \names{\er, h}
\and
w\in J \and J\subseteq I 
}{
M \Par N  \Par  \codah{s}{h}{k}{}
~ \bk 
M' \Par N' \Par \codah{s}{h'}{k'}{} 
}
\\
& \text{where:}
\\
& \begin{array}{ll}
M  = \np{\key{\loc}{\er}}{ \conf { \stack C , \bbra{\ep{s}{\p}}{\lbl_i:P_i}_{i \in I\setminus w}}{ P_w} } 
&
N  = \monir{s_\p}{\myctxr{\ctx T}{\ltbrap{\q}{\lbl_j:S_j\,,\, \lbl_w:\past S_w}{j}{J\setminus w}} }{\mytilde x}{\store}
\\
M'  = \np{\key{\loc}{\er}}{ \conf { \stack C}{\bbra{\ep{s}{\p}}{\lbl_i:P_i}_{i \in I} } } 
&
N'  =\hmoni{s_\p}{\myctxr{\ctx T}{\past\ltbra{\q}{\lbl_j}{S_j}{j}{J}} }{\mytilde x}{\store} 
\\
h  = h' \cons  \valueq{\q}{\p}{\lbl_w} 
&
k'  = k \cons \valueq{\q}{\p}{\lbl_w}
\end{array}
\end{align*}
 \begin{align*}
& \inferrule[\bkcolor{(RBeta)}]
{  
}{
\news{k}\left(  
M \Par N   \Par \mem{\kappa}{(\appl{V}{w})}{\loc} \right)  
\bk 
M' \Par N'
}
\\
& \text{where:}
\\
& \begin{array}{ll}
M  = {\np{\key{\loc}{\p}}{\conf{\ctx{\stack {C}}}{P}}} 
&
N  = \moni{s_\p}{\myctxr{\ctx T}{ \kappa. \past S}}{\mytilde x}{\store}
\\
M'  = \np{\key{\loc}{\p}}{\conf{\stack{C}}{(\appl{V}{w})}}
&
N'  =\moni{s_\p}{\myctxr{\ctx{T}}{\past S}}{\mytilde x}{\store} 
\end{array}
\end{align*}
\begin{align*}
& \inferrule[\bkcolor{(RSpawn)}]
{ 
}{
\news{\loc_1,\loc_2} \left(  
M \Par 
\np{\key{\loc_1}{\p}}{\conf{\inact}{P}}
\Par 
\np{\key{\loc_2}{\p}}{\conf{\inact}{Q}} 
\Par N  \right)  
\bk 
M' \Par N'  
}
\\
& \text{where:}
\\
& \begin{array}{ll}
M  = \np{\key{\loc}{\p}}{\conf{\stack{C}}{\inact}} 
&
N  = \moni{s_\p}{\myctxr{\ctx T}{ (\loc,\loc_1,\loc_2). \past S}}{\mytilde x}{\store} 
\\
M'  = \np{\key{\loc}{\p}}{\conf{\stack{C}}{P \Par Q}} 
&
N'  =\moni{s_\p}{\myctxr{\ctx{T}}{\past S}}{\mytilde x}{\store} 
\end{array}
\end{align*}
\end{mdframed}
\caption{Decoupled semantics for configurations: Backwards reduction ($\bk$) - Part~3/3.}
\label{fig:bk_3}
\end{figure}


\medskip
\noindent
Now we comment on the backward rules (Fig.~\ref{fig:bk_1}, Fig.~\ref{fig:bk_2}, and Fig.~\ref{fig:bk_3}) which, in most cases, change the monitor tags ($\normark$ and $\bkcolor{\rmark}$):

\begin{itemize}[$\bkcolor{\blacktriangleleft}$]
\item Rule \bkcolor{\textsc{(RInit)}} reverses session establishment. 
It requires that local types for every participant are at the beginning of the protocol, and that session queue and process stacks are empty. Run-time elements are discarded; located service accept/requests are reinstated.

\item Rule~\bkcolor{\textsc{(RollS)}} starts to reverse an input-output synchronization between $\p$ and $\q$. 
Enabled when there are complementary session types in the two monitors, this rule changes the monitor tags  from $\normark$ to $\bkcolor{\rmark}$. 
This way, the undoing of input and output actions occurs in a decoupled way. 
Rule \bkcolor{\textsc{(RollC)}} is the analog of~\bkcolor{\textsc{(RollS)}} but for synchronizations originated in labeled choices.

\item Rule \bkcolor{\textsc{(ROut)}} reverses an output. This is only possible for a monitor tagged with $\bkcolor{\rmark}$, exploiting the first message in the input queue. 
After reduction, the process prefix is reinstated, the cursor is adjusted, the message is removed from the queue, 
and the monitor is tagged  with $\normark$.
Rule~\bkcolor{\textsc{(RIn)}} is the analog of Rule \bkcolor{\textsc{(ROut)}}. In this case, we also need to update the state of the store $\store$. 

\item Rule \bkcolor{\textsc{(RBra)}} reverses the input part of a labeled choice: the choice context is reinstated; the cursor is moved; the last message in the input part of the queue is moved to the output part. 
Rule \bkcolor{\textsc{(RSel)}} is the analog of~\bkcolor{\textsc{(RBra)}}, but for the output part of the labeled choice. The non-deterministic   selection is reinstated.

\item Rule \bkcolor{\textsc{(RBeta)}} undoes $\beta$-reduction, reinstating the application. The running function disappears, using the information in the 
  monitor ($k$ in the rule).
Rule~\bkcolor{\textsc{(RSpawn)}} undoes the spawn of a parallel thread, using the identifiers in the  monitor. 
\end{itemize}

\subsubsection{Example: The Three-Buyer Seller Protocol (III)}
\label{sss:examp-tb}

We conclude the example in \S\,\ref{sss:buysel1} and \S\,\ref{sss:buysel2} by illustrating the semantics of configurations ($\fw$ and $\bk$). 

Consider configuration $M$ as in \eqref{eq:buyer}.
The session starts with an application of 
Rule \fwcolor{\textsc{(Init)}}:
\begin{align*}
M  \fw &~ \news{s}\big(\, 
\np{\key{\loc_1}{\pS}}{ \conf{\inact}{S_1\subst{\epS}{x}}} \Par 
\hmoni{s_{\pS}}{\past \tproj{G}{\pS}}{x}{\upd{x}{d}}  
\\
& \Par \np{\key{\loc_2}{\pA}}{ \conf{\inact}{A_1\subst{\epA}{y}}} \Par 
\hmoni{s_\pA}{\past \tproj{G}{\pA}}{y}{\upd{y}{d}} 
\\
& \Par \np{\key{\loc_3}{\pB}}{ \conf{\inact}{B_1\subst{\epB}{z}}} \Par 
\hmoni{s_\pB}{\past \tproj{G}{\pB}}{z}{\upd{z}{d}} 
\\
& \Par \np{\key{\loc_4}{\pC}}{ \conf{\inact}{C_1\subst{\epC}{w}}} \Par 
\hmoni{s_\pC}{\past \tproj{G}{\pC}}{w}{\upd{w}{d}}  
\Par \codah{s}{\emp}{\emp}{}~\big)  = M_1
\end{align*}
where 
$S_1$, $A_1$, $B_1$, and $C_1$ 
stand for the continuation of processes $\text{Seller}$, $\text{Alice}$, $\text{Betty}$, and $\text{Carol}$ after the service 
request/declaration. This way, e.g., 
$
A_1 = \bout{y}{\exBook}\binp{y}{p}\binp{y}{s}\binp{y}{ok}\inact
$.
We use configuration $M_1$ to illustrate some forward and backward reductions.

From $M_1$ we could either undo the reduction (using Rule~\bkcolor{\textsc{(RInit)}})
or execute the communication from $\text{Alice}$ to $\text{Seller}$ (using  Rules~\fwcolor{\textsc{(Out)}} and ~\fwcolor{\textsc{(In)}}). This latter option would proceed as follows:
\begin{align*}
M_1 \fw &  \news{s}(\,  \np{\key{\loc_2}{\pA}}{ \conf{\inact}{\binp{\epA}{p}\binp{\epA}{s}\binp{\epA}{ok}\inact}} 
\\
& 
\Par 
\hmoni{s_\pA}{\ltout{V}{\mathsf{title}}{\past \ltinp{V}{\mathsf{{price}}}{\ltout{B}{\mathsf{share}}{\ltinp{B}{\mathsf{OK}}{\lend}}}}}{y}{\upd{y}{d}} 
\\
& 
\Par N_2 \Par \codah{s}{\emp}{\valueq{\pA}{\pS}{\exBook}}{}~)  = M_2
\end{align*}
where $N_2$ stands for the running processes and monitors for Seller, Betty, and Carol, not involved in the reduction.
We now have:
\begin{align*}
M_2 \fw & \news{s}(\,  \np{\key{\loc_1}{\pS}}{ \conf{\inact}{\bout{\epS}{\mathit{price}(t)} \bout{\epS}{\mathit{price}(t)}\binp{\epS}{ok}
\binp{\epS}{a} \bout{\epS}{date}\inact }} 
\\
& \Par 
\hmoni{s_{\pS}}{\ltinp{A}{\mathsf{title}}{\past \ltout{\{A,B\}}{\mathsf{{price}}}{T_\pS}}}{x,t}{\store_3}  \Par N_3
\Par \codah{s}{\valueq{\pA}{\pS}{\exBook}}{\emp}{}~)  = M_3
\end{align*}
where 
$N_3$ stands for the participants not involved in the reduction,
$\store_3$ stands for the resulting store $\upd{x}{d},\upd{t}{\exBook}$,
and
$T_\pS  = \ltinp{B}{\mathsf{OK}}{\ltinp{B}{\mathsf{address}}{\ltout{B}{\mathsf{date}}{\lend}}}$.
Observe that the cursors in monitors $s_\pS$ and $s_\pA$ have evolved, and that the message from $\pA$ to $\pS$ has now been moved to the input queue.

We illustrate  reversibility by showing how to return to $M_1$ starting from $M_3$.
We need to apply three rules: \bkcolor{\textsc{(RollS)}}, \bkcolor{\textsc{(RIn)}}, and \bkcolor{\textsc{(ROut)}}.
Reversibility is decoupled in the sense that there is no fixed order in which the latter two rules should be applied; below we just illustrate a possible sequence.
First, Rule~\bkcolor{\textsc{(RollS)}} modifies the tags of monitors $s_\pS$ and $s_\pA$,
  leaving the rest unchanged:
\begin{align*}
M_3  \bk &  \news{s}(\,  
\np{\key{\loc_1}{\pS}}{ \conf{\inact}{\bout{\epS}{\mathit{price}(t)}\bout{\epS}{\mathit{price}(t)}\binp{\epS}{ok} 
\binp{\epS}{a} \bout{\epS}{date}\inact }} 
\\
& \Par 
 \monir{s_\pS}{\ltinp{A}{\mathsf{title}}{\past \ltout{\{A,B\}}{\mathsf{{price}}}{T_\pB}}}{x,t}{\store_3} 
\\
& \Par \np{\key{\loc_2}{\pA}}{ \conf{\inact}{\binp{\epA}{p}\binp{\epA}{s}\binp{\epA}{ok}\inact}} 
\\
& \Par 
\monir{s_\pA}{\ltout{S}{\mathsf{title}}{\past \ltinp{S}{\mathsf{{price}}}{\ltout{B}{\mathsf{share}}{\ltinp{B}{\mathsf{OK}}{\lend}}}}}{y}{\upd{y}{d}} 
\\
& \Par N_4 \Par \codah{s}{\valueq{\pA}{\pS}{\exBook}}{\emp}{}~)  = M_4
\end{align*}
where, as before, $N_4$ represents participants not involved in the reduction.
$M_4$ has several possible forward and backward reductions. 
One particular reduction uses Rule~\bkcolor{\textsc{(RIn)}}
to undo the input at \pS:
\begin{align*}
M_4  \bk &  \news{s}(\,  
\np{\key{\loc_1}{\pS}}{ \conf{\inact}{\binp{\epS}{t}\bout{\epS}{\mathit{price}(t)}\bout{\epS}{\mathit{price}(t)} 
\binp{\epS}{ok}\binp{\epS}{a} \bout{\epS}{date}\inact }} 
\\
& \Par 
 \hmoni{s_\pS}{\past \ltinp{A}{\mathsf{title}}{\ltout{\{A,B\}}{\mathsf{{price}}}{T_\pB}}}{x}{\upd{x}{d}} 
\\
& \Par \np{\key{\loc_2}{\pA}}{ \conf{\inact}{\binp{\epA}{p}\binp{\epA}{s}\binp{\epA}{ok}\inact}} 
\\
& \Par 
\monir{s_\pA}{\ltout{V}{\mathsf{title}}{\past \ltinp{V}{\mathsf{{price}}}{\ltout{B}{\mathsf{share}}{\ltinp{B}{\mathsf{OK}}{\lend}}}}}{y}{\upd{y}{d}} 
\\
& \Par N_4 \Par \codah{s}{\emp}{\valueq{\pA}{\pS}{\exBook}}{}~)  = M_5
\end{align*}
Just as an application of Rule~\bkcolor{\textsc{(RollS)}} need not be immediately followed by 
an application of Rule~\bkcolor{\textsc{(RIn)}}, an application of Rule~\bkcolor{\textsc{(RIn)}} need not be immediately followed
by an application of Rule~\bkcolor{\textsc{(ROut)}}. A particular reduction from $M_5$ undoes the output at \pA:
\begin{align*}
M_5  \bk & \news{s}(\,  
\np{\key{\loc_1}{\pS}}{ \conf{\inact}{\binp{\epS}{t}\bout{\epS}{\mathit{price}(t)}\bout{\epS}{\mathit{price}(t)} 
\binp{\epS}{ok}\binp{\epS}{a} \bout{\epS}{date}\inact }} 
\\
& \Par 
 \hmoni{s_\pS}{\past \ltinp{A}{\mathsf{title}}{\ltout{\{A,B\}}{\mathsf{{price}}}{T_\pB}}}{x}{\upd{x}{d}} 
\\
& \Par \np{\key{\loc_2}{\pA}}{ \conf{\inact}{\bout{\epA}{\exBook}\binp{\epA}{p}\binp{\epA}{s}\binp{\epA}{ok}\inact}} 
\\
& \Par 
\hmoni{s_\pA}{\past \ltout{V}{\mathsf{title}}{\ltinp{V}{\mathsf{{price}}}{\ltout{B}{\mathsf{share}}{\ltinp{B}{\mathsf{OK}}{\lend}}}}}{y}{\upd{y}{d}} 
\Par N_4 \Par \codah{s}{\emp}{\emp}{}~)  = M_6
\end{align*}
Clearly, $M_6 = M_1$.
Summing up, the synchronization realized by the (forward) reduction sequence
$M_1 \fw M_2 \fw M_3$ can be reversed by the (backward) reduction sequence
$M_3 \bk M_4 \bk M_5 \bk M_1$.

To illustrate abstraction passing, let us assume that 
$M_3$ above follows a sequence of forward reductions
until the configuration:
\begin{align*}
M_7 = ~ &   \news{s}(\,  \np{\key{\loc_3}{\pB}}{ \conf{\inact}{\bbout{\epB}{\thunkp{\bout{\epB}{\text{`Urbino, 61029'}}\binp{\epB}{d}\inact}}\inact}} 
\\
& \Par \hmoni{s_\pB}{\myctxr{\ctx{T}_7}{\past \ltout{C}{\thunkt}{\ltout{V}{\mathsf{address}}{\ltinp{V}{\mathsf{date}}{\lend}}}}}{z,p,s}{\store_7} 
\\
& \Par \np{\key{\loc_4}{\pC}}{ \conf{\inact}{\binp{\epC}{code}(\appl{code}{\dummyn})}} 
\\
& \Par 
\hmoni{s_\pC}{\myctxr{\ctx{T}_8}{\past \ltinp{B}{\thunkt}{\lend}}}{w,s}{\store_8} 
\Par N_5 \Par \codah{s}{h_7}{\emp}{}~) 
\end{align*}
where 
$120 < \mathit{price}(\exBook)$ is the amount \pB may contribute
and
$\myctxr{\ctx{T}_7}{\bullet}$, $\store_7$, $\myctxr{\ctx{T}_8}{\bullet}$, $\store_8$,
and $h_7$ capture past interactions as follows:
\begin{align*}
\myctxr{\ctx{T}_7}{\bullet} & =
\ltinp{V}{\mathsf{{price}}}{\ltinp{A}{\mathsf{share}}{\ltout{\{A,V\}}{\mathsf{OK}}{
\ltout{C}{\mathsf{share}}{\bullet}}}}
\\
\store_7 & = \upd{z}{d},\upd{p}{\mathit{price}(\exBook)},\upd{s}{120}
\\
\myctxr{\ctx{T}_8}{\bullet} & =\ltinp{B}{\mathsf{share}}{\bullet} \qquad \store_8  = \upd{w}{d},\upd{s}{120}
\\
h_7 & = 
\valueq{\pA}{\pS}{\exBook}
\cons
\valueq{\pS}{\pA}{\mathit{price}(\exBook)}
\cons
\valueq{\pS}{\pB}{\mathit{price}(\exBook)}
\\
& \cons
\valueq{\pA}{\pB}{120}
\cons
\valueq{\pB}{\pA}{\text{`ok'}}
\cons
\valueq{\pB}{\pS}{\text{`ok'}}
\cons
\valueq{\pB}{\pC}{120}
\end{align*}
If $M_7 \fw \fw M_8$ by using Rules~\fwcolor{\textsc{(Out)}} and ~\fwcolor{\textsc{(In)}}
then we would have:
\begin{align*}
M_8  = & ~   \news{s}(\,  \np{\key{\loc_3}{\pB}}{ \conf{\inact}{\inact}} 
\Par \hmoni{s_\pB}{\myctxr{\ctx{T}_7}{\ltout{C}{\thunkt}{\past\ltout{V}{\mathsf{address}}{\ltinp{V}{\mathsf{date}}{\lend}}}}}{z,p,s}{\store_7} 
\\
& \Par \np{\key{\loc_4}{\pC}}{ \conf{\inact}{(\appl{code}{\dummyn}) }} 
\Par 
\hmoni{s_\pC}{\myctxr{\ctx{T}_8}{ \ltinp{B}{\thunkt}{\past\lend}}}{w,s,code}{\store_9} 
\\
& 
\Par N_5 \Par \codah{s}{h_7 \cons \valueq{\pB}{\pC}{\thunkp{\bout{\epB}{\text{`Urbino, 61029'}}\binp{\epB}{d}\inact}}}{\emp}{}~) 
\end{align*}
where
$\store_9 = \store_8 \upd{code}{\thunkp{\bout{\epB}{\text{`Urbino, 61029'}}\binp{\epB}{d}\inact}}$.

We now may apply Rule~\fwcolor{\textsc{(Beta)}} so as to obtain:
\begin{align*}
M_8  \fw &   \news{s}\news{\kappa}(\,  
\np{\key{\loc_4}{\pC}}{\conf{\inact}{\bout{\epB}{\text{`Urbino, 61029'}}\binp{\epB}{d}\inact}} 
\!\!  \Par N_6 
\\
& \!\!\Par 
\mem{\kappa}{(\appl{code}{\dummyn})}{\loc_4} 
\Par 
\hmoni{s_\pC}{\myctxr{\ctx{T}_8}{ \ltinp{B}{\thunkt}{\kappa.\past\lend}}}{w,s,code}{\store_9} 
\\
& 
\!\!\Par \codah{s}{h_7 \cons \valueq{\pB}{\pC}{\thunkp{\bout{\epB}{\text{`Urbino, 61029'}}\binp{\epB}{d}\inact}}}{\emp}{}~) 
= M_9
\end{align*}
where $N_6$ stands for the rest of the system. 
Notice that this reduction has added a running function on a fresh 
$\kappa$, which is also used within the type stored in the monitor $s_\pC$.

The reduction $M_8 \fw M_9$ completes the code mobility from $\pB$ to $\pC$: the now active thunk
will execute $\pB$'s implementation from $\pC$'s location.
This justifies the premise 
$\p = \er \,\vee\, \p \in \names{\er, h}$ present in Rules~\fwcolor{\textsc{(Out)}}, \fwcolor{\textsc{(In)}}, \fwcolor{\textsc{(Sel)}}, and ~\fwcolor{\textsc{(Bra)}} (and in their backward counterparts):
when executing previously received mobile code, the participant mentioned in the location (i.e., $\pC$)
and that mentioned in the located process (i.e., $\pB$) may differ.
Further forward reductions from $M_9$ will  modify the cursor in the type stored in monitor $s_\pB$
based on the process behavior located at $\key{\loc_4}{\pC}$.

\subsection{A Protocol with  Choices}
\label{app:choices}
We close this section by showcasing  reversibility of labeled choices.
Consider the following global type, specifying a simple binary (two-party) protocol between a Buyer~(\pB) and a Seller~(\pS):
\begin{align*}
G = ~&  \gtcom{B}{\pS}{\mathsf{title}}{\gtcom{\pS}{B}{\mathsf{{price}}}{\\
&{\gpart{\pS}\to \gpart{B}\{\mathit{ok}: \gtcom{B}{\pS}{\mathsf{addr}}{\gtcom{\pS}{B}{\mathsf{date}}{\gend} \,; 
 \mathit{quit}: \gend\}
}}}}
\end{align*}
This way, after receiving a title from Buyer, Seller replies with the  {price} of the requested item; subsequently, a  choice indicated by labels  $\mathit{ok}$ and $\mathit{quit}$ takes place: Buyer can select whether to continue with the transaction or to conclude it.
The projection of $G$ onto local types are:
\begin{align*}
\tproj{G}{\pS} ~= ~& \ltinp{B}{\mathsf{title}}{\ltout{B}{\mathsf{{price}}}{\\ &\gpart{B}\&\{\mathit{ok}: \ltinp{B}{\mathsf{addr}}{\ltout{B}{\mathsf{date}}{\lend}} \, ; \, \mathit{quit}: \lend \}}}\\
\tproj{G}{\pB} ~= ~& \ltout{\pS}{\mathsf{title}}{\ltinp{\pS}{\mathsf{{price}}}{\\ &\gpart{\pS}\oplus\{\mathit{ok}: \ltout{\pS}{\mathsf{addr}}{\ltinp{\pS}{\mathsf{date}}{\lend}} \, ; \, \mathit{quit}: \lend \}}}
\end{align*}
Possible implementations for the participants are as follows:
\begin{align*}
\text{Seller} & =  \bout{a}{x:\tproj{G}{\pS}}\binp{x}{title}\bout{x}{quote}
\bbra{x}{\mathit{ok}: \binp{x}{addr}\bout{x}{date} \,;\,\mathit{quit}:\inact}\\
\text{Buyer} & =  \binp{a}{y:\tproj{G}{\pB}}\bout{y}{title}\binp{y}{quote}
\bsel{y}{\mathit{ok}: \bout{y}{addr}\binp{y}{date} \,;\,\mathit{quit}:\inact}
\end{align*}
The whole system, given by configuration $M$ below, is obtained by placing these process implementations in appropriate locations:
$$
M = \myloc{\loc_1}{\text{Seller}} 
\Par
\myloc{\loc_2}{\text{Buyer}}
$$
We may then have:
\begin{align*}
M \fws 
& \news{s}(\,  \np{\key{\loc_1}{\pS}}{ \conf{\inact}{ \bbra{\epS}{\mathit{ok}: \binp{\epS}{addr}\bout{\epS}{date}\inact \,;\,\mathit{quit}:\inact} }}  \\
& \Par \hmoni{s_\pS}{\myctx{T_\text{1}}{\past \gpart{B}\&\{\mathit{ok}: \ltinp{B}{\mathsf{addr}}{\ltout{B}{\mathsf{date}}{\lend}} \, ; \, \mathit{quit}: \lend \}}}{x_1}{\store_1}  \\
&
\Par \np{\key{\loc_2}{\pB}}{ \conf{\inact}{ \bsel{\epB}{\mathit{ok}: \binp{\epB}{addr}\bout{\epB}{date}\inact \,;\,\mathit{quit}:\inact} }}  \\
&\Par \hmoni{s_\pB}{\myctx{S_\text{1}}{\past \gpart{\pS}\oplus\{\mathit{ok}: \ltout{\pS}{\mathsf{addr}}{\ltinp{\pS}{\mathsf{date}}{\lend}} \, ; \, \mathit{quit}: \lend \}}}{x_2}{\store_2} \Par 
\codah{s}{h_1}{\emp}{}~) = M_1
\end{align*}
where $M_1$ is the configuration obtained from $M$ once the two participants have initiated the session and exchanged the title and the corresponding  {price}. Above, $x_1$ and $x_2$ are the free variables of \pS and \pB after the first three interactions; also, $\store_1$ and $\store_2$ represent their respective stores. Queue $h_1$ contains the two messages related to $title$ and $\mathit{price}$.
The context types are: 
$$
\myctxr{\ctx{T}_1}{\bullet} = \ltinp{B}{\mathsf{title}}{\ltout{B}{\mathsf{{price}}}}{\bullet} \qquad
 \myctxr{\ctx{S}_1}{\bullet} = \ltout{S}{\mathsf{title}}{\ltinp{S}{\mathsf{{price}}}}{\bullet}
 $$
 In $M_1$, Buyer can decide either (a) to accept the suggested {price} and continue with the prescribed protocol or (b) to refuse it and exit. The first possibility may proceed using Rule~\fwcolor{\textsc{(Sel)}} as follows:
\begin{align*}
M_1 \fw
&\news{s}(\,\np{\key{\loc_2}{\pB}}{ \conf{\inact, \bsel{\epB}{\mathit{quit}: \inact}}{  \binp{\epB}{addr}\bout{\epB}{date}\inact  }}  \\
&\Par \hmoni{s_\pB}{\myctx{S_\text{1}}{ \gpart{\pS}\oplus\{\mathit{ok}: \past\ltout{\pS}{\mathsf{addr}}{\ltinp{\pS}{\mathsf{date}}{\lend}} \, ; \, \mathit{quit}: \lend \}}}{x_2}{\store_2}  \\
&\Par \codah{s}{h_1}{\valueq{\pB}{\pS}{\mathit{ok}} } \Par N_1~) = M_2
\end{align*} 
where $N_1$  contains the rest of the Seller process and monitor of $M_1$. 
As we can see, in $M_2$ the cursor $\past$ of the Buyer monitor has been moved into the choice. 
Moreover, the process stack of Buyer is updated in order to register the discarded branch of the choice (i.e., the branch involving  label  $\textbf{quit}$). From $M_2$,  Seller can consume the message on top of the queue (which details the choice by \pB), or Buyer can revert its choice. 
In the first case we have the following, using Rule~\fwcolor{\textsc{(Bra)}}:
\begin{align*}
M_2 \fw
&\news{s}(\,\np{\key{\loc_1}{\pS}}{ \conf{\inact , \bbra{\epS}{\mathit{quit}:\inact}}{ \binp{\epS}{addr}\bout{\epS}{date}\inact  }}  \\
&\Par \hmoni{s_\pS}{\myctx{T_\text{1}}{ \gpart{B}\&\{\mathit{ok}: \past\ltinp{B}{\mathsf{addr}}{\ltout{B}{\mathsf{date}}{\lend}} \, ; \, \mathit{quit}: \lend \}}}{x_1}{\store_1}  \\
&
\Par \np{\key{\loc_2}{\pB}}{ \conf{\inact , \bsel{\epB}{\mathit{quit}: \inact}}{  \binp{\epB}{addr}\bout{\epB}{date}\inact  }}  \\
&\Par \hmoni{s_\pB}{\myctx{S_\text{1}}{ \gpart{\pS}\oplus\{\mathit{ok}: \past\ltout{\pS}{\mathsf{addr}}{\ltinp{\pS}{\mathsf{date}}{\lend}} \, ; \, \mathit{quit}: \lend \}}}{x_2}{\store_2}  
\\
&
\Par \codah{s}{h_1\cons \valueq{\pB}{\pS}{\mathit{ok}}}{\emp } ~) = M_3
\end{align*}
In the second case, we can revert the labeled choice by using 
Rule~\bkcolor{\textsc{(RollC)}} from $M_3$ first, 
and then using 
Rules~\bkcolor{\textsc{(RBra)}} and \bkcolor{\textsc{(RSel)}}
in a decoupled fashion.

Having introduced and illustrated our process model and its reversible semantics,
 we now move on to establish its key properties.

%
%

\section{Main Results}
\label{s:res}

We now establish our main result: 
we prove that reversibility in our model of multiparty asynchronous communication is 
\emph{causally consistent}. 
We proceed in three steps:
\begin{enumerate}[a)]
\item First, we introduce an alternative \emph{atomic}  semantics and show that it corresponds, in a tight technical sense, to the decoupled semantics in \S\,\ref{ss:semconf} (Theorems~\ref{theo:correspond} and \ref{t:bf}). 
\item Second, in the light of this correspondence, we establish causal consistency for the atomic semantics, following the  approach of Danos and Krivine~\cite{DanosK04} (Theorem~\ref{t:causal}).
\item Finally, we state a fine-grained, bidirectional connection between the semantics of (high-level) global types with the decoupled semantics of  (low-level) configurations (Theorem~\ref{t:corrgc}). 
\end{enumerate}
These steps allow us to
transfer causal consistency to protocols expressed as global types.

\subsection{An Atomic Semantics}\label{sec:atom}

Our main insight is that  causal consistency for asynchronous communication can be established by considering
a \textit{coarser} synchronous reduction relation.

\begin{defi}[Atomic Reduction]\label{d:atred}
We define \emph{atomic} versions of the forward and backward reduction relations, relying on the rules in Fig.~\ref{fig:atom_fwd} and~\ref{fig:atom_bk}. 
\begin{itemize}
    \item The \emph{forward  atomic reduction}, denoted $\fwa$, is  
the smallest evaluation-closed relation that satisfies Rules \fwcolor{(\textsc{AC})}  and \fwcolor{(\textsc{AS})} (Fig.~\ref{fig:atom_fwd}), together with Rules 
\fwcolor{(\textsc{Init})}, \fwcolor{(\textsc{Beta})}, and \fwcolor{(\textsc{Spawn})} (Fig.~\ref{fig:fw_1} and~\ref{fig:fw_2}). 
\item Similarly, the \emph{backward atomic reduction}, denoted  $\bka$, is  the smallest evaluation-closed relation that satisfies Rules \bkcolor{(\textsc{RAC})}  and \bkcolor{(\textsc{RAS})} (Fig.~\ref{fig:atom_bk}), together with Rules 
\bkcolor{(\textsc{RInit})}, \bkcolor{(\textsc{RBeta})}, and \bkcolor{(\textsc{RSpawn})} (Fig.~\ref{fig:bk_1} and~\ref{fig:bk_3}). 
\end{itemize}
We then define the atomic reduction relation $\reda$ as $\fwa \cup \bka$.
\end{defi}

\begin{figure}[!t]
\begin{mdframed}
\centering
    \begin{tabular}{c|ll}
Symbol & Meaning &  \\ \hline  \hline
$\fwg$ & Forward transition for global types & (Fig.~\ref{f:gts})
\\
$\bkg$ & Backward transition for global types & (Fig.~\ref{f:gts}) 
\\  \hline 
 $\fw$    & Forward decoupled reduction      &  (Fig.~\ref{fig:fw_1} and~\ref{fig:fw_2})\\
$\bk$     & Backward decoupled reduction      & (Fig.~\ref{fig:bk_1}, \ref{fig:bk_2}, and~\ref{fig:bk_3}) \\
$\red$ & $\fw \cup \bk$ \\ \hline 
 $\fwa$    & Forward atomic reduction       & (Fig.~\ref{fig:atom_fwd}) \\
$\bka$     & Backward atomic reduction      & (Fig.~\ref{fig:atom_bk}) \\
$\reda$ & $\fwa \cup \bka$ 
\end{tabular}
\end{mdframed}
\caption{Notations for transition and reduction relations.}\label{f:reds}
\end{figure}

\begin{figure}[!t]
\begin{mdframed}
\begin{align*}
& \inferrule[\fwcolor{(AC)}]
{\p = \er_1 \,\vee\, \p \in \names{\er_1, h} 
\and
\q = \er_2 \,\vee\, \q \in \names{\er_2, h}
 }{
 M_1 \Par N_1 \Par  M_2 \Par N_2 \Par \codah{s}{h}{k}{} 
 ~\fwa~ 
  M'_1 \Par N'_1 \Par  M'_2 \Par N'_2 \Par \codah{s}{h\cons  \valueq{\q}{\p}{V}}{k }{} 
 }
\\
& \text{where:}
\\ 
& \begin{array}{ll}
M_1  =  \np{\key{\loc_1}{\er_1}}{ \conf{\stack{C}_1}{\bout{\ep{s}{\p}}{V}{P}}} 
&
N_1  = \moni{s_\p}{\myctxr{\ctx{T}}{\past  \ltout{\q}{U}{S}}}{\mytilde{x}_1}{\store} 
\\
M'_1  = \np{\key{\loc_1}{\er_1}}{\conf{\stack{C}_1}{P}} 
&
N'_1  = \moni{s_\p}{\myctxr{\ctx{T}}{ \ltoutp{\q}{U}{S}}}{\mytilde{x}_1}{\store}
\\
M_2  = \np{\key{\loc_2}{\er_2}}{\conf{\stack{C}_2}{\binp{\ep{s}{\q}}{y}{Q}}} 
&
N_2  = \moni{s_\q}{\myctxr{\ctx{S}}{\past  \ltinp{\p}{U}{T}}}{\mytilde{x}_2}{\store} 
\\
M'_2  = \np{\key{\loc_2}{\er_2}}{\conf{\stack{C}_2}{Q}} 
&
N'_2  =  \moni{s_\q}{ \myctxr{\ctx{S}}{\ltinpp{\p}{U}{T}}}{\mytilde{x}_2, y}{\store\upd{y}{V}} 
\end{array}
\end{align*}
\begin{align*}
& \inferrule[\fwcolor{(AS)}]
{\p = \er_1 \,\vee\, \p \in \names{\er_1, h} 
\and
\q = \er_2 \,\vee\, \q \in \names{\er_2, h}
 }{
 M_1 \Par N_1 \Par  M_2 \Par N_2 \Par \codah{s}{h}{k}{} 
 ~\fwa~ 
  M'_1 \Par N'_1 \Par  M'_2 \Par N'_2 \Par \codah{s}{h\cons \valueq{\p}{\q}{\lbl_w}}{k }{} 
 }
\\
& \text{where:}
\\ 
& \begin{array}{ll}
M_1  =  \np{\key{\loc_1}{\er_1}}{ \conf{\stack C_1}{\bsel{\ep{s}{\p}}{\lbl_i. P_i}_{i\in I}}}  
&
N_1  = \moni{s_\p}{\myctxr{\ctx{S}}{\past \ltsel{\q}{\lbl_j}{S_j}{j}{J}} }{\mytilde{x}_1}{\store} 
\\
M'_1  = \np{\key{\loc_1}{\er_1}}{ \conf { \stack C_1, \bbra{\ep{s}{\p}}{\lbl_l:P_l}_{l \in I\setminus w}}{ P_w} }
&
N'_1  = \moni{s_\p}{\myctxr{\ctx{S}}{\ltselp{\q}{\lbl_j:S_j\,,\, \lbl_w{:}\past S_w}{j}{J\setminus w}} }{\mytilde {x}_1}{\store}
\\
M_2  = \np{\key{\loc_2}{\er_2}}{ \conf { \stack C_2}{\bbra{\ep{s}{\q}}{\lbl_i:Q_i}_{i \in I} } } 
&
N_2  = \moni{s_\q}{\myctxr{\ctx{T}}{\past \ltbra{\p}{\lbl_j}{T_j}{j}{J}} }{\mytilde{x}_2}{\store} 
\\
M'_2  = \np{\key{\loc_2}{\er_2}}{ \conf{\stack C_2, \bsel{\ep{s}{\q}}{\lbl_l.Q_l}_{l\in I\setminus w} }{Q_w}}
&
N'_2  =  \moni{s_\q}{\myctxr{\ctx{T}}{ \ltbrap{\p}{\lbl_j:T_j \, , \, \lbl_w{:}\past T_w}{j}{J\setminus w} }}{\mytilde{x}_2}{\store} 
\end{array}
\end{align*}
\end{mdframed}
 \caption{Atomic semantics for configurations: Forward reduction ($\fwa$).}
 \label{fig:atom_fwd}
\end{figure}


\begin{figure}[t!]   
\begin{mdframed}
\begin{align*}
& \inferrule[\bkcolor{(RAC)}]
{\p = \er_1 \,\vee\, \p \in \names{\er_1, h} 
\and
\q = \er_2 \,\vee\, \q \in \names{\er_2, h}
 }{
 M_1 \Par N_1 \Par  M_2 \Par N_2 \Par \codah{s}{h\cons  \valueq{\q}{\p}{V}}{k } 
 ~\bka~ 
  M'_1 \Par N'_1 \Par  M'_2 \Par N'_2 \Par \codah{s}{h}{k }{} 
 }
\\
& \text{where:}
\\ 
& \begin{array}{ll}
M_1  = \np{\key{\loc_1}{\er_1}}{\conf{\stack{C}_1}{P}} 
&
N_1  = \moni{s_\p}{\myctxr{\ctx{T}}{ \ltoutp{\q}{U}{S}}}{\mytilde{x}_1}{\store} 
\\
M'_1  = \np{\key{\loc_1}{\er_1}}{ \conf{\stack{C}_1}{\bout{\ep{s}{\p}}{V}{P}}} 
& 
N'_1  = \moni{s_\p}{\myctxr{\ctx{T}}{\past  \ltout{\q}{U}{S}}}{\mytilde{x}_1}{\store} 
\\
M_2  = \np{\key{\loc_2}{\er_2}}{\conf{\stack{C}_2}{Q}} 
&
N_2  =  \moni{s_\q}{ \myctxr{\ctx{S}}{\ltinpp{\p}{U}{T}}}{\mytilde{x}_2, y}{\store\upd{y}{V}} 
\\
M'_2  = \np{\key{\loc_2}{\er_2}}{\conf{\stack{C}_2}{\binp{\ep{s}{\p}}{y}{Q}}}
&
N'_2  =  \moni{s_\q}{\myctxr{\ctx{S}}{\past  \ltinp{\p}{U}{T}}}{\mytilde{x}_2}{\store} 
\end{array}
\end{align*}
\begin{align*}
& \inferrule[\bkcolor{(RAS)}]
{\p = \er_1 \,\vee\, \p \in \names{\er_1, h} 
\and
\q = \er_2 \,\vee\, \q \in \names{\er_2, h}
 }{
 M_1 \Par N_1 \Par  M_2 \Par N_2 \Par\codah{s}{h \cons  \valueq{\p}{\q}{\lbl_w} }{k }{}
 ~\bka~ 
  M'_1 \Par N'_1 \Par  M'_2 \Par N'_2 \Par \codah{s}{h}{k }{} 
 }
\\
& \text{where:}
\\ 
& \begin{array}{ll}
M_1  = \np{\key{\loc_1}{\er_1}}{ \conf { \stack C_1, \bbra{\ep{s}{\p}}{\lbl_l:P_l}_{l \in I\setminus w}}{ P_w} } 
&
N_1  =\moni{s_\p}{\myctxr{\ctx{S}}{\ltbrap{\q}{\lbl_j:S_j\,,\, \lbl_w{:}\past S_w}{j}{J\setminus w}}\!}{\mytilde{x}_1}{\store}
\\
M'_1  =\np{\key{\loc_1}{\er_1}}{ \conf{\stack C_1}{\bsel{\ep{s}{\p}}{\lbl_i. P_i}_{i\in I}}}  
&
N'_1  =\moni{s_\p}{\myctxr{\ctx{S}}{\past \ltbra{\q}{\lbl_j}{S_j}{j}{J}} }{\mytilde{x}_1}{\store} 
\\
M_2  =\np{\key{\loc_2}{\er_2}}{ \conf{\stack C_2, \bsel{\ep{s}{\q}}{\lbl_l.Q_l}_{l\in I\setminus w} }{Q_w}}
&
N_2  =\moni{s_\q}{\myctxr{\ctx{T}}{ \ltselp{\p}{\lbl_j:T_j \, , \, \lbl_w{:}\past T_w}{j}{J\setminus w} }\!}{\mytilde{x}_2}{\store}\!
\\
M'_2  =\np{\key{\loc_2}{\er_2}}{ \conf { \stack C_2}{\bbra{\ep{s}{\q}}{\lbl_i:Q_i}_{i \in I} } } 
&
N'_2  =\moni{s_\q}{\myctxr{\ctx{T}}{\past \ltsel{\p}{\lbl_j}{T_j}{j}{J}} }{\mytilde{x}_2}{\store}  
\end{array}
\end{align*}
\end{mdframed} 
 \caption{Atomic semantics for configurations: Backward reduction ($\bka$).}
 \label{fig:atom_bk}
\end{figure}

Figure~\ref{f:reds} summarizes our notations for reductions.
We start by introducing \emph{reachable} configurations: 

\begin{defi}\label{d:ic}
A configuration $M$ is \emph{initial} if
$M\equiv \news{\tilde n} \prod_{i \in I}\myloc{\loc_i}{P_i}$, for some $I$. A configuration is 
\emph{reachable}, if it is derived from an initial
configuration by using $\red$ (cf. \S\,\ref{ss:semconf}).  A configuration is  
\emph{atomically reachable}, if it is derived from an initial configuration by using $\reda$. 
\end{defi}


\noindent
To relate the decoupled semantics $\red$  with the atomic reduction $\reda$ (just defined), we introduce the concept of \textit{stable} configuration. Roughly speaking, in a stable configuration there are no 
``ongoing'' reduction steps. In the forward case, 
an ongoing step is witnessed by non-empty output queues (which should eventually become empty to complete a synchronization);
in the backward case, an ongoing step is witnessed by a marked monitor
(which should be eventually unmarked when a synchronization is undone).
This way, e.g., in the example of \S\,\ref{sss:examp-tb} configurations $M_3$ and $M_7$ are stable, whereas $M_2$ and $M_4$ are not.
Reduction $\reda$ will move between stable configurations only.
We therefore have:
\begin{defi}
\label{lm:fwstable}
A configuration $M$ is  \emph{stable}, written $\stable(M)$, if 
\begin{align*}
M & \equiv  \prod_{i}\myloc{\loc_i}{P_i} \Par 
\news{s\tilde{a}} \Big( \codah{s}{h_1}{\emp}  \Par\prod_{j}\np{\key{\loc_j}{\p_j}}{\conf{\stack C_j}{P_j}} \Par 
\hmoni{s_{\p_i}}{ T_i }{\mytilde x_i}{\store_i} 
\Big)
\end{align*}

\end{defi}

\noindent
Reduction $\red$ does not preserve stability, but it can be recovered: 

\begin{lem}
Given  a stable configuration $M$ then
\begin{itemize}
	\item if $M\fw N$ with $\lnot\stable(N)$ then there exists an $N'$ such that $N\fw N'$ and $\stable(N')$; 
	\item if $M\bk N$ with $\lnot\stable(N)$ then there exists an $N'$ such that $N\bk\bk N'$ and $\stable(N')$.
\end{itemize}
\end{lem}


We may then have:

\begin{cor}\label{lbl:corollary_finish}
If $\stable(M)$ and $M\trans\red N$ with $\lnot\stable(N)$, then there exists an $N'$ such that $N\trans\red N'$ with $\stable(N')$.
\end{cor}
\begin{proof}
By induction on the reduction sequence $M \trans{\red}N$.
\end{proof}

\noindent
We now show the Loop lemma~\cite{DanosK04}, which 
offers a local consistency guarantee for the interplay of forward and backward reductions: it ensures that every reduction step can be reverted. 
This lemma will be crucial both in proving a correspondence between atomic and decoupled semantics, but also in showing causal consistency of the atomic semantics.

\begin{lem}[Loop]\label{lemma:loop}
Let  $M, N$ be stable and atomic reachable configurations. Then 
$M\fwa N$ if and only if $N\bka M$.

\end{lem}
\begin{proof}
By induction on the derivation of $M\fwa   N$ for the if direction, and on the derivation of $N \bka  M$ for the converse.
\end{proof}

 The following lemma allow us to ``reorder''  decoupled reduction steps so to have the generation of a message (e.g., an application of Rules \fwcolor{(\textsc{Out})} or \fwcolor{(\textsc{Bra})})
followed by its consumption (e.g., an application of Rules \fwcolor{(\textsc{In})} or \fwcolor{(\textsc{Sel})}). This way, the two consecutive decoupled reductions can be mimicked by one atomic step, which will be in turn instrumental to relate the atomic and decoupled semantics.
Below we write $\fwn{i}$ (and $\bkn{i}$) to denote a specific  step in a reduction sequence. 
\begin{lem}[Swap]\label{lemma:swap}
Let $M$ be a reachable configuration.
\begin{itemize}[$\bullet$]
\item If $M \fws\fwn{1}\fws N_1$ and $N_1 \fwn{2} N_2$, where $\fwn{1}$ denotes the use of Rule~\fwcolor{(\textsc{Out})} or ~\fwcolor{(\textsc{Sel})} and  $\fwn{2}$ denotes the respective use of 
Rule~\fwcolor{(\textsc{In})} or ~\fwcolor{(\textsc{Bra})}, then $M \fws \fwn{1} \fwn{2} N \fws N_2$, for some $N$.

\item  If $M \bks\bkn{1}\bks N_1$ and $N_1 \bkn{2} N_2$, where $\bkn{1}$ denotes the use of  Rule~\bkcolor{(\textsc{ROut})} or~\bkcolor{(\textsc{RSel})} and~$\bkn{2}$ denotes the respective use of Rule~\bkcolor{(\textsc{RIn})} or~\bkcolor{(\textsc{RBra})}, then $M \bks \bkn{1} \bkn{2} N \bks N_2$, for some $N$.
\end{itemize}
\end{lem}

\modif{
\begin{example}
	To better understand Lemma~\ref{lemma:swap}, let us consider the following configurations:
\begin{align*}
M_1 = &\np{\key{\loc_1}{\er_1}}{ \conf{\inact}{\bout{\ep{s}{\er_1}}{1}{\bout{\ep{s}{\er_1}}{2}{P_1}}}}
& N_1 & = \moni{s_{\p_1}}{{\past  \ltout{\p_2}{U}{ \ltout{\p_3}{U}{S_1}}}}{\mytilde{x}_1}{\store_1} \\
M_2  = & \np{\key{\loc_2}{\er_2}}{\conf{\inact}{\binp{\ep{s}{\p_2}}{y}{P_2}}} 
& N_2 & = \moni{s_{\p_2}}{{\past  \ltinp{\p_1}{U}{S_2}}}{\mytilde x_2}{\store_2} \\
M_3 = &\np{\key{\loc_3}{\er_3}}{\conf{\inact}{\binp{\ep{s}{\p_3}}{y}{P_3}}} 
& N_3 & = \moni{s_{\p_3}}{{\past  \ltinp{\p_1}{U}{S_3}}}{\mytilde x_3}{\store_3} 
\end{align*} 
Define the configuration $\mathtt{Sys}$ as:
$$\mathtt{Sys} = \news{s}\Big(\codah{s}{\emp}{\emp} \Par  \prod_{i\in \{1,2,3\}} M_i \Par N_i\Big) $$ 
From this configuration, by applying Rule \fwcolor{(\textsc{Out})} twice, we obtain:
\begin{align*}
\mathtt{Sys} \fw \fw \news{s}\Big( \codah{s}{\emp} {\valueq{\p_1}{\p_2}{1}\cons \valueq{\p_1}{\p_3}{2} } \Par M'_1 \Par N'_1 \Par \prod_{i\in \{2,3\}} M_i \Par N_i \Big) = \mathtt{Sys}_{\mathtt{o}} 
\end{align*}
From $\mathtt{Sys}_{\mathtt{o}}$ we can apply twice Rule $\fwcolor{(\textsc{In})}$ and obtain:
\begin{align*}
\mathtt{Sys}_{\mathtt{o}} \fw \fw  \news{s}\Big(\codah{s}{\valueq{\p_1}{\p_2}{1}\cons \valueq{\p_1}{\p_3}{2} }{\emp} \Par \prod_{i\in \{1,2,3\}} M'_i \Par N'_i \Big) = \mathtt{Sys_{\mathtt{end}}}
\end{align*}
where
\begin{align*}
M'_1 = &\np{\key{\loc_1}{\er_1}}{ \conf{\inact}{P_1}}
& N'_1 & = \moni{s_{\p_1}}{{  \ltout{\p_2}{U}{ \ltout{\p_3}{U}{\past S_1}}}}{\mytilde{x}_1}{\store_1} \\
M'_2  = & \np{\key{\loc_2}{\er_2}}{\conf{\inact}{P_2}} 
& N'_2 & = \moni{s_{\p_2}}{{  \ltinp{\p_1}{U}{\past S_2}}}{\mytilde x'_2}{\store'_2} \\
M'_3 = &\np{\key{\loc_3}{\er_3}}{\conf{\inact}{P_3}} 
& N'_3 & = \moni{s_{\p_3}}{{  \ltinp{\p_1}{U}{\past S_3}}}{\mytilde x'_3}{\store'_3} 
\end{align*}
Now, Lemma~\ref{lemma:swap} captures the following observation: starting from $\mathtt{Sys}$, the configuration $\mathtt{Sys}_{\mathtt{end}}$
can also be reached if after the first application of Rule~\fwcolor{(\textsc{Out})} the produced message is immediately consumed, by applying Rule \fwcolor{(\textsc{In})}:
\begin{align*}
\mathtt{Sys} \fw \fw \news{s}\Big(M''_1 \Par N''_1 \Par M'_2 \Par N'_2 \Par M_3 \Par N_3 \Par \codah{s}{\valueq{\p_1}{\p_2}{1}}{k}\Big) \fw \fw \mathtt{Sys}_{\mathtt{end}}
\end{align*}
where
\begin{align*}
M''_1 = &\np{\key{\loc_1}{\er_1}}{ \conf{\inact}{{\bout{\ep{s}{\er_1}}{2}{P_1}}}}
& N''_1 & = \moni{s_{\p_1}}{{  \ltout{\p_2}{U}{\past \ltout{\p_3}{U}{ S_1}}}}{\mytilde{x}_1}{\store_1} 
\end{align*}
\end{example}
}


\smallskip
\noindent
The following theorem provides a first   connection between decoupled and atomic reductions; its proof is immediate from their definitions:
\begin{thm}[Relating $\red$ and $\reda$]\label{theo:correspond}
Let $M$ and $N$ be stable configurations. We have:
\begin{itemize}[$\bullet$]
	\item $M\fwa N$ if and only if either $M{\fw} N$ or $M{\fw\fw} N$;
	\item $M\bka N$ if and only if either  $M{\bk} N$ or $M{\bk\bk\bk} N$.
\end{itemize}
\end{thm}
\noindent
We now embark ourselves in providing a tighter formal connection between $\red$ and $\reda$, using \emph{back-and-forth bisimulations}~\cite{LaneseMS16}.
We shall work with binary relations on configurations, written $\Re  \subseteq \confs \times \confs$.
We now adapt the usual notion of barbs~\cite{DBLP:conf/concur/SangiorgiW01} to our setting:
rather than communication subjects (which are hidden/unobservable names in intra-session communications), 
it suffices to use participant identities as observables:
\begin{defi}[Barbs]\label{lbl:barb}
A reachable configuration $M$ has a barb $\p$, written $M\barb{\p}$, 
if
\begin{itemize}
\item $M\equiv \news{\vect n} (N \Par \np{\key{\loc}{\er}}{\conf{\stack C}{P}} \Par \hmoni{s_\p}{ \ctx{S}[\past T] }{\mytilde x}{\store})$ where either: \\  
(i)~$P\equiv \bout{\ep{s}{\p}}{V}{Q} \Par R$ and $T=\ltout{\q}{U}{T_1} $ or  \\
(ii)~$P\equiv \bsel{\ep{s}{\p}}{\lbl_i.P_i}_{i\in I}\Par R$ and $T= \ltsel{\q}{\lbl_j}{T_j}{j}{J}$. 
\end{itemize}
\end{defi}
\noindent
Notice that our definition of barbs is connected to the notion of stability:
since in 
$M\barb{\p}$ 
we require a monitor with empty tag, this ensures that 
$\p$ is not involved in an ongoing backward step. 
In a way, this allows us to consider just \textit{forward barbs} (as in~\cite{DBLP:journals/jlp/AubertC17}). 

We now  adapt the definition of weak barbed back-and-forth (bf) bisimulation and congruence~\cite{LaneseMS16} in order to work with 
decoupled and atomic reduction semantics:
\begin{defi}\label{d:bf}
A relation $\Re$ is a \emph{(weak) barbed bf simulation} if whenever $M \Re N$
\begin{itemize}
	\item $M\barb{\p}$ implies $N\trans{\red}\barb{\p}$; 
	\item $M\fwa M_1$ implies $N \fws N_1$, with $M_1 \Re N_1$; 
	\item $M\bka M_1$ implies $N \bks N_1$, with $M_1 \Re N_1$.
\end{itemize}
A relation $\Re$ is a  \emph{(weak) barbed bisimulation} if
 $\Re$  and  $\Re^{-1}$ are  weak bf barbed simulations.
 The largest weak barbed bisimulation is  
\emph{(weak) barbed bisimilarity},  noted $\bfbw$. 
We say that
$M$ and $N$ are \emph{(weakly) barbed congruent}, written 
$\congruence{\bfbw}$, 
 if for each context $\ctx{C}$ such that $\ctx{C}[M]$ and  $\ctx{C}[N]$ are 
atomic reachable configurations, then
 $\ctx{C}[M] \bfbw \ctx{C}[N]$.
\end{defi}

\noindent 
\modif{This way, in establishing $M \congruence\bfbw N$ we should consider that \emph{atomic} reduction steps from $M$  are matched by $N$ with \emph{decoupled} reduction steps. This is how $\congruence\bfbw$  enables us to state our second connection between decoupled and atomic reductions. To prove the correspondence between the two semantics, we shall relate the
same configuration $M$ under the two different semantics. Hence:}

\begin{thm}\label{t:bf}
For any atomic reachable configuration $M$, we have that $M \congruence\bfbw M$.
\end{thm}

\begin{proof}
First, notice that showing $\ctx{C}[M]\bfbw\ctx{C}[M]$ is similar to show 
$M_1 \bfbw M_1$ with $M_1 = \ctx{C}[M]$.
This allows us to just focus on the ``hole'' of the context. 
It is then sufficient to show that the following relation is a 
bf weak bisimulation. 
\begin{align*}
	\Re =\big\{ (M,N) \st  M \fws N \text{ via Rules~\fwcolor{\textsc{Out}} or \fwcolor{\textsc{Sel}}} \;\wedge\;   
	M \bks N \text{ via Rules~\bkcolor{\textsc{ROut}} or \bkcolor{\textsc{RSel}}} \big\} 
\end{align*}
Clearly, $(M, M) \in \Re$. We consider the requirements in Def.~\ref{d:bf}.
Let us first consider barbs. 

\begin{itemize}
    \item Suppose that $M$ challenges $N$ with a barb. 
We distinguish two cases: $N$ is stable or not.
If
 $\stable(N)$  then $N$  has the same barb. Otherwise,  if $\lnot\stable(N)$,
 by  Corollary~\ref{lbl:corollary_finish} 
 there exists an $N_1$ such that
 $N\trans{\red} N_1$ and $\stable(N_1)$.
 Since $M \trans{\red} N$ 
  we may derive $M\trans{\red} N_1$ with both stable configurations. 
  By applying Theorem~\ref{theo:correspond} on $M\trans{\red} N_1$ we infer $M\trans{\reda} N_1$;
  then, by
 applying the Loop Lemma (Lemma~\ref{lemma:loop})  we further infer $N_1\trans{\reda} M$. 
 Using again Theorem~\ref{theo:correspond} we infer that $N_1\trans{\red} M$; 
 since we have deduced that $N\trans{\red} N'\trans{\red} M$,
 we know that $N$ weakly matches all the barbs of $M$, as desired. 
 \item  Suppose now that $N$ challenges $M$ with a barb.
 We proceed similarly as above:
  if $\stable(M)$ then $M$ has the same barb; otherwise, if $\lnot\stable(M)$, since $M\trans{\red} N$, by Corollary~\ref{lbl:corollary_finish} we have that $M\trans{\red} N \trans{\red}N_1$, with $\stable(N_1)$.
 Let us note that the reductions in $N\trans{\red} N_1$ do not add barbs to $N_1$: they only finalize ongoing synchronizations; by definition of barbs (Def.~\ref{lbl:barb}) parties involved in ongoing rollbacks do not contribute to barbs. We can conclude by applying Theorem~\ref{theo:correspond} and deriving $M\trans{\reda} N_1$, which has the same barbs of $N$, as desired. 
 \end{itemize}

 
 Let us now consider reductions. We will just focus on synchronizations due to input/output and branching/selection reduction steps, since 
 these are the cases in which $\red$ and $\reda$ differ; indeed, reductions due to Rules~\fwcolor{\textsc{Spawn}} and \fwcolor{\textsc{Beta}} can be trivially matched. 
 
 Let us consider challenges from $M$.
 There are two cases: $M\fwa M_1$ and $M\bka M_1$.
  In the first case, as  we distinguish two sub-cases: either $N$ has already started the synchronization or not. 
    In the first sub-case, $N$ can conclude the step: $N\fw N'$. Now we have that
 $M{\fws} N\fw N'$. Thanks to Lemma~\ref{lemma:swap} we can rearrange such a reduction sequence as follows: $M\fw\fw M_1{\fws} N'$. We then have that the pair $(M_1,N')\in \Re$, as desired. 
 In the second sub-case,
 $N$ can match the step with 2 reductions: $N\fw \fw N'$. Also in this case we can rearrange the reduction 
 sequence so as to obtain $M\fw\fw M_1\fws N'$, with  $(M_1,N')\in \Re$, as desired. 
The case  $M\bka M_1$ (i.e., the challenge is a backward move)  is handled similarly. 
 
We now consider challenges from $N$, focusing only on synchronizations, just as before.
 If $N\fw N'$, we distinguish two cases: whether the reduction ends an ongoing  input/output and branching/selection, or it opens a new one. 
   In the second case $M$ matches the move with an idle move, i.e.,
 $(M,N')\in \Re$. In the other case   we can rearrange the reduction $M{\fws}N\fw N'$ into a similar reduction sequence
 $M\trans{\red} N_1{\fws} N'$ with $\stable(N_1)$, and all reductions  in $N_1\fws N'$ just start new synchronizations. 
 Thanks to Theorem~\ref{theo:correspond}, $M$ can mimick the same reduction to $N_1$, i.e., $M \fwas N_1$, and we have that $(N_1,N')\in \Re$, as desired. 
 The case in which $N \bk N'$ (i.e., the challenge is a backward move) is similar. This concludes the proof.
\end{proof}

\noindent
By observing that the set of 
atomic configurations is a subset of reachable configurations, 
this result can also be  formulated as full abstraction. 
Let $f$ be the (injective, identity) mapping 
from atomic reachable configurations to reachable configurations. We then have:

\begin{cor}[Full Abstraction]
Let $f$ be the injection from atomic reachable configurations to reachable configurations, and let $M, N$ 
be two atomic reachable configurations. 
Then we have $f(M) \congruence\bfbw f(N)$ if and only if $M \congruence\bfbw N$.
\end{cor}
\begin{proof}
From Theorem~\ref{t:bf} we have $M\congruence\bfbw f(M)$ and 
$N\congruence\bfbw f(N)$. The thesis follows then by transitivity of $\congruence\bfbw$.
\end{proof}
The results above ensure that the loss of atomicity  preserves the reachability of configurations yet does not make undesired configurations reachable.

\subsection{Causal Consistency}

 Theorems~\ref{theo:correspond} and~\ref{t:bf} allow us to focus on the atomic reduction $\reda$  for the purposes of establishing causal consistency. 
We follow the approach we developed in our prior work~\cite{DBLP:journals/corr/MezzinaP16,MezzinaP17}, here considering the more general multiparty  setting with asynchronous, higher-order communication; in turn, our prior approach adapts to the proof technique by Danos and Krivine~\cite{DBLP:conf/concur/DanosK05} (developed for a reversible CCS).

In a nutshell, causal consistency concerns traces of \emph{transitions}, each one endowed with an appropriate \emph{stamp}. The proof of causal consistency relies on \emph{square}, \emph{rearranging}, and \emph{shortening} lemmas, which together express properties of traces and transitions that characterize flexible and consistent reversible steps. We start by defining transitions:

\begin{defi}[Transitions and Stamps]
\label{d:stamp}
A \emph{transition} $t$ is a triplet of the form
$t : M\lreda{\memstamp}N$ where $M\reda N$ (cf. \defref{d:atred}) and 
the \emph{transition stamp}
$\memstamp$ is defined as follows:
\begin{itemize}[$\bullet$]
	\item $\memstamp = \{ \loc_1,\p_1,\cdots,\loc_n, \p_n \}$, if  Rule~\fwcolor{(\textsc{Init})} or~\bkcolor{(\textsc{RInit})}  is used;
	\item $\memstamp = \{ \p,\q \}$, if one of Rules~\fwcolor{(\textsc{AC})}, \fwcolor{(\textsc{AS})}, \bkcolor{(\textsc{RAC})} and \bkcolor{(\textsc{RAS})} is used;
	\item $\memstamp = \{ \loc, \p \}$, if one of Rules~\fwcolor{(\textsc{Beta})}, \fwcolor{(\textsc{Spawn})}, 
	\bkcolor{(\textsc{RBeta})} or \bkcolor{(\textsc{RSpawn})} is used.
\end{itemize} 
\end{defi}
\noindent

Some terminology on transitions, taken from~\cite{DBLP:journals/corr/MezzinaP16,MezzinaP17}, is in order.

\begin{defi}[Terminology for Transitions]
Suppose a transition $t : M\lfwa{\memstamp}N$.
\begin{itemize}
    \item We say 
$M$ 
and $N$ are the 
source 
and target of $t$
(written $\tsource{t}$ and $\ttarget{t}$, respectively). 

\item Transition
$t : M\lreda{\memstamp}N$ is \textit{forward} if $M\fwa N$ and \textit{backward} if $M\bka N$. 

\item The \emph{inverse} of $t$, denoted $\nottrace{t}$, is the transition $\nottrace{t}:N\lreda{\memstamp}M$.

\item Two transitions are  \emph{coinitial} if they have the same source; 
\emph{cofinal} if they have the same target;
\emph{composable} if the target of the first one is the source of the other. 
\item Given coinitial transitions $t_1 : M \lreda{\memstamp_1} N_1$ and 
$t_2 : M \lreda{\memstamp_2} N_2$, we define $t_2/t_1$ (read ``$t_2$ after $t_1$'') as 
$N_1\lreda{\memstamp_2} N_2$, i.e., the transition with stamp $\memstamp_2$ that starts from the target of $t_1$ 
\modif{and ends
in the target of $t_2$.}
\end{itemize}

\end{defi}

Two important classes of transitions are \emph{conflicting} and \emph{concurrent} ones:

\begin{defi}\label{def:confl}
Two coinitial transitions $t_1: M\lreda{\memstamp_1} M_1$ and 
$t_2: M\lreda{\memstamp_2} M_2$ are said to be in \emph{conflict} if
$\memstamp_1 \cap \memstamp_2 \neq \emptyset$. 
Two transitions are  \emph{concurrent} if they are not in conflict.
\end{defi}
\noindent
We now consider the so-called Square Lemma~\cite{DanosK04}, which may be informally described as follows. Assume a configuration from which two transitions are possible: if these transitions are concurrent then the order in which they are executed does not matter, and the same configuration is reached. 
\begin{lem}[Square]\label{lemma:square}
If $t_1: M\lreda{\memstamp_1} M_1$ and $t_2: M\lreda{\memstamp_2} M_2$ are coinitial and concurrent transitions, then there exist cofinal transitions $t_2/t_1= M_1 \lreda{\memstamp_2} N$ and
$t_1/t_2= M_2 \lreda{\memstamp_1} N$.
\end{lem}
{
\begin{proof}
By case analysis on the possible rules used to derive $M\lreda{\memstamp_1} M_1$ and 
$M\lreda{\memstamp_2} M_2$. 
To define the valid combinations of rules, we define sets 
\begin{align*}
 \textsc{Rule}  = \{\fwcolor{\textsc{Init}}, \fwcolor{\textsc{AC}}, \fwcolor{\textsc{AS}}, \fwcolor{\textsc{Beta}}, \fwcolor{\textsc{Spawn}}\}
 \qquad 
 \nottrace{\textsc{Rule}}  = 
 \{\bkcolor{\textsc{RInit}}, \bkcolor{\textsc{RAC}}, \bkcolor{\textsc{RAS}}, \bkcolor{\textsc{RBeta}}, \bkcolor{\textsc{RSpawn}}\}
\end{align*}
The licit combinations are given by pairs of rules in the set 
 $\{(r_i,r_j) \,| \, \{r_i,r_j\} \subset \textsc{Rule} \cup \nottrace{\textsc{Rule}} \}$.
Let us note that the definition of concurrent transitions (Definition~\ref{def:confl}) ensures that the pairs  $(r_i, r_j)$ concern rule applications that involve different participants.

All valid cases are treated similarly; we content ourselves by considering only the case (\bkcolor{\textsc{RAS}}, \fwcolor{\textsc{AC}}), in which queue equivalence (Definition~\ref{eq:queue}) plays an important role. 
We have that:
  \begin{align*}
  M\equiv& \news{\vect n} (N \Par  \np{\key{\loc_1}{\er_1}}{ \conf{\stack{C_1}}{\bout{\ep{s}{\p_1}}{V}{P}}} 
 \Par \moni{s_{\p_1}}{\myctxr{\ctx{T}}{\past  \ltout{\q_1}{U}{S}}}{\mytilde x_1}{\store_1} \Par \\ 
  & \np{\key{\loc_2}{\er_2}}{\conf{\stack{C_2}}{\binp{\ep{s}{\q_1}}{y}{Q}}} \Par \moni{s_{\q_1}}{\myctxr{\ctx{S}}{\past  \ltinp{\p_1}{U}{T}}}{\mytilde x_2}{\store_2}  \Par \\
 &\np{\key{\loc_3}{\er_3}}{ \conf { \stack C_3 , \bbra{\ep{s}{\p_2}}{\lbl_l:P_l}_{l \in I\setminus w}}{ P_w} }  \Par 
 \moni{s_{\p_2}}{\myctxr{\ctx{S}}{\ltbrap{\q_2}{\lbl_j:S_j\,,\, \lbl_w{:}\past S_w}{j}{J\setminus w}} }{\mytilde x_3}{\store_3}\Par \\
& \np{\key{\loc_4}{\er_4}}{ \conf{\stack C_4 , \bsel{\ep{s}{\q_2}}{\lbl_l.Q_l}_{l\in I\setminus w} }{Q_w}} \Par 
\moni{s_{\q_2}}{\myctxr{\ctx{T}}{ \ltselp{\p_2}{\lbl_j:T_j \, , \, \lbl_w{:}\past T_w}{j}{J\setminus w} }}{\mytilde x_4}{\store_4} \Par \\
& \codah{s}{h \cons \valueq{\p_2}{\q_2}{l_w} }{  k }  
  \end{align*}
  From $M$ we have  two possible reductions: either $\p_1, \q_1$ communicate or
  $\p_2, \q_2$ undo the selection. 
  By Definition~\ref{d:stamp},  
$\memstamp_1 = \{\p_1,\q_1\}$ and 
  $\memstamp_2 = \{\p_2,\q_2\}$. By Rule~\fwcolor{\textsc{AC}} we have:
    \begin{align*}
  M \lreda{\memstamp_1} & \news{\vect n} (N \Par \np{\key{\loc_1}{\er_1}}{\conf{\stack{C_1}}{P}} \Par \moni{s_{\p_1}}{\myctxr{\ctx{T}}{ \ltoutp{\q_1}{U}{S}}}{\mytilde x_1}{\store_1} \Par \\ 
  & \np{\key{\loc_2}{\er_2}}{\conf{\stack{C_2}}{Q}} \Par \moni{s_{\q_1}}{ \myctxr{\ctx{S}}{\ltinpp{\p_1}{U}{T}}}{\mytilde x_2, y}{\store_2\upd{y}{V}} \Par \\
 &\np{\key{\loc_3}{\er_3}}{ \conf { \stack C_3 , \bbra{\ep{s}{\p_2}}{\lbl_l:P_l}_{l \in I\setminus w}}{ P_w} }  \Par 
 \moni{s_{\p_2}}{\myctxr{\ctx{S}}{\ltbrap{\q_2}{\lbl_j:S_j\,,\, \lbl_w{:}\past S_w}{j}{J\setminus w}} }{\mytilde x_3}{\store_3}\Par \\
& \np{\key{\loc_4}{\er_4}}{ \conf{\stack C_4 , \bsel{\ep{s}{\q_2}}{\lbl_l.Q_l}_{l\in I\setminus w} }{Q_w}} \Par 
\moni{s_{\q_2}}{\myctxr{\ctx{T}}{ \ltselp{\p_2}{\lbl_j:T_j \, , \, \lbl_w{:}\past T_w}{j}{J\setminus w} }}{\mytilde x_4}{\store_4} \Par \\
& \codah{s}{h \cons  \valueq{\p_2}{\q_2}{l_w}  \cons \valueq{\p_1}{\q_1}{V} }{ k} = M_1
  \end{align*}
 and by Rule~\bkcolor{\textsc{RAS}} we have:
 \begin{align*}
  M\lreda{\memstamp_2}&  \news{\vect n} (N \Par  \np{\key{\loc_1}{\er_1}}{ \conf{\stack{C_1}}{\bout{\ep{s}{\p_1}}{V}{P}}} 
 \Par \moni{s_{\p_1}}{\myctxr{\ctx{T}}{\past  \ltout{\q_1}{U}{S}}}{\mytilde x_1}{\store_1} \Par \\ 
  & \np{\key{\loc_2}{\er_2}}{\conf{\stack{C_2}}{\binp{\ep{s}{\q_1}}{y}{Q}}} \Par \moni{s_{\q_1}}{\myctxr{\ctx{S}}{\past  \ltinp{\p}{U}{T}}}{\mytilde x_2}{\store_2}  \Par \\
 &\np{\key{\loc_3}{\er_3}}{ \conf { \stack C_3}{ \bbra{\ep{s}{\p_2}}{\lbl_l:P_l}_{l \in I}} }  \Par 
 \moni{s_{\p_2}}{\myctxr{\ctx{S}}{\past\ltbrap{\q_2}{\lbl_j:S_j}{j}{J}} }{\mytilde x_3}{\store_3}\Par \\
& \np{\key{\loc_4}{\er_4}}{ \conf{\stack C_4}{\bsel{\ep{s}{\q_2}}{\lbl_l.Q_l}_{l\in I} }} \Par 
\moni{s_{\q_2}}{\myctxr{\ctx{T}}{\past \ltselp{\p_2}{\lbl_j:T_j }{j}{J} }}{\mytilde x_4}{\store_4} \Par \\
& \codah{s}{h }{  k }  = M_2
  \end{align*}
Now it is easy to see that  there is an $N$ such that $M_1 \lreda{\memstamp_2} N$  and   $M_2 \lreda{\memstamp_1} N$:
\begin{align*}
M_1 \lreda{\memstamp_2} &  \news{\vect n} (N \Par \np{\key{\loc_1}{\er_1}}{\conf{\stack{C_1}}{P}} \Par \moni{s_{\p_1}}{\myctxr{\ctx{T}}{ \ltoutp{\q_1}{U}{S}}}{\mytilde x_1}{\store_1} \Par \\ 
  & \np{\key{\loc_2}{\er_2}}{\conf{\stack{C_2}}{Q}} \Par \moni{s_{\q_1}}{ \myctxr{\ctx{S}}{\ltinpp{\p_1}{U}{T}}}{\mytilde x_2, y}{\store_2\upd{y}{V}} \Par \\
  &\np{\key{\loc_3}{\er_3}}{ \conf { \stack C_3} {\bbra{\ep{s}{\p_2}}{\lbl_l:P_l}_{l \in I}} }  \Par 
 \moni{s_{\p_2}}{\myctxr{\ctx{S}}{\past \ltbrap{\q_2}{\lbl_j:S_j}{j}{J}} }{\mytilde x_3}{\store_3}\Par \\
& \np{\key{\loc_4}{\er_4}}{ \conf{\stack C_4}{\bsel{\ep{s}{\q_2}}{\lbl_l.Q_l}_{l\in I} }} \Par 
\moni{s_{\q_2}}{\myctxr{\ctx{T}}{ \past\ltselp{\p_2}{\lbl_j:T_j}{j}{J} }}{\mytilde x_4}{\store_4} \Par \\
& \codah{s}{h \cons \valueq{\p_1}{\q_1}{V} }{  k } = N
\end{align*}
\begin{align*}
M_2 \lreda{\memstamp_1} &  \news{\vect n} (N \Par \np{\key{\loc_1}{\er_1}}{\conf{\stack{C_1}}{P}} \Par \moni{s_{\p_1}}{\myctxr{\ctx{T}}{ \ltoutp{\q_1}{U}{S}}}{\mytilde x_1}{\store_1} \Par \\ 
  & \np{\key{\loc_2}{\er_2}}{\conf{\stack{C_2}}{Q}} \Par \moni{s_{\q_1}}{ \myctxr{\ctx{S}}{\ltinpp{\p_1}{U}{T}}}{\mytilde x_2, y}{\store_2\upd{y}{V}} \Par \\
  &\np{\key{\loc_3}{\er_3}}{ \conf { \stack C_3}{\bbra{\ep{s}{\p_2}}{\lbl_l:P_l}_{l \in I}} }  \Par 
 \moni{s_{\p_2}}{\myctxr{\ctx{S}}{\past \ltbrap{\q_2}{\lbl_j:S_j}{j}{J}} }{\mytilde x_3}{\store_3}\Par \\
& \np{\key{\loc_4}{\er_4}}{ \conf{\stack C_4}{\bsel{\ep{s}{\q_2}}{\lbl_l.Q_l}_{l\in I} }} \Par 
\moni{s_{\q_2}}{\myctxr{\ctx{T}}{ \past\ltselp{\p_2}{\lbl_j:T_j}{j}{J} }}{\mytilde x_4}{\store_4} \Par \\
& \codah{s}{h \cons \valueq{\p_1}{\q_1}{V} }{  k } = N
\end{align*}
Let us note that in $M_1$ the equivalence on queues (cf. Definition~\ref{eq:queue}) allows the swapping of the two messages $ \valueq{\p_2}{\q_2}{l_w}  \cons \valueq{\p_1}{\q_1}{V}$ so to enact the Rule~\bkcolor{\textsc{RAS}}.
\end{proof}
}

\noindent

We now turn our attention to \emph{traces}, sequences of pairwise composable transitions. 
We let 
$\trace$ 
range over  traces. 
Notions of target, source, composability and inverse extend naturally from transitions to traces. 
We write $\etrace_M$ to denote the empty trace with source $M$, 
and $\trace_1;\trace_2$ to denote the composition of two composable traces $\trace_1$ and $\trace_2$. 

\begin{defi}
We define $\causeq$ as the least equivalence between traces that is closed under composition and that obeys:
i)~$t_1; t_2/t_1 \causeq t_2; t_1/t_2$; ii)~ $t;\nottrace{t} \causeq \etrace_{\tsource{t}}$; iii)~$\nottrace{t};t \causeq \etrace_{\ttarget{t}}
$.
\end{defi}
\noindent
Intuitively, $\causeq$ says that: 
(a) given two concurrent transitions, the traces obtained by swapping their execution order are equivalent; 
(b) a trace consisting of opposing transitions is equivalent to the empty trace.

The proof of causal consistency follows that in~\cite{DanosK04}, but with simpler arguments (because of our simpler transition stamps), which mirror those in \cite{DBLP:journals/corr/MezzinaP16,MezzinaP17}.

The following lemma says that, up to causal equivalence, traces can be rearranged so as to reach the maximum freedom of choice, first going only backwards, and then going only forward. 

\begin{lem}[Rearranging]\label{lemma:rearranging}
 Let $\trace$ be a trace. There are forward traces $\trace', \trace''$ such that 
 $\trace \causeq \nottrace{ \trace' }; \trace''$.
\end{lem}
\begin{proof}
 By lexicographic induction on the length of $\trace$ and on the distance between the beginning of 
$\trace$ and the earliest pair of opposing transitions in $\trace$. The analysis uses both the Loop Lemma 
(Lemma~\ref{lemma:loop}) and the Square Lemma (Lemma~\ref{lemma:square}).
\end{proof}

\noindent
If trace $\trace_1$ and forward trace 
$\trace_2$ start from the same configuration and end up in the same configuration, then $\trace_1$ may contain some ``local steps'', not present in $\trace_2$, which must be eventually reversed---otherwise there would be a difference with respect to   $\trace_2$. 
Hence, $\trace_1$ could be \emph{shortened} by removing such local steps and their corresponding reverse steps. 
\begin{lem}[Shortening]\label{lemma:shortening}
 Let $\trace_1, \trace_2$ be coinitial and cofinal traces, 
 with $\trace_2$ forward. Then, there exists a forward trace 
 $\trace'_1$ of length at most that of $\trace_1$ such that 
 $\trace'_1 \causeq \trace_1$.
\end{lem}
\begin{proof}
By induction on the length of $\trace_1$, using Square and Rearranging Lemmas (Lemmas~\ref{lemma:square} and~\ref{lemma:rearranging}). 
The proof uses the forward trace $\trace_2$ as guideline for shortening $\trace_1$ into a forward trace, relying  on the fact that $\trace_1$ and $\trace_2$ share the same source and target.
\end{proof}
We may now state our main result:
\begin{thm}[Causal consistency]\label{t:causal}
 Let $\trace_1$ and $\trace_2$ be coinitial traces, then 
 $\trace_1 \causeq \trace_2$ if and only if $\trace_1$ and $\trace_2$ are cofinal.
\end{thm}
\begin{proof}
The `if' direction follows by definition of $\causeq$ and trace composition. The `only if' direction uses Square, Rearranging, and Shortening Lemmas (Lemmas~\ref{lemma:square},~\ref{lemma:rearranging}, and~\ref{lemma:shortening}). 
\end{proof}


At this point one may object that causal-consistency has been proved on the atomic semantics and not on the decoupled semantics, and wonder whether the chosen behavioural equivalence (cf. Definition~\ref{d:bf}) is causal-preserving. First, we observe that our notion of equivalence is a congruence, and that the only visible event in a message-passing system is the receipt of a message itself. Then it is easy to see that our equivalence preserves the order of messages sent, and hence that the decoupled semantics respects the same notion of causality of the atomic semantics.

\subsection{Connecting (Reversible) Protocols and (Reversible) Configurations}\label{ss:conn}
We now connect the two levels of abstraction in our reversible model by relating protocols and configurations.
This is the content of Theorem~\ref{t:corrgc}, which relies on a few auxiliary definitions.

We introduce a notion of \emph{well-formed} processes and configurations that implement a given local type.
Figure~\ref{f:wft} reports a set of rules for decreeing well-formed processes: 
it is inspired by the type system for higher-order session processes defined in~\cite{KPY2016}.

\begin{figure}[!t]
\begin{mdframed}
{\small
\begin{mathpar}
\inferrule{ }
{\emptyset; \emptyset \vdash \mathtt{true} :: \bool  }
\qquad 
\inferrule{ }
{\emptyset; \emptyset \vdash \mathtt{false} :: \bool  }
\qquad 
\inferrule{\Gamma ; \Delta, x:T \vdash P}
{\Gamma; \Delta \vdash \abs{x}{P} :: \shot{T}}
\\
\inferrule
{\Gamma; \Delta \vdash P \quad u \not\in \dom{\Delta} }{
\Gamma; \Delta,u: \lend   \vdash P}
\qquad
\inferrule
{ }{
\Gamma; \emptyset \vdash  \inact 
}
\qquad 
\inferrule{
\Gamma; \Delta \vdash V :: \shot{T}
\quad
\vdash u :: T
}{
\Gamma; \Delta \vdash \appl{V}{u} 
}
\\
\inferrule
{
u:T \in \Delta_1, \Delta_2
\quad 
\Gamma; \Delta_1   \vdash P  \quad  \Gamma; \Delta_2 \vdash V::U}
{\Gamma;((\Delta_1, \Delta_2) \setminus u:T), u:\typeOut{\p}{U}.T \vdash \bout{u}{V}{P}}
\qquad
\inferrule{\Gamma; \Delta, u:T, x:U \vdash P}
{\Gamma; \Delta, u:\typeIn{\p}{U}.T \vdash \binp{u}{x}{P}}
\\
\inferrule{\forall i \in  \{1, \ldots, n\}.(\Gamma; \Delta, u:T_i \vdash P_i)}
{\Gamma; \Delta, u:\ltsel{\q}{\lbl_i}{T_i}{i}{\{1, \ldots, n\}} \vdash  \bsel{u}{\lbl_i. P_i}_{i\in \{1, \ldots, n\}}}
\\
\inferrule{\forall i \in  \{1, \ldots, n\}.(\Gamma; \Delta, u:T_i \vdash P_i)}
{\Gamma; \Delta, u:\ltbra{\q}{\lbl_i}{T_i}{i}{\{1, \ldots, n\}} \vdash  \bbra{u}{\lbl_i : P_i}_{i\in \{1, \ldots, n\}} }
\\
\inferrule
{  }{
\Gamma, X{:}\Delta;  \Delta \vdash X}
\qquad
\inferrule
{
\Gamma, X{:}\Delta; \Delta  \vdash P}
{\Gamma; \Delta  \vdash \mu X.P}
\end{mathpar}
}
\vspace{-4mm}
\end{mdframed}
\caption{Well-formed processes.\label{f:wft}}
\end{figure}

Our system for well-formedness is simple, and relies on two contexts: $\Gamma$ (for recursion variables) and $\Delta$ (for assignments of variables to local types). 
We omit these contexts when empty and/or unimportant. 
Well-formedness uses the following judgments: 
\begin{itemize}
\item $\vdash u :: T$ says that $u$ is a name of local type $T$
\item $\Gamma; \Delta \vdash V:: U$ says that $V$ is a well-formed value of type $U$
\item $\Gamma; \Delta \vdash P$ says that $P$ is a well-formed process 
\end{itemize}
The first three rules in Figure~\ref{f:wft} are for values: booleans and abstractions (rules for other base values are similar).
Then we have a rule enforcing a weakening principle, and  rules for inaction and application,  which are as expected. 
The rule for output enables us to account for processes in which a communicated abstraction specifies a protocol which is continued outside the output action; this is case for process $\text{Betty}$
in \secref{sss:buysel2}, in which part of the protocol on $z$ is sent around as a thunk.
The remaining rules, for input, selection, branching, and recursive processes, are self-explanatory. Notice that for the sake of simplicity, we consider restriction-free, single-threaded  processes (i.e., no processes of the form $\news{n} P$ and $P_1 \Par P_2$).

We are interested in well-formed processes that implement a single session with local type $T$ along  $u$ (a session name or a variable):
\begin{defi}[Well-Formed Processes]
We say process $P$ is \emph{well-formed} 
if  
$\emptyset; \{u:T\} \vdash P$
in the system of Figure~\ref{f:wft}, for some $u$ and $T$.
This is denoted $\ladeq{P}{T}{u}$.
\end{defi}

We may then define the configurations that implement a global type with history 
(cf.~Def.~\ref{d:gth}). First, an auxiliary definition:

\begin{defi}[Reachable Global Types]
\label{d:reachgth}
We say the global type with history $\gth{H}$ is \emph{reachable} if it can be obtained from a global type $G$
via a 
sequence of $\fwg$ and $\bkg$ transitions (cf. Fig.~\ref{f:gts}).
\end{defi}

\begin{defi}[Configurations Implementing Global Types]
\label{d:ii}
Let ${G}$ 
be a global type, 
with $\parties{G}= \{\p_1,\cdots,\p_n\}$.
\begin{itemize}
    \item We say that
configuration $M$  \emph{initially implements} $G$, written $\initadeq{M}{G}$, if we have
\begin{align*}
M\equiv  
\news{s} \Big(\codah{s}{\epsilon}{\epsilon}  \Par 
\prod_{i \in\{1,\cdots,n\}} \np{\key{\loc_i}{\p_i}}{ \conf{\inact}{P_i\subst{\key{s}{\p_i}}{x_i}}} \Par 
\moni{s_{\p_i}}{\past\tproj{G}{\p_i}}{x_i}{\store_i}  \Big)
\end{align*}
with $\ladeq{P_i}{\tproj{G}{\p_i}}{x_i}$, for all $i \in\{1,\!\cdots\!,n\}$, 
for some stores $\store_1, \ldots, \store_n$.

\item A configuration $N$ \emph{implements} the global type with history $\gth{H}$, written $\adeq{N}{\gth{H}}$, if
there exist $M, G$ such that (i)~$\gth{H}$ is reachable from $G$, (ii) $\initadeq{M}{G}$, and (iii) $N$ is reachable from $M$.
\end{itemize}

\end{defi}

\begin{figure}[!t]
\begin{mdframed}
{\small
\begin{mathpar}
\inferrule
{\ladeq{P}{T}{u}}
{\ladeq{\conf{\inact}{P}}
{\past T}{u}}
\qquad
\inferrule{\ladeq{\conf{\stack{C}}{\bout{u}{V}{P}}}{\myctxr{\ctx{T}}{\past  \ltout{\q}{U}{S}}}{u}}
{\ladeq{\conf{\stack{C}}{{P}}}{\myctxr{\ctx{T}}{\ltoutp{\q}{U}{S}}}{u}}
\qquad
\inferrule{\ladeq{\conf{\stack{C}}{\binp{u}{y}{P}}}{\myctxr{\ctx{T}}{\past  \ltinp{\q}{U}{S}}}{u}}
{\ladeq{\conf{\stack{C}}{{P}}}{\myctxr{\ctx{T}}{\ltinpp{\q}{U}{S}}}{u}}
\\
\inferrule{\ladeq{\conf{\stack C}{\bsel{u}{\lbl_i. P_i}_{i\in I}}}{\myctxr{\ctx{T}}{\past \ltsel{\q}{\lbl_j}{S_j}{j}{J}}}{u} \quad w \in I,J}
{\ladeq{\conf{\stack C , \bsel{u}{\lbl_l.P_l}_{l\in I\setminus w} }{P_w}}{\myctxr{\ctx{T}}{ \ltselp{\q}{\lbl_j:S_j \, , \, \lbl_w:\past S_w}{j}{J\setminus w}} }{u}}
\\
\inferrule{\ladeq{\conf{\stack C}{\bbra{u}{\lbl_i. P_i}_{i\in I}}}{\myctxr{\ctx{T}}{\past \ltbra{\q}{\lbl_j}{S_j}{j}{J}}}{u} \quad w \in I,J}
{\ladeq{\conf{\stack C , \bbra{u}{\lbl_l.P_l}_{l\in I\setminus w} }{P_w}}{\myctxr{\ctx{T}}{ \ltbrap{\q}{\lbl_j:S_j \, , \, \lbl_w:\past S_w}{j}{J\setminus w}} }{u}}
\\
\inferrule{\ladeq{\conf{\stack{C}}{{P}}}{\myctxr{\ctx{T}}{\ltoutp{\q}{U}{S}}}{u} \quad \vdash V :: U}
{\ladeq{\conf{\stack{C}}{\bout{u}{V}{P}}}{\myctxr{\ctx{T}}{\past  \ltout{\q}{U}{S}}}{u}}
\qquad
\inferrule{\ladeq{\conf{\stack{C}}{{P}}}{\myctxr{\ctx{T}}{\ltinpp{\q}{U}{S}}}{u} \quad y:U \vdash P :: S}
{\ladeq{\conf{\stack{C}}{\binp{u}{y}{P}}}{\myctxr{\ctx{T}}{\past  \ltinp{\q}{U}{S}}}{u}}
\\
\inferrule
{\ladeq{\conf{\stack C , \bsel{u}{\lbl_l.P_l}_{l\in I} }{P_w}}{\myctxr{\ctx{T}}{ \ltselp{\q}{\lbl_j:S_j \, , \, \lbl_w:\past S_w}{j}{J}} }{u}}
{\ladeq{\conf{\stack C}{\bsel{u}{\lbl_i. P_i}_{i\in I \cup w}}}{\myctxr{\ctx{T}}{\past \ltsel{\q}{\lbl_j}{S_j}{j}{J \cup w}}}{u}}
\\
\inferrule
{\ladeq{\conf{\stack C , \bbra{u}{\lbl_l.P_l}_{l\in I} }{P_w}}{\myctxr{\ctx{T}}{ \ltbrap{\q}{\lbl_j:S_j \, , \, \lbl_w:\past S_w}{j}{J}} }{u}}
{\ladeq{\conf{\stack C}{\bbra{u}{\lbl_i. P_i}_{i\in I \cup w }}}{\myctxr{\ctx{T}}{\past \ltbra{\q}{\lbl_j}{S_j}{j}{J \cup w}}}{u}}
\end{mathpar}
}
\end{mdframed}
\caption{Well-formed configurations with respect to a local type with history.\label{f:wfc}}
\end{figure}

Observe how $\initadeq{M}{G}$ formalizes $M$ as the result of initializing the configuration, following Rule \fwcolor{\textsc{(Init)}} (cf. Fig.~\ref{fig:fw_1}).
This way, $\adeq{N}{\gth{H}}$ reflects the evolution from an initial implementation, 
with $\gth{H}$ being reachable from $G$ and $N$ being reachable from $M$ following forward and backward rules. 
The following proposition details the shape of a configuration that is reachable from $\initadeq{M}{G}$:

\begin{prop}\label{p:shape}
Let $\adeq{N}{\gth{H}}$ 
with $\parties{\gth{H}}= \{\p_1,\cdots,\p_n\}$.
Then we have 
\begin{align*}
M\equiv  
\news{s,\widetilde{n}} \Big(\, 
\prod_{i \in\{1,\cdots,n\}} \np{\key{\loc_i}{\p_i}}{ \conf{\stack{C}_i}{Q_i}} \Par 
\monig{s_{\p_i}}{\myctxr{\ctx{T}_i}{ \past S_i}}{\widetilde{x_i}}{\store_i}  \Par \codah{s}{h_1^i}{h_2^i}\Big)
\end{align*}
where, for all $i \in\{1,\!\cdots\!,n\}$,
$\ladeq{{\conf{\stack{C}_i}{Q_i}}}{\myctxr{\ctx{T}_i}{ \past S_i}}{\key{s}{\p_i}}$ holds as in Figure~\ref{f:wfc}.
\end{prop}
\begin{proof}
Immediate from Definition~\ref{d:ic} (reachable configuration), Definition~\ref{d:ii} (``initially implements''),
and the reduction semantics $\red$.
\end{proof}

Recall that $\mytagg$ can be either full $\bkcolor{\rmark}$ or empty $\normark$; thus, if 
$\adeq{M}{\gth{H}}$ then $M$ may not be stable.

The last ingredient required 
is a \emph{swapping relation} over global types, denoted \swap, which enables 
behavior-preserving
transformations among causally independent communications. 

\begin{defi}[Swapping]
We define \swap as the smallest congruence on $G$ that satisfies the rules in Fig.~\ref{f:swap}
(where we omit the symmetric versions of (\textsc{Sw1}), (\textsc{Sw2}), and (\textsc{Sw3})).
We extend \swap  to global types with history $\gth{H}$ as follows:
$\ctx{G}[\past G_1] \swap \ctx{G'}[\past G_2]$ 
if 
$\ctx{G}[\gend] \swap \ctx{G'}[\gend]$
and 
$G_1 \swap G_2$.
\end{defi}

\begin{figure}[!t]
\begin{mdframed}
{\small
\begin{mathpar}
\inferrule*[left=(Sw1)]{\{\p_1,\q_1\}\#\{\p_2,\q_2\}}
{
\gtcom{\p_1}{\q_1}{U_1}{(\gtcom{\p_2}{\q_2}{U_2}{G})}
\swap 
\gtcom{\p_2}{\q_2}{U_2}{(\gtcom{\p_1}{\q_1}{U_1}{G})}
}
\\
\inferrule*[left=(Sw2)]{\{\p_1,\q_1\}\#\{\p_2,\q_2\}}
{
\gtcom{\p_1}{\q_1}{U_1}{(\gtcho{\p_2}{\q_2}{\lbl_i}{G_i})}
\swap 
\gtcho{\p_2}{\q_2}{\lbl_i}{(\gtcom{\p_1}{\q_1}{U_1}{G_i})}
}
\\
\inferrule*[left=(Sw3)]{\{\p_1,\q_1\}\#\{\p_2,\q_2\}}
{
\gtchoi{\p_1}{\q_1}{\lbl_i}{(\gtchoi{\p_2}{\q_2}{\lbl_j}{G_j}{j \in J})}{i \in I}
\swap
\gtchoi{\p_2}{\q_2}{\lbl_j}{(\gtchoi{\p_1}{\q_1}{\lbl_i}{G_i}{i \in I})}{j \in J}
}
\end{mathpar}
}
\end{mdframed}
\caption{Swapping on global types. We write $A \# B$ if $A$ and $B$ are disjoint sets.\label{f:swap}}
\end{figure}


 {Notice that Definition \ref{eq:queue} and swapping play similar r\^{o}les but at different levels: queues/con\-fi\-gu\-ra\-tions and global types, respectively.}

We comment on the statement of  Theorem~\ref{t:corrgc}, given next,  which relates (i)~transitions in the  semantics of (high-level) global types (with history) with 
(ii)~reductions in the  semantics of their (low-level) process implementations. 
It is in two parts, which capture an asymmetry between a global type $\gth{H}$ and a configuration $M$, with $\adeq{M}{\gth{H}}$:
while Part~(a) shows that the behavior of $\gth{H}$
can be closely mimicked by $M$, 
Part~(b) shows that $M$ may have more immediate behaviors than  $\gth{H}$:
this is because $M$ may include several 
independent (and immediate) reductions (written $M \fw N_i$ and  $M \bk N_i$ below), which 
are matched by $\gth{H}$ up to swapping.
The asymmetry can then be interpreted as configurations being more concurrent (less sequential) than a global type.


Below, $M \bkk M'$ denotes a sequence of $j \geq 0$ reduction steps (if $j=0$ then $M = M'$).
Also, we write 
$\gth{H}^{\swap} \fwg \gth{H}''$
to mean that 
$\gth{H} \swap \gth{H}' \land \gth{H}' \fwg \gth{H}''$, 
for some $\gth{H}'$ (and similarly for $\bkg$).

\begin{thm}\label{t:corrgc}
Let $\mathsf{H}$  be a reachable global type with history (cf. Def.~\ref{d:reachgth}).
Suppose $\adeq{M}{\gth{H}}$.
\begin{enumerate}[a)]
\item If   $\gth{H} \fwg \gth{H}'$  then  $M \fw M'$  and $\adeq{M'}{\gth{H'}}$, for some $M'$. \\
Also, if $\gth{H} \bkg \gth{H}'$ then $M \bkk M'$ (with $j =1$ or $j=2$) and $\adeq{M'}{\gth{H'}}$, for some $M'$.

\item  For all $N_i$ such that $M \fw N_i$, there exist 
$\gth{H}_i,  \gth{H}'$,
and $M'$,
such that
 $\gth{H}^{\swap} \fwg \gth{H}_i$,
 $\adeq{N_i}{\gth{H}_i}$,
 $N_i \fws M'$,
 $\gth{H}_i \fwgs \gth{H}'$,
 and 
 $\adeq{M'}{\gth{H}'}$
 (and similarly for $\bk$, $\bkg$).

\end{enumerate} 
\end{thm}



\begin{proof}
We consider both parts separately:
\begin{description}
    \item[Part (a)] By induction on the transitions $\gth{H} \fwg \gth{H}'$ and $\gth{H} \bkg \gth{H}'$, with a case analysis on the last applied rule (Fig.~\ref{fig:prot_sem}).
    
For $\gth{H} \fwg \gth{H}'$ there are four possible transitions, we have a one-to-one correspondence:
\begin{itemize}
    \item a transition derived using Rule~\textsc{(FVal1)} is matched by $M$   using Rule~\textsc{(Out)};  
    \item a transition derived using Rule~\textsc{(FVal2)} is matched by $M$  using Rule~\textsc{(In)};  
    \item a transition derived using Rule~\textsc{(FCho1)} is matched by $M$   using Rule~\textsc{(Sel)};  
    \item  a transition derived using Rule~\textsc{(FCho2)} is matched by $M$  using Rule~\textsc{(Bra)}.
\end{itemize}

The analysis for $\gth{H} \bkg \gth{H}'$ is similar, but  
 we may require an additional reduction step from $M$, depending on the tag 
 of the corresponding monitor (cf. Prop.~\ref{p:shape}).
 If the tag of $M$ is $\rmark$  then the transition can be immediately matched as follows ($j=1$):
 \begin{itemize}
     \item a transition derived using Rule~\textsc{(BVal1)} is matched by $M$  using Rule~\textsc{(ROut)};  
     \item a transition derived using Rule~\textsc{(BVal2)} is matched by $M$   using Rule~\textsc{(RIn)};  
     \item a transition derived using Rule~\textsc{(BCho1)} is matched by $M$   using Rule~\textsc{(RSel)};  
     \item a transition derived using Rule~\textsc{(BCho2)} is matched by $M$   using Rule~\textsc{(RBra)}.
 \end{itemize}
Otherwise, if  the tag of $M$ is $\normark$, then $j=2$ because an additional reduction  (using Rule~\textsc{(RollS)} or \textsc{(RollC)})
 is required 
 in order to reach a configuration with tag $\rmark$.

\item[Part (b)] By induction on transitions $M \fw N$ and $M \bk N$, with a case analysis on the last applied rule, following similar lines.
There are two main cases: 
\begin{enumerate}[(i)]
\item There is exactly one reduction $M \fw N$ (and $M \bk N$), which involves participants that appear at the top-level in $\gth{H}$. 
\item There are one or more reductions $M \fw N_i$ (and $M \bk N_i$) whose involved participants cannot be found at the top-level in $\gth{H}$.
\end{enumerate}

We discuss case~(i) first. 
Suppose $M \fw N$: then the reduction was obtained using one of the following 
Rules~\textsc{(Out)}, \textsc{(In)}, \textsc{(Sel)}, \textsc{(Bra)},  and ~\textsc{(Beta)}.
Notice that a reduction with Rule~\textsc{(Spawn)} is not possible under our  definition of well-formed processes (and configurations).
For the first four cases, a corresponding transition  $\gth{H} \fwg \gth{H}'$  can be easily obtained, as in the analysis for Part~(a); a reduction obtained with Rule~\textsc{(Beta)} does not involve the global type, and so  $\gth{H}' = \gth{H}$. In either case, $\adeq{N}{\gth{H}'}$ holds easily.
Now suppose $M \bk N$. Here the analysis depends on the tag in $M$: if it is $\rmark$ then the reduction was derived using Rules~\textsc{(ROut)}, \textsc{(RIn)}, \textsc{(RSel)}, \textsc{(RBra)}, or \textsc{(RBeta)}. Here again a reduction via Rule~\textsc{(RSpawn)} is not possible.
As before, in the first four cases the reduction can be mimicked directly by one transition $\gth{H} \bkg \gth{H}'$; in the last one, there is no global type transition ($\gth{H}' = \gth{H}$). 
If the tag is $\normark$ then the reduction is derived using Rules~\textsc{(RollS)} or \textsc{(RollC)}. This reduction is not mimicked by $\gth{H}$ but enables a reduction  $N \bk M'$, using one of the Rules~\textsc{(Out)}, \textsc{(In)}, \textsc{(Sel)}, \textsc{(Bra)},  and ~\textsc{(Beta)}, which can be mimicked as discussed for the case $\rmark$, ensuring $\adeq{M'}{\gth{H}'}$.

In case (ii) we use $\swap$  to obtain 
behavior-preserving transformations $\gth{H}^*_i$ of $\gth{H}$ in which the participants involved in the reductions
 ($M \fw N_i$ or $M \bk N_i$) appear at the top-level.
 Such transformations exist, because of assumption $\adeq{M}{\gth{H}}$.
This way, the reductions from $M$ can be matched by $\gth{H}^*_i$, following the analysis described in case~(i); after all the independent communication actions 
have been performed and matched (there are finitely many of them), it is easy to obtain $M'$ and $\gth{H}'$ such that  $\adeq{M'}{\gth{H}'}$.
\qedhere
\end{description}
\end{proof}

\bigskip

\noi \textbf{Summing up,} we have that
Theorem~\ref{t:causal} ensures that reversibility in the atomic semantics is causally consistent.
 Theorem~\ref{t:bf} transfers this result to decoupled semantics; since by 
Theorem~\ref{t:corrgc} decoupled semantics defines a sound local implementation, we conclude that reversibility for global types is also causally consistent.



\section{Discussion and Related Work}\label{s:rw}


\begin{figure}[!t]
\begin{mdframed}
\begin{align*}
& \inferrule[\fwcolor{(Init$^*$)}]
{\parties{G}= \{\p_1,\cdots,\p_n\} \and 
T_1 = \tproj{G}{\gpart{p}_1} ~~ \cdots~~   T_n = \tproj{G}{\gpart{p}_n}
}{
\prod_{i \in \{1 .. n\}} L_i ~
\fwalt 
\news{s}\Big( \prod_{i \in \{1 .. n\}} 
\codah{s_{\p_i}}{\emp}{\emp}{}
\Par M_i \Par N_i 
\Big)}
\\
& \text{where:}
\\
& \begin{array}{ll}
L_1  = \myloc{\loc_1}{\bout{a}{x_1:T_1}{P_{1}}}
& 
M_i  = \np{\key{\loc_i}{\p_i}}{ \conf{\inact}{P_i\subst{\ep{s}{\p_i}}{x_i}}} \text{~for $i = 1..n$} 
\\
L_j  = \myloc{\loc_j}{\binp{a}{x_j:T_j}{P_{j}}} \text{~for $j = 2..n$} 
&
N_i  = \moni{s_{\p_i}}{\past T_{i}}{x_i}{\upd{x_i}{a }} \text{~for $i = 1..n$} 
\end{array}
\end{align*}
\begin{align*}
& \inferrule[\fwcolor{(Out$^*$)}]
{\p = \er \,\vee\, \p \in \names{\er, h}
}{
M \Par N  \Par 
\codah{s_{\p}}{h}{k}{}
~ \fwalt
M' \Par N' \Par 
\codah{s_{\p}}{h}{k \cons \langle \q,\myeval{V}{\sigma}\rangle}{}
}
\\
& \text{where:} 
\\
& \begin{array}{ll}
M  = \np{\key{\loc}{\er}}{ \conf{\stack{C}}{\bout{\ep{s}{\p}}{V}{P}}}
&
N  = \moni{s_\p}{\myctxr{\ctx{T}}{\past  \ltout{\q}{U}{S}}}{\mytilde x}{\store}
\\
M'  = \np{\key{\loc}{\er}}{\conf{\stack{C}}{P}}
&
N'  =\moni{s_\p}{\myctxr{\ctx{T}}{ \ltoutp{\q}{U}{S}}}{\mytilde x}{\store}
\end{array}
\end{align*}
\begin{align*}
& \inferrule[\fwcolor{(In$^*$)}]
{\p = \er \,\vee\, \p \in \names{\er, h_1}
}{
\begin{array}{c}
M \Par N  
\Par  \codah{s_{\p}}{h_1}{k_1}{}
\Par\codah{s_{\q}}{h_2}{\langle \p,V\rangle  \cons k_2}{}
~ \fwalt \qquad \qquad 
\\
M' \Par N'  
\Par  \codah{s_{\p}}{h_1 \cons \langle \q,V\rangle }{k_1}{}
\Par\codah{s_{\q}}{h_2}{k_2}{}
\end{array}
}
\\
& \text{where:} 
\\
& \begin{array}{ll}
M  = \np{\key{\loc}{\er}}{\conf{\stack{C}}{\binp{\ep{s}{\p}}{y}{P}}} 
&
N  = \moni{s_\p}{\myctxr{\ctx{T}}{\past  \ltinp{\q}{U}{S}}}{\mytilde x}{\store}
\\
M'  = \np{\key{\loc}{\er}}{\conf{\stack{C}}{P}} 
&
N'  =\moni{s_\p}{ \myctxr{\ctx{T}}{\ltinpp{\q}{U}{S}}}{\mytilde x, y}{\store\upd{y}{V}} 
\end{array}
\end{align*}
\end{mdframed}
\caption{An alternative decoupled semantics for configurations ($\fwalt$, $\bkalt$).}
\label{fig:fw_exp1}
\end{figure}

\begin{figure}[!t]
\begin{mdframed}
\begin{align*}
& \inferrule[\bkcolor{(RInit$^*$)}]
{\parties{G}= \{\p_1,\cdots,\p_n\} \and 
T_1 = \tproj{G}{\gpart{p}_1} ~~ \cdots~~   T_n = \tproj{G}{\gpart{p}_n}
}{
\news{s}\Big( \prod_{i \in \{1 .. n\}} \codah{s_{\p_i}}{\emp}{\emp}{} \Par  M_i \Par N_i 
\Big)
\bkalt
\prod_{i \in \{1 .. n\}} L_i 
}
\\
& \text{where:}
\\
& 
\begin{array}{ll}
M_i  = \np{\key{\loc_i}{\p_i}}{ \conf{\inact}{P_i\subst{\ep{s}{\p_i}}{x_i}}} \text{~for $i = 1..n$} 
&
L_1  = \myloc{\loc_1}{\bout{a}{x_1:T_1}{P_{1}}} 
\\
N_i  = \hmoni{s_{\p_i}}{\past T_{i}}{x_i}{\upd{x_i}{a }} \text{~for $i = 1..n$} 
&
L_i  = \myloc{\loc_i}{\binp{a}{x_i:T_i}{P_{i}}} \text{~for $i = 2..n$} 
\end{array}
\end{align*}
\begin{align*}
& \inferrule[\bkcolor{(RollS$^*$)}]
{ 
}{
N_1^{\normark} \Par N_2^{\normark} 
\Par 
M
~ \bkalt 
N_1^{\rmark} \Par N_2^{\rmark} 
\Par 
M  }
\\
& \text{where:}
\\
& \begin{array}{ll}
M  = \prod_{i \in \{1 .. n\}} \codah{s_{\p_i}}{h_i}{k_i}{}
&
N_1  = 
\moni{s_\p}{ \myctxr{\ctx T}{\ltinpp{\q}{U}{T}}}{\mytilde x}{\store_1} 
\\
& N_2  = \moni{s_\q}{ \myctxr{\ctx S}{\ltoutp{\p}{U}{S}}}{\mytilde y}{\store_2}  
\end{array}
\end{align*}
\begin{align*}
& \inferrule[\bkcolor{(ROut$^*$)}]
{\p = \er \,\vee\, \p \in \names{\er, h}
}{
M \Par N \Par \codah{s_{\p}}{h}{k \cons \langle \q,\myeval{V}{\sigma}\rangle }{}
~\bkalt 
M' \Par N'  \Par \codah{s_{\p}}{h}{k}{}  
}
\\
& \text{where:}
\\
& \begin{array}{ll}
M  = \np{\key{\loc}{\er}}{\conf{\stack{C}}{P}}
&
N  =\monir{s_\p}{\myctxr{\ctx{T}}{ \ltoutp{\q}{U}{S}}}{\mytilde x}{\store}
\\
M'  = \np{\key{\loc}{\er}}{ \conf{\stack{C}}{\bout{\ep{s}{\p}}{V}{P}}}
&
N'  = \hmoni{s_\p}{\myctxr{\ctx{T}}{\past  \ltout{\q}{U}{S}}}{\mytilde x}{\store}
\end{array}
\\[2mm]
& \inferrule[\bkcolor{(RIn$^*$)}]
{\p = \er \,\vee\, \p \in \names{\er, h_1}
}{
\begin{array}{c}
M \Par N 
\Par  \codah{s_{\p}}{h_1 \cons \langle \q,V\rangle }{k_1}{}
\Par\codah{s_{\q}}{h_2}{k_2}{}
\\
~ \bkalt 
M' \Par N'  
\Par  \codah{s_{\p}}{h_1}{k_1}{}
\Par\codah{s_{\q}}{h_2}{\langle \p,V\rangle  \cons k_2}{}
\end{array}
}
\\
& \text{where:}
\\
& \begin{array}{ll}
M  = \np{\key{\loc}{\er}}{\conf{\stack{C}}{P}} 
&
N  =\monir{s_\p}{ \myctxr{\ctx T} {\ltinpp{\q}{U}{S}} }{\mytilde x,y}{\store} 
\\
M'  = \np{\key{\loc}{\er}}{\conf{\stack{C}}{\binp{\ep{s}{\p}}{y}{P}}} 
&
N'  = \hmoni{s_\p}{ \myctxr{\ctx T}{\past\ltinp{\q}{U}{S}}}{\mytilde x}{\rup{\store}{y}}
\end{array}
\end{align*}
\end{mdframed}
\caption{An alternative decoupled semantics for configurations ($\fwalt$, $\bkalt$).}
\label{fig:fw_exp2}
\end{figure}


\subsection{An Alternative Semantics}
Our decoupled semantics for asynchronous communication, given by $\fw \cup \bk$, relies on a single (global) queue for all the participants in the  session. This is different from the semantics in works such as~\cite{HYC08,KY13,KY2015}, where there is a queue per channel/participant.
 Here we discuss an alternative decoupled semantics with a dedicated (local) queue per participant, and argue that our main results hold also in such an alternative semantics.

The syntax of processes and configurations, given in Fig.~\ref{fig:syntax}, is kept largely unchanged; we only need to consider multiple queues $\codah{s_{\p}}{h}{k}~$  for each participant $\p$ in the session (where $h$ is the input part $h$ and $k$ is the output part, as before). Intuitively, the communication of a message $m$ from $\p$ to $\q$ now operates as follows. First, the process implementation for $\p$ enqueues a message $\langle \q,m\rangle$ to the output part of its own queue; subsequently, the message is moved to the input part of $\q$'s queue, where it is renamed as $\langle \p,m \rangle $ to make its provenance explicit. 

More formally, the rules in Fig.~\ref{fig:fw_exp1} and Fig.~\ref{fig:fw_exp2} define the alternative evaluation-closed forward and backward reduction relations, denoted $\fwalt$ and $\bkalt$.\footnote{For  conciseness, Fig.~\ref{fig:fw_exp1} contains only rules for input-output communication; other rules follow similar lines.}
Rule \fwcolor{\textsc{(Init$^*$)}}
is similar to  Rule~\fwcolor{\textsc{(Init)}} in Fig.~\ref{fig:fw_1} but it initialises a local queue for each participant in the given protocol $G$.
Rules \fwcolor{\textsc{(Out$^*$)}}
and 
\fwcolor{\textsc{(In$^*$)}}
make the above intuitions  formal by defining how communication operates when using separate queues for participants, ensuring that output and input steps must be supported by appropriate types in the monitors for $\p$ and $\q$. 
In particular, notice how Rule~\fwcolor{\textsc{(In$^*$)}} involves the two different local queues (for $\p$ and for $\q$) and renames the participant mentioned in the message.
Rules \bkcolor{\textsc{(RInit$^*$)}},
\bkcolor{\textsc{(RollS$^*$)}}, 
\bkcolor{\textsc{(ROut$^*$)}}, 
and 
\bkcolor{\textsc{(RIn$^*$)}} are the corresponding backward rules, which are defined similarly as before because the tags 
`$\normark$' and `$\rmark$' are
attached to monitors, rather than to queues. The equivalence on queues given by Definition~\ref{eq:queue} needs to be revised as follows:

\begin{defi}[Equivalence on message queues, revised]\label{eq:queuemod}
We define the structural equivalence on queues, denoted $\equivq$, as follows:
\begin{align*}
h
\cons 
\langle \p_1,m_1\rangle
\cons 
\langle \p_2,m_2\rangle
\cons 
h'
\equivq 
h
\cons 
\langle \p_2, m_2\rangle
\cons 
\langle \p_1,m_1\rangle
\cons 
h'
\end{align*}
whenever $\p_1 \neq \p_2$.
The equivalence $\equivq$ extends  to configurations $M$ as expected.
\end{defi}

The difference between the two decoupled semantics is in the shape of the queues.  The following definition makes this difference precise.

\begin{defi}[From Single to Local Queues]\label{d:singletolocal}
Given $\codah{s}{h}{k}{}$, a single queue as defined in Fig.~\ref{fig:syntax}, the local queue of participant $\p$ is $\codah{s_{\p}}{h\lfloor \p }{k\rfloor \p}{}$, where:
	\begin{align*}
				h\lfloor \p &  =	\begin{cases}
			\emp & \text{if $h=\emp$}
			\\
			\langle \q_i, V_i \rangle \cons (h'\lfloor \p) & \text{if $h=  \valueq{\q_i}{\p'}{V_i} \cons h'$ and $\p = \p'$}
						\\ 
						h'\lfloor \p & \text{if $h=  \valueq{\q_i}{\p'}{V_i} \cons h'$ and $\p \not= \p'$}
		\end{cases}
		\\
		k\rfloor \p & = 
		\begin{cases}
			\emp & \text{if $k=\emp$}
			\\
			\langle \q_i, V_i \rangle \cons (k'\rfloor \p) & \text{if $k=  \valueq{\p'}{\q_i}{V_i} \cons k'$ and $\p = \p'$}
						\\ 
						k'\rfloor \p & \text{if $k=  \valueq{\p'}{\q_i}{V_i} \cons k'$ and $\p \not= \p'$}
		\end{cases}
	\end{align*}
	
	\noindent 
Given  a reachable configuration $M$ (\defref{d:ic}) for a protocol with participants $\{\p_1,\cdots,\p_n\}$, we write $\singleQ{M}$ to denote the configuration obtained from $M$ by (i)~removing the single queue $\codah{s}{h}{k}{}$ and (ii)~adding in parallel a local queue $\codah{s_{\p_i}}{h\lfloor \p_i }{k\rfloor  \p_i}{}$ for each $\p_i \in \{\p_1,\cdots,\p_n\}$.
\end{defi}

\begin{example}
Recall configuration $M_7$ as in \S\,\ref{sss:examp-tb}:
\begin{align*}
M_7 = ~ &   \news{s}(\,  \np{\key{\loc_3}{\pB}}{ \conf{\inact}{\bbout{\epB}{\thunkp{\bout{\epB}{\text{`Urbino, 61029'}}\binp{\epB}{d}\inact}}\inact}} 
\\
& \Par \hmoni{s_\pB}{\myctxr{\ctx{T}_7}{\past \ltout{C}{\thunkt}{\ltout{V}{\mathsf{address}}{\ltinp{V}{\mathsf{date}}{\lend}}}}}{z,p,s}{\store_7} 
\\
& \Par \np{\key{\loc_4}{\pC}}{ \conf{\inact}{\binp{\epC}{code}(\appl{code}{\dummyn})}} 
\\
& \Par 
\hmoni{s_\pC}{\myctxr{\ctx{T}_8}{\past \ltinp{B}{\thunkt}{\lend}}}{w,s}{\store_8} 
\Par N_5 
\\
& \Par \codah{s}{h_7}{\emp}{}~) 
\end{align*}
where 
$\myctxr{\ctx{T}_7}{\bullet}$, $\store_7$, $\myctxr{\ctx{T}_8}{\bullet}$, $\store_8$,
are as before and $h_7$ is as follows:
\begin{align*}
h_7 & = 
\valueq{\pA}{\pS}{\exBook}
\cons
\valueq{\pS}{\pA}{\mathit{price}(\exBook)}
\cons
\valueq{\pS}{\pB}{\mathit{price}(\exBook)}
\\
& \cons
\valueq{\pA}{\pB}{120}
\cons
\valueq{\pB}{\pA}{\text{`ok'}}
\cons
\valueq{\pB}{\pS}{\text{`ok'}}
\cons
\valueq{\pB}{\pC}{120}
\end{align*}
Then, applying \defref{d:singletolocal}, we obtain:
\begin{align*}
\singleQ{M_7}~ = ~ &   \news{s}(\,  \np{\key{\loc_3}{\pB}}{ \conf{\inact}{\bbout{\epB}{\thunkp{\bout{\epB}{\text{`Urbino, 61029'}}\binp{\epB}{d}\inact}}\inact}} 
\\
& \Par \hmoni{s_\pB}{\myctxr{\ctx{T}_7}{\past \ltout{C}{\thunkt}{\ltout{V}{\mathsf{address}}{\ltinp{V}{\mathsf{date}}{\lend}}}}}{z,p,s}{\store_7} 
\\
& \Par \np{\key{\loc_4}{\pC}}{ \conf{\inact}{\binp{\epC}{code}(\appl{code}{\dummyn})}} 
\\
& \Par 
\hmoni{s_\pC}{\myctxr{\ctx{T}_8}{\past \ltinp{B}{\thunkt}{\lend}}}{w,s}{\store_8} 
\Par N_5 \\
& \Par \codah{s_\pA}{h_\pA}{\emp}{} \Par \codah{s_\pB}{h_\pB}{\emp}{} \Par \codah{s_\pC}{h_\pC}{\emp}{} \Par \codah{s_\pS}{h_\pS}{\emp}{}~) 
\end{align*}
where the difference with respect to $M_7$ is in the last line, with the following local queues:
\begin{align*}
h_\pA & = 
\langle \pS, \mathit{price}(\exBook) \rangle 
 \cons
\langle \pB, \text{`ok'} \rangle 
\\
h_\pB & = 
\langle \pS, \mathit{price}(\exBook) \rangle 
 \cons
\langle \pA, \text{`ok'}\rangle 
\\
h_\pC & = 
\langle  \pB, 120 \rangle 
\\
h_\pS & = 
\langle  \pA, \exBook \rangle \cons 
\langle \pB, \text{`ok'} \rangle 
\end{align*}
\end{example}
We now relate the decoupled semantics (given by $\fw \cup \bk$) with the alternative decoupled semantics (given by $\fwalt \cup \bkalt$)  by means of the following correspondence:
\begin{prop} Let $M$ be a reachable configuration. We have:
\begin{enumerate}
	\item $M \fw M'$ if and only if $\singleQ{M} \fwalt \singleQ{M'}$.
	\item $\singleQ{M} \bk \singleQ{M'}$ if and only if $M \bkalt M'$.
\end{enumerate}	
\end{prop}

\begin{proof}
	Immediate from the definitions of $\singleQ{\cdot}$, 
	$\fw$, and $\fwalt$ (resp. $\bk$ and $\bkalt$).
\end{proof}

This tight correspondence between the original and alternative decoupled semantics ensures that our main results (in particular, causal consistency) carry over to a setting in which each participant handles its own queue for messages.

\subsection{Related Work}
Reversibility in concurrency has received much attention in the last decade. 
A detailed overview of the literature on the intersection between reversibility and 
behavioral contracts/types appears in~\cite{wg2}. 
Within this research line, 
the works most related to ours are~\cite{DBLP:conf/rc/TiezziY16,DBLP:journals/corr/Dezani-Ciancaglini16a,CastellaniDG17,CastellaniDG19,FrancalanzaMT18,FrancalanzaMT20,BarbaneraLd18,NeykovaY17,Kapus-Kolar20}.

Tiezzi and Yoshida~\cite{DBLP:conf/rc/TiezziY16} study
the cost of implementing 
different ways of reversing binary and multiparty sessions; 
since they work in a \emph{synchronous} setting, 
these alternatives are simpler or incomparable to our asynchronous, decoupled rollback.

In a series of works,  Dezani-Ciancaglini et al. have developed multiparty session types with \emph{checkpoints}~\cite{DBLP:journals/corr/Dezani-Ciancaglini16a,CastellaniDG17,CastellaniDG19}. 
These checkpoints are choice points in the global protocol to which the computation may return. The initial theory has been presented in \cite{DBLP:journals/corr/Dezani-Ciancaglini16a} and further developed in~\cite{CastellaniDG17}, with improvements that include a more liberal syntax for processes and types and refined representations for past communications.
 {The work in~\cite{CastellaniDG19} extends~\cite{CastellaniDG17} with {flexible choices} and connecting communications, allowing for different sets of participants in each branch. The intuition is that in some parts of the protocol
(as delimited by a choice construct) some participants are required to take part in
the interaction, while some others may be optional. }

We briefly compare our approach  with respect to the framework in~\cite{CastellaniDG17}. 
While our reversible actions are embedded in/guaranteed by the semantics, rollbacks in~\cite{CastellaniDG17} should specify the name of the checkpoint to which computation should revert.
Defining reversibility in~\cite{CastellaniDG17} requires modifying both processes and types. 
In contrast, we consider \emph{untyped} processes governed by local types (with cursors) as monitors. 
While we show causal consistency with a direct proof, in~\cite{CastellaniDG17} causal consistency follows indirectly, as a consequence of typing. 
Another difference with respect to~\cite{CastellaniDG17} is that reversibility in our model is fine-grained, in that we allow reversible actions concerning exactly two of the protocol participants; in~\cite{CastellaniDG17}, when a checkpoint is taken, also parties not related with that choice are forced to return to a checkpoint. 
A distinctive aspect of~\cite{CastellaniDG17} is that when a branch of a choice is reversed, it is discarded: this way, the same choice is not redone in the future.
We have chosen to be more liberal, as the same action can be done and redone infinite times. 
To encode this behavior, we could use the \emph{reversibility modes}---annotations that describe the reversibility capabilities of the processes governed by types)---that we introduced in~\cite{MezzinaP17}.


Similarly to our work, Francalanza et al.~\cite{FrancalanzaMT18,FrancalanzaMT20} use monitors that enact reversibility by storing the decision points (e.g., distributed choices) that the different participants may take, and by coordinating with each other in order to bring the system back  to a previous consistent state when certain conditions are met. To do so, they extend the \emph{global graphs} by Tuosto and Guanciale~\cite{TuostoG18} (a formalism that expresses the behaviour of a message passing system from a global point of view) with a decoration on choices, which includes a condition dictating when a computation on a particular branch of a distributed choice should be reversed. 
Then, these extended global descriptions can be used to (i)~synthesise actors implementing the normal (forward) local behaviour of the system prescribed by the global graph, but also (ii)~synthesise monitors that are able to coordinate a  distributed rollback when certain conditions (denoting abnormal behaviour) are met. 
Their synthesis algorithm produces Erlang code, with two actors per participant derived from the  global graph: one implements the normal/forward behaviour of the participant; the second one (a monitor) implements its backward behaviour.
Reversibility is confined into distributed choices, and triggered by conditions on the internal state of some participants. 
In contrast, in our framework every communication step can be undone;  the mechanism devised in~\cite{FrancalanzaMT18,FrancalanzaMT20} can be encoded in an extension of our framework with  conditional reversibility.

Neykova and Yoshida~ \cite{NeykovaY17} develop a recovery algorithm for Erlang programs by exploiting causal information induced by global protocols. They then show that their recovery algorithm outperforms Erlang's   built-in recovery algorithm. 

Behavioural contracts are abstract descriptions of expected communication patterns followed by either clients or servers during their interaction. They come naturally equipped with a notion of compliance: when a client and a server follow compliant contracts, their interaction is guaranteed to progress or successfully complete. Barbanera et al.~\cite{BarbaneraLd18} study two extensions of behavioural contracts: (i)~retractable contracts dealing with backtracking and (ii)~speculative contracts dealing with speculative execution. These two extensions give rise to the same notion of compliance. As a consequence, they also give rise to the same subcontract relation, which determines when one server can be replaced by another while preserving compliance.


 From a global specification (e.g., a global graph) an abstract semantics can be derived \cite{GuancialeT19}. The semantics is abstract since it is given in terms of a partial order of events, representing the causality induced by the global specification. 
 Recently, Kapus-Kolar~\cite{Kapus-Kolar20} has enhanced this abstract semantics to account for reversibility.
  Here, the assumption  is that  a global graph is realizable (i.e., the projection function can be defined on all participants) and that each automata implementing the behaviour of a single participant has an inverse. Then, it is shown that the causality induced by the global specification is preserved also while going backwards.  

\section{Concluding Remarks}\label{s:conc}
We have presented a process framework of reversible, multiparty protocols built upon session-based concurrency. 
As illustrated throughout the paper,
the distinguishing features of 
our framework 
(decoupled rollbacks and abstraction passing, including delegation) endow it with substantial 
expressiveness, improving on and distinguishing it from prior works.

Our processes/configurations are untyped, but their (reversible) behaviour is governed by monitors derived from local (session) types.
In our view, our monitored approach to reversibility  
is particularly appropriate for specifying and reasoning about systems with 
components whose behaviour may not be statically analyzed (e.g., legacy components or services available as black-boxes). 
A monitored approach is general enough to support also the analysis of reversible systems that combine typed and
untyped components.

We proved that our reversible semantics is \emph{causally consistent}, which ensures that reversing a computation leads  to a  state that could have been reached by performing only forward steps.
The proof is challenging (and, in our view, also interesting), as we must resort to an alternative atomic semantics for rollbacks (Fig.~\ref{fig:atom_fwd} and~\ref{fig:atom_bk}). 
We then connected reversibility at the level of process/configurations with reversibility at the level of global types, therefore linking the operational and declarative levels of abstraction typical of multiparty sessions communication-centric software systems.

\subsubsection*{Extensions and Future Work}
As already mentioned, our framework does not include name passing, which is known to be representable, in a fully abstract way, using 
name abstractions~\cite{KPY2016}. Primitive support for name passing is not difficult, but would come at the cost of additional notational burden. 
An extension with name passing would allow us to relate our framework with known
typed frameworks for monitored networks (without reversibility) based on multiparty sessions~\cite{DBLP:journals/tcs/BocchiCDHY17}.

In future work, we plan to 
 extend our framework with  {reversibility modes}~\cite{MezzinaP17},
which implement \emph{controlled reversibility} \cite{LaneseMSS11} by specifying 
 how many times a particular protocol step can be reversed---zero (e.g., it is an atomic action), one, or infinite times.
 (Currently, all actions can be reversed infinite times.) 
 In a related vein, we plan to explore variants of our model in which certain protocol branches are ``forgotten'' after they have been reversed; this modification is delicate, because it would weaken the notion of causal consistency. 
 
 On the practical side,  the work~\cite{VriesP18} describes 
 a  {Haskell} implementation of our reversible model, where algebraic types are used to represent all the various formal ingredients defined in \secref{sec:calculus}. We plan to keep improving this implementation, as we believe that pure functional languages support natively reversibility. 
 In this direction, it would be beneficial to have a ``reversible workbench'' to test and compare all the different semantics of the aforementioned reversible behavioural types.

\subsubsection*{Acknowledgements}
  We are grateful to Peter D. Mosses for useful exchanges and suggestions.
  We would also like to thank the anonymous reviewers for their valuable suggestions and constructive feedback. \\
    Mezzina has been partially supported by
   the French ANR project DCore ANR-18-CE25-0007 and by the Italian INdAM - GNCS 2020 project `Sistemi Reversibili Concorrenti: dai Modelli ai Linguaggi'.
  P\'{e}rez has been partially supported by the Dutch Research Council (NWO) under project No. 016.Vidi.189.046 (Unifying Correctness for Communicating Software).



%

\bibliographystyle{alpha}
\bibliography{session,biblio}


\end{document}